\documentclass[11pt]{iopart}

\usepackage{graphicx,color,rotating,pifont}
\usepackage{amssymb,bm}
\usepackage{iopams}
\usepackage{ae}
\usepackage{pstricks}
\usepackage{array}
\usepackage[numbers,sort&compress]{natbib}
\usepackage{url}
\expandafter\let\csname equation*\endcsname\relax
\expandafter\let\csname endequation*\endcsname\relax
\usepackage{amsmath}

\allowdisplaybreaks

\DeclareMathSymbol{\NS}{\mathord}{AMSb}{"4E}

\newcommand{\ket}[1]{\ensuremath{\,|{#1}\rangle}}
\newcommand{\braket}[2]{\ensuremath{\langle{#1}|{#2}\rangle}}
\newcommand{\matrixe}[3]{\ensuremath{\langle{#1}|\,{#2}\,|{#3}\rangle}}

\newcommand{\dmatrixe}[2]{\matrixe{#1}{#2}{#1}}

\newcommand{\sgn}{\mathrm{sgn}\,}

\newcommand{\comm}[2]{\ensuremath{[{#1},{#2}]}}
\newcommand{\acomm}[2]{\ensuremath{ \big\{ {#1}, {#2} \big\} }}

\newcommand{\op}[1]{\ensuremath{#1}}
\newcommand{\adj}[1]{\ensuremath{{{#1}}^{\dag}}}

\newcommand{\totd}[2]{\ensuremath{ \frac{d {#1}} {d {#2}} }}

\newcommand{\nord}[1]{\ensuremath{:\!#1:}}

\newcommand{\aO}{\ensuremath{\op{a}}}

\newcommand{\alphaO}{\ensuremath{\op{\alpha}}}

\newcommand{\etaO}{\ensuremath{\op{\eta}}}

\newcommand{\aaO}{\ensuremath{\adj{\op{a}}}}

\newcommand{\aalphaO}{\ensuremath{\adj{\op{\alpha}}}}

\newcommand{\AO}{\ensuremath{\op{A}}}
\newcommand{\HO}{\ensuremath{\op{H}}}
\newcommand{\NO}{\ensuremath{\op{N}}}
\newcommand{\OO}{\ensuremath{\op{O}}}
\newcommand{\PO}{\ensuremath{\op{P}}}
\newcommand{\TO}{\ensuremath{\op{T}}}
\newcommand{\UO}{\ensuremath{\op{U}}}

\newcommand{\ZO}{\ensuremath{\op{Z}}}

\newcommand{\UUO}{\ensuremath{\adj{\op{U}}}}
\newcommand{\ZZO}{\ensuremath{\adj{\op{Z}}}}

\newcommand{\kOV}{\ensuremath{\vec{\op{k}}}}
\newcommand{\pOV}{\ensuremath{\vec{\op{p}}}}
\newcommand{\qOV}{\ensuremath{\vec{\op{q}}}}

\renewcommand{\AC}{\ensuremath{\mathcal{A}}}

\newcommand{\OC}{\ensuremath{\mathcal{O}}}
\newcommand{\PC}{\ensuremath{\mathcal{P}}}

\newcommand{\nn}{\ensuremath{\bar{n}}}

\newcommand{\Trel}{\ensuremath{\TO_\text{rel}}}

\newcommand{\Hint}{\ensuremath{\HO_\text{int}}}

\newcommand{\Rch}{\ensuremath{R_\text{ch}}}
\newcommand{\Hfinal}{\ensuremath{\mkern 3mu\overline{\mkern-3muH}}}

\newcommand{\Vlowk}{\ensuremath{V_{\text{low-k}}}}

\newcommand{\nuc}[2]{\ensuremath{^{#2}\mathrm{#1}}}

\newcommand{\fm}{\ensuremath{\,\text{fm}}}
\newcommand{\fmi}{\ensuremath{\,\text{fm}^{-1}}}

\newcommand{\keV}{\ensuremath{\,\text{keV}}}
\newcommand{\MeV}{\ensuremath{\,\text{MeV}}}
\newcommand{\MeVi}{\ensuremath{\,\text{MeV}^{-1}}}

\newcommand{\GeV}{\ensuremath{\,\text{GeV}}}

\newcommand{\lambdaSRG}{\ensuremath{\lambda}}

\newcommand{\hw}{\ensuremath{\hbar\omega}}
\newcommand{\eMax}{\ensuremath{e_{\text{max}}}}

\newcommand{\EMax}{\ensuremath{E_{3\text{max}}}}
\newcommand{\Nmax}{\ensuremath{N_\text{max}}}

\newcommand{\NNNLO}{N$^3$LO}
\newcommand{\NNLO}{NNLO}
\newcommand{\NNLOsat}{$\text{NNLO}_\text{sat}$}

\definecolor{FGViolet}{rgb}{0.61,0.32,0.61}
\definecolor{FGDarkBlue}{rgb}{0,0,0.6}
\definecolor{FGBlue}{rgb}{0,0,0.8}
\definecolor{FGLightBlue}{rgb}{0.2, 0.6, 0.8}
\definecolor{FGGreen}{rgb}{0.2,0.7,0.2}
\definecolor{FGLightGreen}{rgb}{0.4,1,0.4}
\definecolor{FGYellow}{rgb}{1,0.95,0}
\definecolor{FGOrange}{rgb}{0.95,0.5,0.1}
\definecolor{FGRed}{rgb}{0.8,0,0}
\definecolor{FGWhite}{rgb}{1,1,1}
\definecolor{FGLightGray}{rgb}{0.8,0.8,0.8}
\definecolor{FGGray}{rgb}{0.5,0.5,0.5}
\definecolor{FGDarkGray}{rgb}{0.3,0.3,0.3}
\definecolor{FGBlack}{rgb}{0,0,0}

\begin{document}
\title{In-Medium Similarity Renormalization Group for Closed and Open-Shell Nuclei}

\author{H.~Hergert}
\address{NSCL/FRIB Laboratory and Department of Physics \& Astronomy, Michigan State University, East Lansing, MI 48824, United States}
\ead{hergert@nscl.msu.edu}

\begin{abstract}
We present a pedagogical introduction to the In-Medium Similarity Renormalization
Group (IMSRG) framework for \emph{ab initio} calculations of nuclei. The IMSRG
performs continuous unitary transformations of the nuclear many-body Hamiltonian
in second-quantized form, which can be implemented with polynomial computational
effort. Through suitably chosen generators, it is possible to extract eigenvalues
of the Hamiltonian in a given nucleus, or drive the Hamiltonian matrix 
in configuration space to specific structures, e.g., band- or block-diagonal form. 

Exploiting this flexibility, we describe two complementary approaches for the
description of closed- and open-shell nuclei: The first is the Multireference 
IMSRG (MR-IMSRG), which is designed for the efficient calculation of nuclear
ground-state properties. The second is the derivation of nonempirical valence-space
interactions that can be used as input for nuclear Shell model (i.e., configuration 
interaction (CI)) calculations. This IMSRG+Shell model approach provides immediate 
access to excitation spectra, transitions, etc., but is limited in applicability 
by the factorial cost of the CI calculations.

We review applications of the MR-IMSRG and IMSRG+Shell model approaches
to the calculation of ground-state properties for the oxygen, calcium, and nickel 
isotopic chains or the spectroscopy of nuclei in the lower $sd$ shell, respectively,
and present selected new results, e.g., for the ground- and excited state properties 
of neon isotopes.
\end{abstract}

\submitto{\PS}

\section{Introduction\label{sec:intro}}
More than 60 years have passed since Rainwater, Bohr, and Mottelson published
the seminal works that led to them winning the 1975 Nobel Prize in Physics
\cite{Rainwater:1950ly,Bohr:1951eu,Bohr:1953fy,Bohr:1953pd,Bohr:1953qd}. The 
collective model developed in these publications is an essential tool 
for nuclear physicists, and it qualifies as one of the most successful 
data-driven approaches to nuclear structure. A variety of approaches
exist in this category, ranging from local to global applicability,
and from microscopic to macroscopic views of nuclei: Collective
Bohr-Mottelson Hamiltonians are fine-tuned to specific nuclei or
limited regions of the nuclear chart, and do not treat the dynamics
of all nucleons on a fully microscopic level. The traditional nuclear 
configuration interaction (CI) approach (see, e.g., \cite{Brown:2001rg,Caurier:2005qf}) 
uses phenomenological interactions that are highly optimized, 
e.g., to $sd$-shell data \cite{Brown:2006fk}, and treats only the dynamics
of valence nucleons on top of an inert core in fully microscopic fashion.
Finally, nuclear Density Functional Theory (DFT) takes a global perspective,
and aims for a microscopic description of the entire nuclear chart based
on energy density functionals (EDFs) that are optimized to data 
\cite{Erler:2012dz,Kortelainen:2010ad,Kortelainen:2012fu,McDonnell:2015di}.

The philosophy behind data-driven models is complementary to that of 
\emph{ab initio} nuclear many-body theory, although the lines are somewhat
blurry. An \emph{ab initio} approach attempts to describe nuclear
structure and dynamics based on fundamental degrees of freedom and their
interactions. In the Standard Model, the fundamental theory of strong
interactions is Quantum Chromodynamics (QCD), but a description of nuclear 
observables on the level of quarks and gluons is not feasible, 
except for the lightest few-nucleon systems (see, e.g., \cite{Detmold:2015xw}).
Instead, we start from nuclear interactions that describe low-energy QCD observables in the $NN$ and $3N$
systems, like scattering data or binding energies. Nowadays, such interactions 
are derived in Chiral Effective Field Theory (EFT),
which provides a constructive framework and organizational hierarchy
for $NN$, $3N$, and higher many-nucleon forces, as well as consistent
electroweak operators (see, e.g., \cite{Epelbaum:2009ve,Machleidt:2011bh,Epelbaum:2015gf,Entem:2015qf,Entem:2015hl,Gezerlis:2014zr,Lynn:2016ec,Pastore:2009zr,Pastore:2011dq,Piarulli:2013vn,Kolling:2009yq,Kolling:2011bh}).
Since Chiral EFT is a low-momentum expansion, high-momentum (short-range)
physics is not explicitly resolved by the theory, but parametrized
by the so-called low-energy constants (LECs). 

In principle, the LECs can be determined by matching calculations of
the same observables in chiral EFT and (Lattice) QCD in the overlap
region of the two theories. Since such a calculation is currently not 
feasible, they are
fit to experimental data, typically in the $\pi{}N$, $NN$, and $3N$ sectors. 
Recently, Ekstr\"om \emph{et al.} have developed an optimization protocol
for chiral interactions that gives up on the reductionist approach of
fixing the LECs in the few-nucleon system, and includes certain many-body
data in the fit as well. The many-body data, e.g., selected radii, are 
chosen in order to improve the deficient saturation behavior of chiral interactions 
that are used as input for nuclear many-body calculations. The first interaction
optimized with this protocol is \NNLOsat{} \cite{Ekstrom:2015fk}, which is
able to accurately describe the ground-state energies and charge radii of 
$\nuc{Ca}{40,48}$ at the same time. Following 
the same philosophy but not the same approach, Shirokov \emph{et al.} have 
produced Daejeon16, a softened chiral $NN$ interaction that has been tuned 
for the description of light nuclei without explicit $3N$ forces 
\cite{Shirokov:2004fe,Shirokov:2005bv,Shirokov:2007by,Shirokov:2016wo}.

Renormalization group (RG) methods are natural companions for
EFTs, because they make it possible to smoothly connect theories with 
different resolution scales and degrees of freedom. Since they 
were introduced in low-energy nuclear physics around the start 
of the millennium \cite{Bogner:2003os,Bogner:2007od,Bogner:2010pq,Furnstahl:2013zt}, 
they have provided a systematic framework for formalizing many ideas on the 
renormalization of nuclear interactions and many-body effects that had been 
discussed in the nuclear structure community since the 1950s. For instance,
soft and hard-core $NN$ interactions can reproduce scattering data equally
well, but have significantly different saturation properties, which caused
the community to move away from the former in the 1970s (see, e.g., \cite{Bethe:1971qf}).
What was missing at that time was the recognition of the intricate link
between the off-shell $NN$ interaction and $3N$ forces that was formally
demonstrated for the first time by Polyzou and Gl\"ockle in 1990 \cite{Polyzou:1990fk}.
From the modern RG perspective, soft- and hard-core interactions emerge
as representations of low-energy QCD at different resolution scales, 
and the dialing of the resolution scale necessarily leads to induced
$3N$ forces, in such a way that observables (including saturation
properties) remain invariant under the RG flow (see section \ref{sec:srg} 
and \cite{Bogner:2010pq,Furnstahl:2013zt}). In conjunction, chiral EFT
and nuclear RG applications demonstrate that one cannot treat the $NN, 3N, \ldots$
sectors in isolation from each other.

During the 1960s, Kuo and Brown pioneered work on the \emph{ab initio} 
derivation of effective interactions for nuclear valence-space CI calculations,
culminating in the publication of Hamiltonians for the $sd$ and
$pf$ shells \cite{Kuo:1967qf,Kuo:1968ty}. Their approach relied on
Brueckner's $G$ matrix to treat short-range correlations induced
by the free-space $NN$ interaction, and employed the so-called hole-line
expansion to second order \cite{Brueckner:1954qf,Brueckner:1955rw,Day:1967zl}.
After some initial successes, Barret, Kirson, and others demonstrated
a lack of order-by-order convergence of this expansion 
\cite{Barrett:1970jl,Kirson:1971la,Barrett:1972bs,Kirson:1974oq,Goode:1974pi},
and Vary, Sauer and Wong found a disturbingly strong 
model-space dependence in intermediate-state summations, with larger
model spaces actually degrading the agreement with experimental data 
\cite{Vary:1973dn}. Bogner \emph{et al.} revisited this issue with
the help of the Similarity Renormalization Group (SRG) \cite{Glazek:1993il,Wegner:1994dk}, 
and demonstrated that the $G$ matrix retains significant coupling between low- 
and high-momentum
nodes of the underlying interaction \cite{Bogner:2010pq}, so the convergence
issues are not surprising from a modern perspective. In the SRG and other
modern RG approaches, low- and high-momentum physics are decoupled properly,
and the resulting low-momentum $NN+3N$ interactions are indeed perturbative
 \cite{Bogner:2006qf,Bogner:2010pq}. For such interactions, results from 
finite-order many-body perturbation theory (MBPT) are in good agreement 
with non-perturbative results if the expansion is based on a Hartree-Fock 
reference state \cite{Tichai:2016vl,Roth:2010ys}.

Of course, low-momentum $NN+3N$ interactions are well-suited inputs not
just for MBPT, but for all methods that work in truncated configuration
spaces. The decoupling of low- and high-momentum modes of the interaction
leads to a greatly improved convergence behavior, which in turn extends 
the range of nuclei a many-body method can be applied to. With SRG-evolved
interactions, the No-Core Shell Model (NCSM) and related large-scale 
diagonalization methods can be extended into the lower $sd-$shell 
\cite{Barrett:2013oq,Jurgenson:2013fk,Hergert:2013ij,Roth:2014fk}, and 
methods with systematic many-body truncations like Coupled Cluster (CC)
are nowadays applied to nuclei as heavy as tin \cite{Binder:2014fk,Hagen:2014ve,Hagen:2016rb}.

Instead of merely using the SRG as a tool to ``pre-process'' the nuclear
interactions that are used as inputs for other many-body methods, we can 
turn it into a method for solving the many-body Schr\"odinger 
equation itself. This leads us to the so-called In-Medium SRG (IMSRG), which is
the main focus of the present work \cite{Tsukiyama:2011uq,Hergert:2013mi,Hergert:2016jk}. 
In a nutshell, we want to use SRG-like flow equations to 
decouple physics at different excitation energy scales of the nucleus, 
and render the Hamiltonian matrix in configuration space block or 
band diagonal in the process. This can also be viewed as a re-organization 
of the many-body expansion, in which correlations that are described 
explicitly by the configuration space are absorbed into an 
\emph{RG-improved} Hamiltonian. With an appropriately chosen decoupling 
strategy, it is even possible to extract eigenvalues and eigenstates
of the nuclear Hamiltonian, and therefore, the IMSRG qualifies as an 
\emph{ab initio} method for solving quantum many-body problems.

In this work, we will discuss two distinct implementations of the 
IMSRG ideas. The first is the so-called Multireference IMSRG (MR-IMSRG),
which is designed for calculations of the ground-state properties of 
closed- and open-shell nuclei.  Like most many-body
approaches, it relies on the organization of the many-body basis
in terms of a reference state and its excitations. 
Contrary to approaches like MBPT or CI, which employ Slater
determinant reference states, the MR-IMSRG is built for arbitrary
correlated reference states. This gives us the greatest possible
flexibility in the description of correlations: Static correlations,
e.g., due to intrinsic deformation, can be built into the reference
state, while dynamic correlations due to the excitation of nucleon pairs, 
triples, etc. are described by the MR-IMSRG transformation. 

The second approach uses the IMSRG to construct RG-improved
Hamiltonians for nuclear valence CI calculations. These interactions
are subsequently used as input for existing Shell model codes
like \texttt{NuShellX} \cite{Brown:2014fk}. Such a combined approach
gives us immediate access to a much larger number of observables than
the MR-IMSRG, but it is limited by the computational effort of the
CI part of the calculation.

The idea of using flow equations to solve quantum many-body problems
was already discussed in Wegner's initial work on the SRG \cite{Wegner:1994dk} 
(also see \cite{Kehrein:2006kx} and references therein). In the 
solid-state physics literature, the approach is also known as 
continuous unitary transformation (CUT) theory, see  
\cite{Heidbrink:2002kx,Drescher:2011kx,Krull:2012bs,Fauseweh:2013zv,Krones:2015ft}.
When we discuss our decoupling strategies for the nuclear many-body
problem, it will become evident that the IMSRG is related to CC 
\cite{Shavitt:2009,Hagen:2014ve}, canonical transformation theory (CT) 
\cite{White:2002fk,Yanai:2006uq,Yanai:2007kx}, and the Irreducible (or
Anti-Hermitian) Contracted Schr\"odinger Equation (ICSE) approach 
\cite{Nakatsuji:1976yq,Valdemoro:1987zl,Mukherjee:2001uq,Kutzelnigg:2002kx,
Kutzelnigg:2004vn,Kutzelnigg:2004ys,Mazziotti:2006fk}, and there is
even some overlap with purely variational methods (see section \ref{sec:generators}).
What sets the IMSRG apart from these methods is that the Hamiltonian 
instead of the wave function is at the center of attention, in the
spirit of RG methodology. This seems to be a trivial distinction, but
there are practical advantages of this viewpoint, e.g., the simultaneous 
decoupling of ground and excited states (see section \ref{sec:decoupling}), 
the avoidance of $N$-representability issues \cite{Mazziotti:2007qe}, and
more. Inspired in part by our work on the IMSRG in nuclear physics, Evangelista 
and co-workers have recently presented the Driven SRG for \emph{ab initio}
calculations in quantum chemistry, which implements IMSRG transformations 
in terms of inhomogeneous nonlinear equations rather than flow equations 
\cite{Evangelista:2014rq,Li:2015nq,Li:2016rm,Hannon:2016lp}.

\subsubsection*{Organization of this work:}
Let us conclude our introduction of the IMSRG with a look ahead at 
the main body of this work. In section \ref{sec:srg}, we briefly review 
the essential concepts of the free-space SRG, and how it is used to dial the resolution 
scale of (nuclear) interactions and operators. To set up the IMSRG 
formalism, we first discuss normal-ordering techniques and Wick's theorem for arbitrary reference states in 
section \ref{sec:no}. This will be followed by the presentation of 
the MR-IMSRG flow equations in section \ref{sec:mrimsrg}, which also discusses 
the choice of decoupling scheme 
and generators. The single-reference IMSRG equations are obtained as a 
limit of the more general MR-IMSRG framework. In section \ref{sec:features}, 
general features of (MR-)IMSRG flows are discussed, and section 
\ref{sec:groundstate} reviews applications of the MR-IMSRG to the 
ground-state properties of closed- and open-shell nuclei. In section 
\ref{sec:sm}, we describe the derivation of nonempirical interactions for the nuclear 
valence-space CI approach. Salient points are summarized in
section \ref{sec:epilogue}, and we look ahead at new developments. Expressions
for products and commutators of normal-ordered operators are collected
in \ref{app:NO}, and \ref{app:PNP} recapitulates elements of
the Hartree-Fock Bogoliubov theory and particle-number projection,
which are used in the construction of reference states for the MR-IMSRG.

\section{\label{sec:srg}The Similarity Renormalization Group}

\subsection{General Concept}

The Similarity Renormalization Group (SRG) was first formulated independently by Wegner 
\cite{Wegner:1994dk} and Glazek and Wilson \cite{Glazek:1993il}, for applications in condensed 
matter physics and light-front quantum field theory, respectively. The general concept of 
the method is to simplify the structure of the Hamiltonian in a suitable representation through 
the use of a continuous unitary transformation, 
\begin{equation}\label{eq:cut}
  \HO(s)=\UO(s)\HO(0)\UUO(s)\,.
\end{equation}
Here, $H(s=0)$ is the starting Hamiltonian and $s$ denotes the so-called flow parameter, 
which parameterizes the unitary transformation. Taking the derivative of equation \eqref{eq:cut}
with respect to $s$, we obtain the operator flow equation
\begin{equation}\label{eq:opflow}
  \totd{}{s}\HO(s) = \comm{\etaO(s)}{\HO(s)}\,,
\end{equation} 
where the anti-Hermitian generator $\etaO(s)$ is related to $\UO(s)$ by
\begin{equation}
  \eta(s)=\totd{U(s)}{s}U^{\dag}(s) = -\eta^{\dag}(s)\,.
\end{equation}
By rearranging this relation, we obtain a differential equation for $\UO(s)$ whose formal solution 
is given by the \emph{path-}or \emph{S-ordered} exponential \cite{Dyson:1949fj,Blanes:2009fk}
\begin{equation}
  U(s) = \mathcal{S}\exp \int^s_0 ds' \eta(s')
       \equiv \sum_n \frac{1}{n!}\int^s_0 ds_1 \int^s_0 ds_2 \ldots 
          \int^s_0 ds_n \mathcal{S}\{\eta(s_1)\ldots\eta(s_n)\}\,.
\end{equation}
Here, the $S$-ordering operator $\mathcal{S}$ ensures that the flow parameters appearing 
in the integrands are always in descending order, $s_1 > s_2 > \ldots$.
Since our continuous unitary transformation preserves the spectrum of the Hamiltonian and 
any other observable of interest, it is an example of a so-called isospectral flow, a class 
of transformations which has been studied extensively in the mathematics literature (see, 
e.g., \cite{Brockett:1991kx,Chu:1994vn,Chu:1995ys}).

With a suitable choice of generator $\etaO(s)$, we can smoothly transform the Hamiltonian to 
almost arbitrary structures as we integrate the flow equation \eqref{eq:opflow} for $s\to\infty$.
Inspired by the work of Brockett \cite{Brockett:1991kx} on the so-called double-bracket flow, 
Wegner \cite{Wegner:1994dk} proposed the generator
\begin{equation}\label{eq:def_Wegner_general}
  \etaO(s) \equiv \comm{H_d(s)}{H_{od}(s)}\,,
\end{equation}
which is constructed by splitting the Hamiltonian into suitably chosen \emph{diagonal} ($H_d(s)$) 
and \emph{offdiagonal} ($H_{od}(s)$) parts. It can be shown analytically that the generator 
\eqref{eq:def_Wegner_general} will monotonically suppress $H_{od}(s)$ as the Hamiltonian is 
evolved via equation \eqref{eq:opflow} (see, e.g., \cite{Wegner:1994dk,Kehrein:2006kx,Hergert:2016jk}). 
Note that the label diagonal does not need to mean strict diagonality here, but rather refers
to a desired structure that the Hamiltonian will assume in the limit $s\to\infty$. By working
in bases that are ordered by momenta or energies, the capability to impose structure on the 
Hamiltonian allows us to make an explicit connection with renormalization group (RG) ideas.

\subsection{SRG Evolution of Nuclear Interactions}

\begin{figure*}[t]
  \begin{center}
    \includegraphics[width=0.9\textwidth]{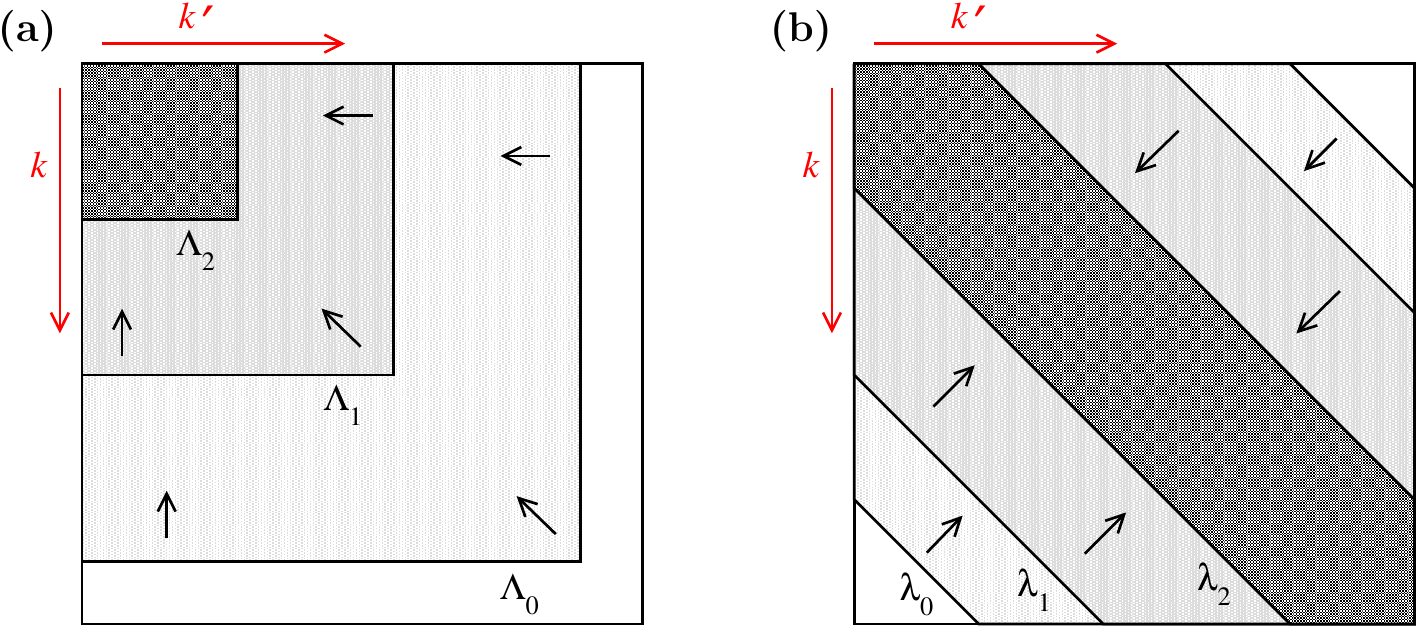}
  \end{center}
  \caption{\label{fig:schematic}Schematic illustration of two types of RG evolution 
    for $NN$ potentials in momentum space: (a) \Vlowk{} running in $\Lambda$, and 
    (b) SRG running in $\lambdaSRG$ (see main text). Here, $k$ and $k'$ denote the 
    relative momenta of the initial and final state, respectively. At each $\Lambda_i$ 
    or $\lambdaSRG_i$, the matrix elements outside of the corresponding 
    blocks or bands are negligible, implying that high- and low-momentum 
    states are decoupled.
  }
\end{figure*}

In figure \ref{fig:schematic}, we show schematic examples of RG evolutions that are applied to
nucleon-nucleon interactions in momentum-space representation. Figure \ref{fig:schematic}(a) 
implements the RG as a decimation: The interaction is evolved to decreasing
cutoff scales $\Lambda_0 > \Lambda_1 > \Lambda_2$, and we end up with a low-momentum interaction
\Vlowk{} that only has non-zero matrix elements between states with initial and final 
relative momenta $k,k'\leq\Lambda$ \cite{Bogner:2003os,Bogner:2010pq}. In contrast, 
figure \ref{fig:schematic}(b) results from a continuous unitary transformation via the flow 
equation \eqref{eq:opflow}, using a Wegner-type generator built from the relative kinetic 
energy in the two-nucleon system:
\begin{equation}\label{eq:def_srg_generator}
  \eta(\lambda) \equiv \Big[\,\frac{\kOV^2}{2\mu}, v(\lambda)\Big]\,.
\end{equation}
Here, $\kOV=\tfrac{1}{2}(\pOV_1-\pOV_2)$, and $\mu$ is the reduced nucleon mass. We have
parametrized the transformation with $\lambdaSRG=s^{-1/4}$, which has the 
dimensions of a momentum (in natural units). As suggested by figure \ref{fig:schematic}(b),
$\lambdaSRG$ is a measure for the ``width'' of the band-diagonal Hamiltonian in momentum 
space, i.e., it controls the scale of momentum transfers between nucleons. Because  
\begin{equation}
  |\qOV|=|\kOV' - \kOV|\lesssim \lambdaSRG
\end{equation}
low- and high-lying momenta are decoupled in a proper RG sense as $\lambdaSRG$ is 
decreased.

The decoupling of low- and high-lying momenta significantly improves the convergence
properties of configuration-space based many-body methods, because it prevents the 
Hamiltonian from scattering nucleon pairs from low to high momentum states. Methods 
like the NCSM or the IMSRG discussed below yield converged results in much smaller 
many-body Hilbert spaces, which in turn makes it possible to
apply these methods to heavier nuclei \cite{Roth:2011kx,Barrett:2013oq,Jurgenson:2013fk,Roth:2014fk,Hergert:2013ij,Hergert:2013mi,Hergert:2014vn,Hergert:2016jk,Hagen:2010uq,Roth:2012qf,Binder:2013zr,Binder:2014fk,Soma:2011vn,Soma:2013ys,Soma:2014fu,Soma:2014eu}.
 However, this improvement 
comes at a cost, which is best illustrated by considering the Hamiltonian in a 
second-quantized form, assuming only a two-nucleon interaction for simplicity:
\begin{equation}\label{eq:H}
  \Hint = \Trel + V = \frac{1}{4}\sum_{pqrs} \matrixe{pq}{\frac{\kOV_{12}^2}{2\mu}+v_{12}}{rs}\aaO_p\aaO_q\aO_s\aO_r\,.
\end{equation}
If we plug $\Trel$ and $V$ into the commutators in equations \eqref{eq:def_srg_generator} and 
\eqref{eq:opflow}, we obtain
\begin{equation}\label{eq:comm2B_vac}
  \comm{\aaO_i\aaO_j\aO_l\aO_k}{\aaO_p\aaO_q\aO_s\aO_r}=
  \delta_{lp}\aaO_i\aaO_j\aaO_q\aO_s\aO_r\aO_k + \{\aaO\aaO\aaO\aO\aO\aO\} 
  -\delta_{lp}\delta_{kq}\aaO_i\aaO_j\aO_s\aO_r + \{\aaO\aaO\aO\aO\}\,, 
\end{equation}
where the bracketed terms with suppressed indices schematically stand for additional two- and
three-body operators. Thus, even if we start from a pure two-body interaction, the SRG 
flow will induce operators of higher rank, i.e., three-, four-, and in general 
up to $A$-nucleon interactions. Of course, these induced interactions 
are only probed if we study an $A$-nucleon system. If we truncate the SRG flow equations
at the two-body level, we preserve the properties of the two-nucleon system, in particular
phase shifts and the deuteron binding energy. A truncation at the three-body level 
ensures the invariance of observables in $A=3$ nuclei, 
e.g.~$\nuc{H}{3}$ and $\nuc{He}{3}$ ground-state energies, and so on. Truncations in the 
SRG flow equation cause a violation of unitarity that manifests as
a (residual) dependence of many-body results on $\lambdaSRG$. By varying this parameter, 
the size of the missing contributions can be assessed (see, e.g., \cite{Bogner:2010pq,
Jurgenson:2009bs,Hebeler:2012ly,Roth:2011kx,Hergert:2013mi,Hergert:2013ij,Binder:2014fk,
Soma:2014eu,Griesshammer:2015dp}).

State-of-the-art SRG evolutions of nuclear interactions are nowadays performed in 
the three-body system, using relative (Jacobi) harmonic oscillator
\cite{Jurgenson:2009bs,Jurgenson:2011zr,Jurgenson:2013fk}, relative momentum plane
wave \cite{Hebeler:2012ly}, or momentum-space hypherspherical harmonics representations 
\cite{Wendt:2013uq}. Pioneering work on implementing the SRG evolution in the lowest partial 
waves of the four-body system has been carried out by A.~Calci and 
co-workers \cite{Calci:2014xy}, again working in Jacobi HO representation.

\begin{figure*}[t]
  \begin{center}
    \includegraphics[width=0.95\textwidth]{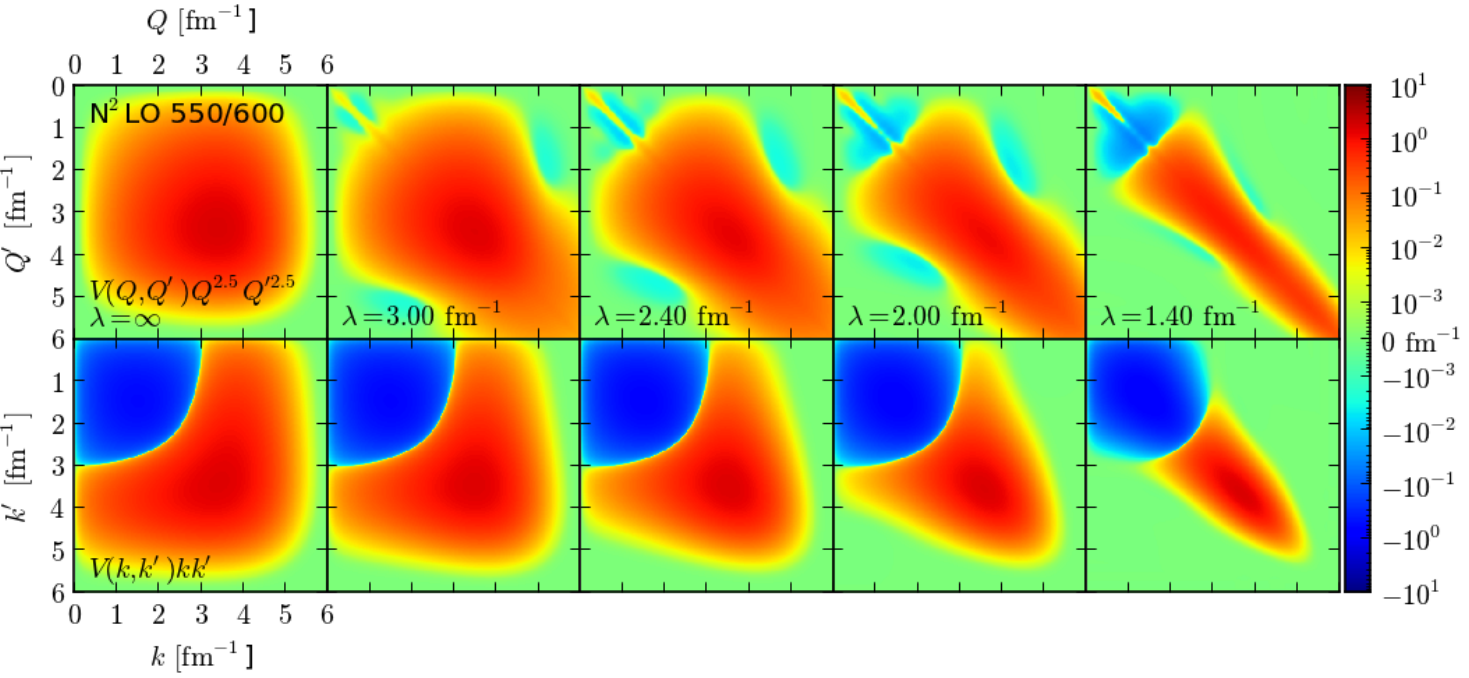}
  \end{center}    
  \caption{\label{fig:vsrg_momentum}SRG evolution of a chiral \NNLO{} $NN\!+\!3N$
  Hamiltonian with cutoffs $550/600$ MeV \cite{Epelbaum:2002nr,Epelbaum:2006mo}
  in a three-body hyperspherical momentum basis. The figure shows contour
  plots of the matrix elements as a function of $\lambdaSRG$ in the lowest 
  hyperspherical partial wave, both for the $3N$ interaction (top panel)
  and the embedded $NN$ interaction in that
  partial wave (lower panel). See \cite{Wendt:2013ys} for
  additional details. Figure courtesy of K.~Wendt.}
\end{figure*}

\begin{figure*}[t]
  \setlength{\unitlength}{\textwidth}
  \begin{center}
    \includegraphics[width=0.6\unitlength]{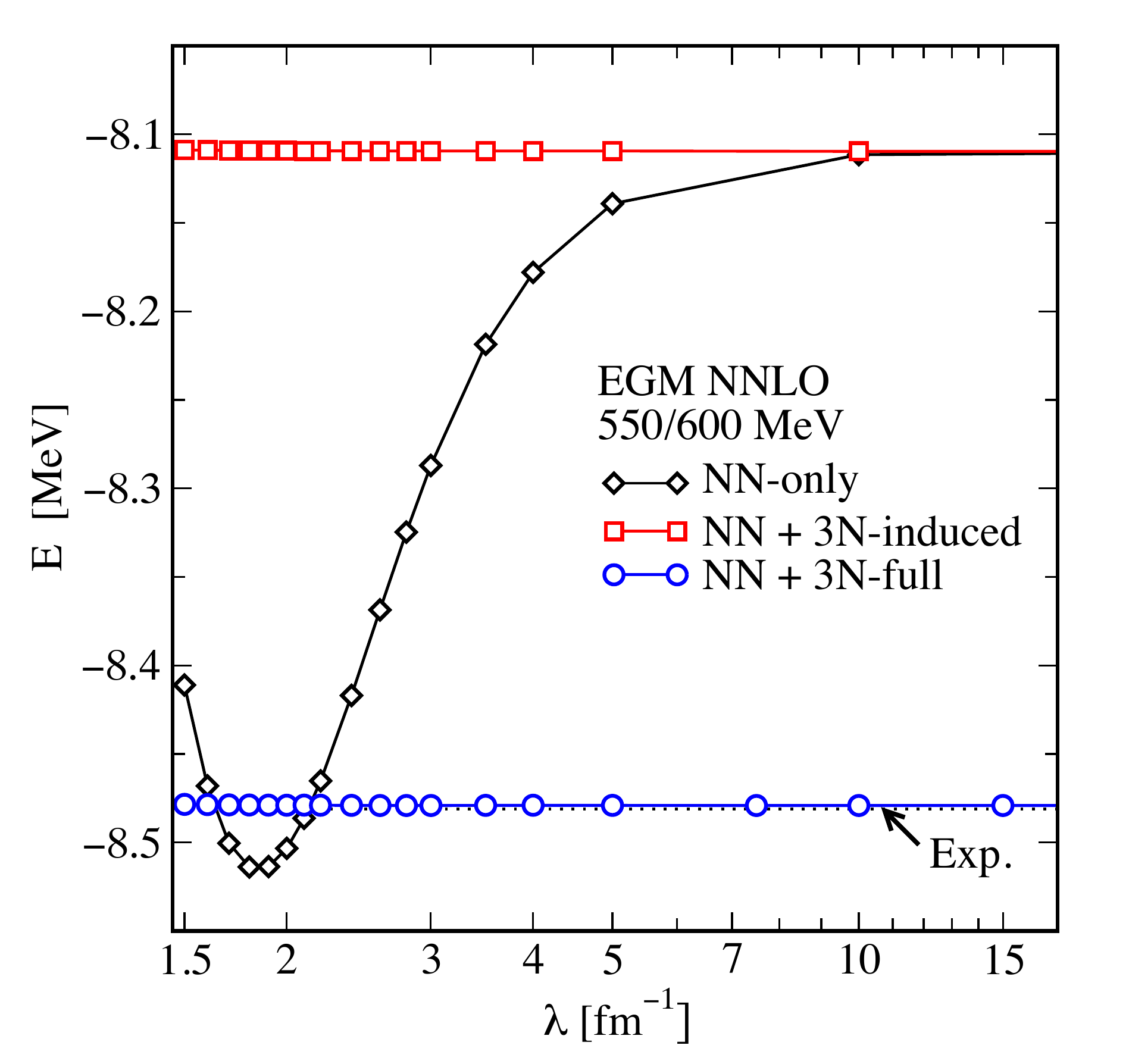}
  \end{center}  
  \vspace{-10pt}
  \caption{\label{fig:triton}Ground state energy of $\nuc{H}{3}$ as a function 
  of the flow parameter $\lambdaSRG$ for a chiral \NNLO{} $NN\!+\!3N$ interaction
  with cutoffs $550/600$ MeV (cf.~Fig.~\ref{fig:vsrg_momentum} and 
  \cite{Epelbaum:2002nr,Epelbaum:2006mo}). $NN$-only means initial and 
  induced $3N$ interactions are discarded, $NN\!+\!3N$-induced takes only 
  induced $3N$ interactions into account, and $3N$-full contains initial
  $3N$ interactions as well. The black dotted line shows the experimental 
  binding energy \cite{Wang:2012uq}. Data courtesy of K.~Hebeler.}
\end{figure*}

Figure \ref{fig:vsrg_momentum} shows the evolution of 
NN and $3N$ matrix elements of a chiral \NNLO{} interaction by Epelbaum, Gl\"ockle,
and Mei\ss{}ner \cite{Epelbaum:2002nr,Epelbaum:2006mo}, with cutoffs 550/600 MeV. As discussed for our 
schematic example, both the $NN$ and $3N$ interactions become band diagonal and the SRG 
decouples the low- and high-momentum regimes as we evolve to lower values of $\lambdaSRG$. 
In figure \ref{fig:triton}, the same family 
of SRG-evolved interactions is used to calculate the ground-state energy of the triton,
as a function of $\lambdaSRG$. If only the $NN$ part of the chiral interaction is used as 
input, and the SRG generator and flowing Hamiltonian are truncated at the two-body level 
(curve `$NN$-only'), the SRG evolution is not unitary in the three-body system. The 
energy exhibits a significant dependence on $\lambdaSRG$, on the order of 5--6\%. If the 
flow equations are truncated at the three-body level instead, induced $3N$ interactions are
properly included and the unitarity of the transformation is restored (`$NN\!+\!3N$-induced'): 
The energy does not change as $\lambdaSRG$ is varied. Finally, the curve `$NN\!+\!3N$-full' 
shows the result for a calculation with initial $NN$ and $3N$ forces that are consistently 
SRG-evolved at the three-body level. The triton ground-state energy is again invariant 
under the SRG flow, and closely reproduces the experimental value that is used as 
a constraint in the adjustment of the $3N$ force's low-energy constants (see, e.g., 
\cite{Epelbaum:2009ve,Machleidt:2011bh,Gazit:2009qf}).

The SRG flow equations force us to manipulate large sections (or the 
entirety) of the Hamiltonian's spectrum in order to avoid basis 
truncation artifacts (also cf.~\cite{Roth:2014fk,Binder:2014fk})
We may ask, then, if it might be possible to avoid the use of matrix 
representations entirely by solving the operator flow equation 
\eqref{eq:opflow} directly in the algebra of operators. This is the 
strategy that we will explore in the following, which will ultimately 
lead us to the formulation of the In-Medium SRG. First, we have to 
lay some groundwork on normal ordering techniques and Wick's theorem.

\section{\label{sec:no}Normal Ordering and Wick's Theorem for Arbitrary Reference States}
\subsection{References States and Many-Body Bases}
To describe the structure and dynamics of an atomic nucleus of
mass $A$, we need to work in an $A$-body Hilbert space\footnote{To include
continuum degrees of freedom, i.e., resonant and scattering states,
we would have to treat the nucleus as an open quantum system in
a so-called rigged Hilbert space
\cite{Madrid:2005rt,Michel:2009oz}. This setting also creates
opportunities for a completely microscopic description of nuclear 
reactions, see, e.g., \cite{Jaganathen:2014zp,Fossez:2015if}.},
and choose a suitable $A$-body basis. Since we are dealing with a
system of fermions, a straightforward choice are \emph{antisymmetrized
product states}, or \emph{Slater determinants}. Introducing 
fermionic creation and annihilation operators $\aaO_i$ and $\aO_j$ that 
satisfy the canonical anticommutation relations 
\begin{equation}
  \acomm{\aaO_i}{\aaO_j} = \acomm{\aO_i}{\aO_j} = 0 \,, \acomm{\aaO_i}{\aO_j}=\delta_{ij}\,,
\end{equation}
we can write a generic $A$-particle Slater determinant as
\begin{equation}
  \ket{\Phi} = \prod_{k=1}^A \aaO_{i_k}\ket{\text{vac}}\,,\label{eq:Slater_generic}
\end{equation}
where $\ket{\text{vac}}$ refers to the particle vacuum. Here, the indices run 
over a suitably chosen single-particle basis, e.g., spatially 
localized orbitals if we plan to describe a finite system like a nucleus.
A complete basis for the many-body Hilbert space can be obtained by 
distributing $A$ nucleons over the available single-particle states in
all possible ways. 

Of course, not all of the states in this naively chosen basis are created equal. As
alluded to in section \ref{sec:srg}, nuclear interactions and the nucleus itself
have characteristic energy or momentum scales. The ground state and 
low-lying excitation spectrum of an $A$-body nucleus is typically dominated by excitations 
of particles in the vicinity of its Fermi energy. Thus, we can find a Slater 
determinant $\ket{\Phi}$ that is a fair approximation to the nuclear ground state, 
and use it as a \emph{reference state} for the construction and organization of 
our many-body basis. Slater determinants that are variationally
optimized through a Hartree-Fock (HF) procedure have been shown to be reasonable 
reference states for interactions with low resolution scales around $\lambda=2.0\fm^{-1}$ 
(see, e.g., Refs.~\cite{Bogner:2010pq,Roth:2010vp,Barrett:2013oq,Hagen:2014ve,Hergert:2016jk,Tichai:2016vl} 
and references therein), allowing post-HF methods like MBPT, 
CC, or the IMSRG discussed below to converge rapidly to the exact FCI result. 
Starting from such a HF reference state $\ket{\Phi_\text{HF}}$, we can obtain 
a basis consisting of the state itself and up to $A$-particle, $A$-hole ($ApAh$) 
excitations:
\begin{equation}
  \ket{\Phi_\text{HF}},\,\aaO_{p_1}\aO_{h_1}\ket{\Phi_\text{HF}},\;\ldots\;,\,\aaO_{p_1}\ldots\aaO_{p_A}\aO_{h_A}\ldots\aO_{h_1}\ket{\Phi_\text{HF}}\,.
\end{equation}
Here, indices $p_i$ and $h_i$ run over all one-body basis states with energies above 
(\emph{particle} states) and below the Fermi level (\emph{hole} states), respectively.

Many-body bases built from such a single Slater determinant and its particle-hole
excitations work best for systems with large gaps in the single-particle 
spectrum, e.g., closed-shell nuclei. If the gap is small, particle-hole excited 
basis states can be near-degenerate with the reference state, which usually results
in spontaneous symmetry breaking and strong configuration mixing. At best, these
phenomena impede the convergence of a many-body calculation by forcing us to use
model spaces that contain $npnh$ excitations with large $n$, e.g., in a CI framework. 
At worst, the behavior of a truncated many-body method like IMSRG or CC may be 
completely uncontrolled. We want to overcome these problems by building correlations
from configuration mixing into the reference state, and constructing a basis of
generalized $ApAh$ excitations on top of this state. A key element of such an approach 
are generalized normal ordering techniques.

\subsection{Normal-Ordered Operators and Wick Contractions}
In Ref.~\cite{Kutzelnigg:1997fk}, Kutzelnigg and Mukherjee developed a 
generalized normal ordering for arbitrary reference states. Here, we present
the essential elements of their discussion that we will need in the following, 
but use the slightly different notation of Kong \emph{et al.}~\cite{Kong:2010kx}.

First, we introduce a pseudo-tensorial notation for strings of creation 
and annihilation operators, to facilitate book-keeping and make the formalism
more compact. A particle-number conserving product of $k$ creators and 
annihilators each is written as
\begin{equation}
  \AO^{i_1\ldots i_k}_{j_1\ldots j_k}\equiv
  \aaO_{i_1}\ldots\aaO_{i_k}\aO_{j_k}\ldots\aO_{j_1}\,.
\end{equation}
We do not consider particle-number changing operators in the present work,
because they cause ambiguities in the contraction and sign rules for the
$A$ operators that are defined in the following. The anticommutation relations 
imply
\begin{equation}\label{eq:def_A_permutations}
  \AO^{\PC(i_1 \ldots i_k)}_{\PC'(j_1 \ldots j_k)} 
  = (-1)^{\pi(\PC)+\pi(\PC')} \AO^{i_1\ldots i_k}_{j_1\ldots j_k}\,,
\end{equation}
where $\pi(\PC)=\pm1$ indicates the parity (or signature) of a permutation 
$\PC$. A general $k$-body operator in second quantization can now be written 
in terms of the basis operators as 
\begin{equation}
 O^{(k)} = \frac{1}{(k!)^2}\sum_{\substack{i_1\ldots i_k\\j_1\ldots j_k}}
  o^{i_1\ldots i_k}_{j_1\ldots j_k}\AO^{i_1\ldots i_k}_{j_1\ldots j_k}\,,
\end{equation}
where we assume that the coefficients $o^{i_1\ldots i_k}_{j_1\ldots j_k}$
are antisymmetrized, and therefore also obey equation \eqref{eq:def_A_permutations}
under index permutations. 

Next, we introduce \emph{irreducible $k$-body density matrices $\lambda^{(k)}$}. In
the one-body case, we have the usual density matrix
\begin{equation}
  \lambda^{i}_{j} \equiv \matrixe{\Phi}{\AO^{i}_{j}}{\Phi}\,,
\end{equation}
and for future use, we also define
\begin{equation}
  \xi^{i}_{j} \equiv \lambda^{i}_{j} - \delta^{i}_{j}\,.
\end{equation}
Up to a factor $(-1)$ that unifies the sign rules for one-body contractions
presented below, $\xi^{(1)}$ is simply the generalization of the hole 
density matrix for a correlated state. In the natural orbital basis, 
i.e., the eigenbasis of $\lambda^{(1)}$, both one-body density matrices 
are diagonal:
\begin{equation}\label{eq:def_natorb}
  \lambda^{i}_{j}=n_i \delta^{i}_{j}\,,\quad 
  \xi^{i}_{j}=-\nn_i\delta^{i}_{j}\equiv-(1-n_i)\delta^{i}_{j}\,
  \,.
\end{equation}
The fractional occupation numbers $0\leq n_i \leq 1$ are the eigenvalues 
of $\lambda^{(1)}$. 

For $k\geq 2$, we denote full density matrices by
\begin{align}
  \rho^{i_1 \ldots i_k}_{j_1 \ldots j_k} &= 
    \dmatrixe{\Phi}{\AO^{i_1 \ldots i_k}_{j_1 \ldots j_k}}\,,
\end{align}
and define
\begin{align}
  \lambda^{ij}_{kl} &\equiv \rho^{ij}_{kl} - \AC\{\lambda^{i}_{k}\lambda^{j}_{l}\}\,, \label{eq:def_Lambda2}\\
  \lambda^{ijk}_{lmn}&\equiv \rho^{ijk}_{lmn} - \AC\{\lambda^{i}_{l}\lambda^{jk}_{mn}\} -
                      \AC\{\lambda^{i}_{l}\lambda^{j}_{m}\lambda^{k}_{n}\}\,,\label{eq:def_Lambda3}
\end{align}
etc., where $\AC\{\ldots\}$ fully antisymmetrizes the indices of the expression
within the brackets, e.g.,
\begin{equation}
  \AC\{\lambda^{i}_{k}\lambda^{j}_{l}\} = \lambda^{i}_{k}\lambda^{j}_{l} - 
    \lambda^{i}_{l}\lambda^{j}_{k}\,.
\end{equation}
From equation \eqref{eq:def_Lambda2}, it is easy to see that $\lambda^{(2)}$ encodes the
two-nucleon correlation content of the reference state $\ket{\Phi}$. If 
the reference state is a Slater determinant, i.e., an independent-particle
state, the full two-body density matrix factorizes, and $\lambda^{(2)}$
vanishes:
\begin{equation}
  \lambda^{ij}_{kl} = \rho^{ij}_{kl} - \AC\{\lambda^{i}_{k}\lambda^{j}_{l}\}
  = \lambda^{i}_{k}\lambda^{j}_{l} -  \lambda^{i}_{k}\lambda^{j}_{l} - 
    \left(\lambda^{i}_{k}\lambda^{j}_{l} -  \lambda^{i}_{k}\lambda^{j}_{l}\right) = 0\,.
\end{equation}
Equation \eqref{eq:def_Lambda3} shows that $\lambda^{(3)}$ is constructed
by subtracting contributions from three independent particles as well as
two correlated nucleons in the presence of an independent spectator particle
from the full three-body density matrix, and therefore encodes the genuine
three-nucleon correlations. This construction and interpretation  
generalizes to irreducible density matrices of rank $k$.

Now we consider the expansion of a (number-conserving) string of creation 
and annihilation operators in normal-ordered components. First, we define a
normal-ordered one-body operator by subtracting from a given one-body operator its
expectation value in the reference state:
\begin{equation}
  \nord{A^{a}_{b}} \equiv A^{a}_{b} - \matrixe{\Phi}{A^{a}_{b}}{\Phi} = A^{a}_{b} - \lambda^{a}_{b}\,.
\end{equation}
This implies that the expectation value of the normal-ordered operator in the
reference state vanishes by construction:
\begin{equation}\label{eq:ex_nord_1B}
  \matrixe{\Phi}{\nord{A^{a}_{b}}}{\Phi} = 0\,.
\end{equation}
For a two-body operator, we have the expansion
\begin{align}
  A^{ab}_{cd} &= \nord{A^{ab}_{cd}} 
                  + \lambda^{a}_{c}\nord{A^{b}_{d}} 
                  - \lambda^{a}_{d}\nord{A^{b}_{c}} 
                  + \lambda^{b}_{d}\nord{A^{a}_{c}}
                  - \lambda^{b}_{c}\nord{A^{a}_{d}}
                  + \lambda^{a}_{c}\lambda^{b}_{d}
                  - \lambda^{a}_{d}\lambda^{b}_{c}
                  + \lambda^{ab}_{cd}\,.\label{eq:nord_2B}
\end{align}
As a consequence of equation \eqref{eq:def_A_permutations}, the sign of each
term is determined by the product of the parities of the permutations 
that map upper and lower indices to their ordering in the initial operator. 
Except for the last term, this expression looks like the result for the
regular normal ordering, with pairwise contractions of indices giving
rise to one-body density matrices. The last term, a contraction
of four indices, appears because we are dealing with an arbitrary, 
correlated reference state here. 

Taking the expectation value of equation \eqref{eq:nord_2B} and using
equation \eqref{eq:ex_nord_1B}, we obtain
\begin{align}
  \rho^{ab}_{cd} &= \matrixe{\Phi}{\nord{A^{ab}_{cd}}}{\Phi}                  
                    + \underbrace{\lambda^{a}_{c}\lambda^{b}_{d}
                    - \lambda^{a}_{d}\lambda^{b}_{c}
                    + \lambda^{ab}_{cd}}_{=\rho^{ab}_{cd}}\,,
\end{align}
and see that
\begin{equation}\label{eq:nord_ex}
  \matrixe{\Phi}{\nord{A^{ab}_{cd}}}{\Phi} = 0\,.
\end{equation}
The normal ordering procedure can be extended in analogy to the one-
and two-body cases, e.g.,
\begin{align}
  A^{abc}_{def} &= \nord{A^{abc}_{def}} 
                  + \AC\{\lambda^{a}_{d}\nord{A^{bc}_{ef}} \}
                  + \AC\{\lambda^{a}_{d}\lambda^{b}_{e}\nord{A^{c}_{f}} \}
                  + \AC\{\lambda^{ab}_{de}\nord{A^{c}_{f}} \} \notag\\
                &\hphantom{=} + \lambda^{abc}_{def}
                  + \AC\{\lambda^{a}_{d}\lambda^{bc}_{ef} \}
                  + \AC\{\lambda^{a}_{d}\lambda^{b}_{e}\lambda^{c}_{f} \}\,,
\end{align}
yielding normal-ordered operators of arbitrary rank $k$ that satisfy
\begin{equation}
 \matrixe{\Phi}{\nord{\AO^{i_1\ldots i_k}_{j_1\ldots j_k}}}{\Phi}=0\,.
\end{equation}

Finally, a generalized Wick's theorem for arbitrary reference states
can be formulated: Any product of two normal-ordered operators can be
expanded in a sum of normal-ordered terms, with Wick contractions
and operators containing at least one index from each of the original
operators. For example, the basic contractions appearing in the
expansion of a product of normal-ordered two-body operators are
(notice the signs)

{\setlength{\fboxsep}{1pt}
\begin{align}
  \nord{A^{a{\fbox{\scriptsize $b$}}}_{cd}}\nord{A^{ij}_{\fbox{\scriptsize $k$}l}} &= - \lambda^{\fbox{\scriptsize $b$}}_{\fbox{\scriptsize $k$}}\nord{A^{aij}_{cdl}}\,,\label{eq:contract_lambda1B}\\ 
  \nord{A^{ab}_{\fbox{\scriptsize $c$}d}}\nord{A^{i\fbox{\scriptsize $j$}}_{kl}} &= - \xi^{\fbox{\scriptsize $j$}}_{\fbox{\scriptsize $c$}}\nord{A^{bia}_{dkl}}\,,\label{eq:contract_chi1B}\\[10pt]
  \nord{A^{\fbox{\scriptsize $ab$}}_{cd}}\nord{A^{ij}_{\fbox{\scriptsize $kl$}}} &= + \lambda^{\fbox{\scriptsize $ab$}}_{\fbox{\scriptsize $kl$}}\nord{A^{ij}_{cd}}\,,\label{eq:contract_lambda2Ba}\\ 
  \nord{A^{a{\fbox{\scriptsize $b$}}}_{cd}}\nord{A^{\fbox{\scriptsize $i$}j}_{\fbox{\scriptsize $kl$}}} &= - \lambda^{\fbox{\scriptsize $ib$}}_{\fbox{\scriptsize $kl$}}\nord{A^{aj}_{cd}}\,,\label{eq:contract_lambda2Bb}\\ 
  \nord{A^{ab}_{\fbox{\scriptsize $c$}d}}\nord{A^{\fbox{\scriptsize $ij$}}_{\fbox{\scriptsize $k$}l}} &= - \lambda^{\fbox{\scriptsize $ij$}}_{\fbox{\scriptsize $ck$}}\nord{A^{ab}_{dl}}\,,\label{eq:contract_lambda2Bc}\\ 
  \nord{A^{\fbox{\scriptsize $ab$}}_{c\fbox{\scriptsize$d$}}}\nord{A^{\fbox{\scriptsize $i$}j}_{\fbox{\scriptsize $kl$}}} &= - \lambda^{\fbox{\scriptsize $abi$}}_{\fbox{\scriptsize $dkl$}}\nord{A^{j}_{c}}\,,\label{eq:contract_lambda3B}\\ 
  \nord{A^{\fbox{\scriptsize $ab$}}_{\fbox{\scriptsize$cd$}}}\nord{A^{\fbox{\scriptsize $ij$}}_{\fbox{\scriptsize $kl$}}} &= +\lambda^{\fbox{\scriptsize $abij$}}_{\fbox{\scriptsize $cdkl$}}\,.\label{eq:contract_lambda4B}
\end{align}
}
Only the first two contraction types, equations \eqref{eq:contract_lambda1B} 
and \eqref{eq:contract_chi1B}, appear in the regular Wick's theorem for 
uncorrelated reference states. The additional contractions 
\eqref{eq:contract_lambda2Ba}--\eqref{eq:contract_lambda4B} increase 
the number of terms when we expand operator products using the generalized 
Wick's theorem --- examples are shown in appendix \ref{app:NO}. Fortunately,
we will see in section \ref{sec:mrimsrg} that the overall increase in complexity is 
manageable.

\subsection{Normal-Ordered Hamiltonian and Normal-Ordered Two-Body Approximation}
We conclude this section by applying the generalized normal ordering to
an intrinsic nuclear $A$-body Hamiltonian containing both $NN$ and $3N$ 
interactions, which will be relevant for the applications discussed later
in this work. Let
\begin{equation}\label{eq:def_Hint}
  H = \left(1-\frac{1}{A}\right)T^{(1)} + \frac{1}{A}T^{(2)} + V^{(2)} +V^{(3)}\,,
\end{equation}
where 
\begin{equation}
  T^{(1)} \equiv \sum_{i}\frac{\pOV_i^2}{2m}\,,\quad T^{(2)} \equiv -\sum_{i<j}\frac{\pOV_i\cdot\pOV_j}{m}
\end{equation}
(see, e.g., \cite{Hergert:2009wh}). Choosing a generic correlated reference state 
$\ket{\Phi}$, we rewrite the Hamiltonian as
\begin{align}\label{eq:Hno}
  \HO &= E 
        + \sum_{ij}f^{i}_{j}\nord{A^{i}_{j}} 
        + \frac{1}{4}\sum_{ijkl}\Gamma^{ij}_{kl}\nord{A^{ij}_{kl}}\notag\\
      &\hphantom{=}
        + \frac{1}{36}\sum_{ijklmn}W^{ijk}_{lmn}\nord{A^{ijk}_{lmn}}\,,
\end{align}
where the labels have been chosen for historical reasons. The individual 
normal-ordered contributions in equation \eqref{eq:Hno} are given by
\begin{align}
  E &= \left(1-\frac{1}{A}\right)\sum_{ab}t^{a}_{b}\lambda^{a}_{b}
        + \frac{1}{4}\sum_{abcd}\left(\frac{1}{A}t^{ab}_{cd} + v^{ab}_{cd}\right)\rho^{ab}_{cd}\notag\\
      &\hphantom{=}+ \frac{1}{36}\sum_{abcdef}v^{abc}_{def} \rho^{abc}_{def}\,,
      \label{eq:E0}\\
  f^{i}_{j} &= \left(1-\frac{1}{A}\right)t^{i}_{j} + \sum_{ab}\left(\frac{1}{A}t^{ia}_{jb} + v^{ia}_{jb}\right)\lambda^{a}_{b}\notag\\
      &\hphantom{=}+ \frac{1}{4}\sum_{abcd}v^{iab}_{jcd}\rho^{ab}_{cd}\,,\label{eq:f}   \\
  \Gamma^{ij}_{kl} &= \frac{1}{A}t^{ij}_{kl} + v^{ij}_{kl} + \sum_{ab}v^{ija}_{klb}\lambda^{a}_{b}\,,\label{eq:Gamma}\\
  W^{ijk}_{lmn}&=v^{ijk}_{lmn}\,.
\end{align}
Here, we use the full density matrices for compactness, but it is easy to express
equations \eqref{eq:E0}--\eqref{eq:Gamma} completely in terms of irreducible density
matrices by using equations \eqref{eq:def_Lambda2} and \eqref{eq:def_Lambda3}. Note that
the normal-ordered zero-, one-, and two-body parts of the Hamiltonian all contain 
in-medium contributions from the free-space $3N$ interaction. It has been shown 
empirically that the omission of the normal-ordered three-body piece of the Hamiltonian 
causes a deviation of merely 1--2\% in ground-state and (absolute) excited state 
energies of light and medium-mass nuclei \cite{Hagen:2007zc,Roth:2011kx,Roth:2012qf,Binder:2013fk,Gebrerufael:2016fe}. 
This \emph{normal-ordered two-body approximation} (NO2B) to the Hamiltonian is
widely used nowadays, because it provides an efficient means to account for $3N$
force effects in nuclear many-body calculations without incurring the computational
expense of explicitly treating three-body operators. In the next section, we will also
see that the NO2B approximation meshes in a natural way with the framework of 
the MR-IMSRG, which makes it especially appealing for our purposes.

\section{\label{sec:mrimsrg}The Multireference In-Medium Similarity Renormalization Group}
\subsection{MR-IMSRG Flow Equations}
We are now ready to formulate the MR-IMSRG flow equations by applying 
the tools discussed in the previous section to the operator differential 
equation \eqref{eq:opflow}. We express all operators in terms of 
normal-ordered strings of creation and annihilation operators. As 
discussed in section \ref{sec:srg}, each evaluation of the commutator on
the right-hand side will induce operators of higher rank,
\begin{equation}\label{eq:induced}
  \comm{\nord{A^{ab}_{cd}}}{\nord{A^{ij}_{kl}}}= \delta_{ci}\nord{A^{abj}_{dkl}} +\ldots,
\end{equation}
and we would need to include operators up to rank $A$ if we want the MR-IMSRG flow
to be unitary in an $A$-body system, which is not feasible. However, 
in contrast to equation \eqref{eq:comm2B_vac}, we are now working with 
\emph{normal-ordered} operators whose in-medium contributions have been 
absorbed into terms of lower rank. Consequently, we expect the induced operators 
to be much weaker than in the free-space SRG case. The empirical success of the 
NO2B approximation discussed at the end of section \ref{sec:no} certainly
seems to justify this expectation in the case of nuclear $NN$+$3N$ 
Hamiltonians with low resolution scales. 

Following this line of reasoning further, we choose to truncate all flowing
operators at a given particle rank $n\leq A$ in order to obtain a closed
system of flow equations. For $n=2$, we demand that for all values of the
flow parameter $s$
\begin{align} 
  \etaO(s) &\approx\etaO^{(1)}(s)+\etaO^{(2)}(s)\,,\label{eq:imsrg2_eta}\\
  \HO(s) &\approx E(s) + f(s) + \Gamma(s)\,,\label{eq:imsrg2_H}\\
  \totd{}{s}\HO(s) &\approx \totd{}{s}E(s) + \totd{}{s}f(s) + \totd{}{s}\Gamma(s)\label{eq:imsrg2_dH}\,.
\end{align}
This is the so-called MR-IMSRG(2) truncation, which has been our primary
workhorse in past applications \cite{Tsukiyama:2011uq,Tsukiyama:2012fk,Hergert:2013mi,Hergert:2013ij,Hergert:2014vn,Morris:2015ve,Hergert:2016jk}. It is the basis for all results in this work. We note that the MR-IMSRG(2) at this specific
level of truncation is a cousin to a variety of other truncated many-body schemes,
e.g., Canonical Transformation theory with Singles and Doubles excitations (CTSD)
\cite{White:2002fk,Yanai:2006uq,Yanai:2007kx}, the two-body Antisymmetrized or 
Irreducible Contracted Schr\"odinger Equation approach (ICSE(2)) \cite{Nakatsuji:1976yq,Valdemoro:1987zl,Mukherjee:2001uq,Kutzelnigg:2002kx,Kutzelnigg:2004vn,Kutzelnigg:2004ys,Mazziotti:2006fk} and of course CCSD (Coupled Cluster 
with Singles and Doubles) (see, e.g., \cite{Shavitt:2009}), although the latter is 
based on non-unitary similarity transformations.

Plugging equations \eqref{eq:imsrg2_eta}--\eqref{eq:imsrg2_dH} into the operator
flow equation \eqref{eq:opflow} and organizing contributions by particle rank,
we obtain the system of MR-IMSRG(2) flow equations:
\begin{align}
  \totd{E}{s} &=     
    \sum_{ab}(n_{a}-n_{b})\eta^{a}_{b}f^{b}_{a}
    +\frac{1}{4}\sum_{abcd}
        \left(\eta^{ab}_{cd}\Gamma^{cd}_{ab}-\Gamma^{ab}_{cd}\eta^{cd}_{ab}\right)
        n_{a}n_{b}\bar{n}_{c}\bar{n}_{d}
    \notag\\
  &\hphantom{=}
    +\frac{1}{4}\sum_{abcd}\left(\totd{}{s}\Gamma^{ab}_{cd}\right)\lambda^{ab}_{cd}
    +\frac{1}{4}\sum_{abcdklm}\left(\eta^{ab}_{cd}\Gamma^{kl}_{am}-\Gamma^{ab}_{cd}\eta^{kl}_{am}\right)\lambda^{bkl}_{cdm}\,,
    \label{eq:mr_flow_0b_tens}
  \\[10pt]
  \totd{}{s}f^{i}_{j} &=
    \sum_{a}\left(\eta^{i}_{a}f^{a}_{j}-f^{i}_{a}\eta^{a}_{j}\right)
    +\sum_{ab}\left(\eta^{a}_{b}\Gamma^{bi}_{aj}-f^{a}_{b}\eta^{bi}_{aj}\right)(n_{a}-n_{b})
  \notag\\
  &\hphantom{=}
  +\frac{1}{2}\sum_{abc}
    \left(\eta^{ia}_{bc}\Gamma^{bc}_{ja}-\Gamma^{ia}_{bc}\eta^{bc}_{ja}\right)\left(n_{a}\bar{n}_{b}\bar{n}_{c}+\bar{n}_{a}n_{b}n_{c}\right)
  \notag\\
  &\hphantom{=}
      +\frac{1}{4}\sum_{abcde}\left(\eta^{ia}_{bc}\Gamma^{de}_{ja}-\Gamma^{ia}_{bc}\eta^{de}_{ja}\right)\lambda^{de}_{bc} 
    +\sum_{abcde}\left(\eta^{ia}_{bc}\Gamma^{be}_{jd}-\Gamma^{ia}_{bc}\eta^{be}_{jd}\right)\lambda^{ae}_{cd}
  \notag\\
  &\hphantom{=}
      -\frac{1}{2}\sum_{abcde}\left(\eta^{ia}_{jb}\Gamma^{cd}_{ae}-\Gamma^{ia}_{jb}\eta^{cd}_{ae}\right)\lambda^{cd}_{be}
    +\frac{1}{2}\sum_{abcde}\left(\eta^{ia}_{jb}\Gamma^{bc}_{de}-\Gamma^{ia}_{jb}\eta^{bc}_{de}\right)\lambda^{ac}_{de}\,,
  \label{eq:mr_flow_1b_tens}\\[10pt]
  \totd{}{s}\Gamma^{ij}_{kl}&=  
  \sum_{a}\left(\eta^{i}_{a}\Gamma^{aj}_{kl}+\eta^{j}_{a}\Gamma^{ia}_{kl}-\eta^{a}_{k}\Gamma^{ij}_{al}-\eta^{a}_{l}\Gamma^{ij}_{ka}
  -f^{i}_{a}\eta^{aj}_{kl}-f^{j}_{a}\eta^{ia}_{kl}+f^{a}_{k}\eta^{ij}_{al}+f^{a}_{l}\eta^{ij}_{ka}\right)
  \notag\\
  &\hphantom{=}
    +\frac{1}{2}\sum_{ab}\left(\eta^{ij}_{ab}\Gamma^{ab}_{kl}-\Gamma^{ij}_{ab}\eta^{ab}_{kl}\right)
     \left(1-n_{a}-n_{b}\right)
  \notag\\
  &\hphantom{=}
    +\sum_{ab}(n_{a}-n_{b})\left(\left(\eta^{ia}_{kb}\Gamma^{jb}_{la}-\Gamma^{ia}_{kb}\eta^{jb}_{la}\right)-\left(\eta^{ja}_{kb}\Gamma^{ib}_{la}-\Gamma^{ja}_{kb}\eta^{ib}_{la}\right)\right)\,.
  \label{eq:mr_flow_2b_tens}
\end{align}
All single-particle indices and occupation numbers (cf.~section \ref{sec:no}) 
refer to natural orbitals, and the $s$-dependence has been suppressed for 
brevity. The \emph{single-reference limit} is readily obtained by setting the 
irreducible density matrices $\lambda^{(2)}$ and $\lambda^{(3)}$ to
zero in the previous expressions.

We solve the flow equations \eqref{eq:mr_flow_0b_tens}--\eqref{eq:mr_flow_2b_tens} 
by integrating from $s=0$ to $s\to\infty$, using the components of the normal-ordered 
input Hamiltonian (equations \eqref{eq:E0}--\eqref{eq:Gamma}) as initial values.
In this process, the flow equations will re-shuffle the correlations in 
the $A$-body system, generating a highly nonperturbative resummation of the 
many-body expansion (see section \ref{sec:flow_gs} for numerical examples).

To interpret the multireference flow equations, we associate the 
fractional occupation numbers $\nn_i$ and $n_i$ with particle- and 
hole-like states, respectively (cf.~equation \eqref{eq:def_natorb}), and 
note that
\begin{align}
 1-n_a-n_b = \nn_a\nn_b - n_a n_b\,, \\
 n_a-n_b = n_a \nn_b - \nn_a n_b\,.
\end{align}
For the typical ans\"atze that we use for $\etaO(s)$ (see section \ref{sec:generators}),
the generator is proportional to the (offdiagonal) Hamiltonian, and we see
the first two terms of the zero-body flow equation have the structure of second-order 
energy corrections, but evaluated for the \emph{flowing} Hamiltonian $\HO(s)$. 
Furthermore, we recognize that the second and third lines of equation \eqref{eq:mr_flow_2b_tens}
have the structure of ladder (particle-particle / hole-hole) and ring 
(particle-hole) skeleton diagrams, respectively. They generate ladder
and ring summations in the limit $s\to\infty$, but also ring-ladder interference
diagrams with rich topologies that go far beyond traditional re-summation 
methods \cite{Day:1967zl,Brandow:1967tg,Fetter:2003ve}. A detailed perturbative
analysis is presented in \cite{Hergert:2016jk}.

For general reference states, the MR-IMSRG flow equations also include couplings 
to correlated pairs and triples of nucleons through the irreducible density matrices 
$\lambda^{(2)}$ and $\lambda^{(3)}$. It is noteworthy that the MR-IMSRG(2) flow 
equations do not depend on $\lambda^{(4)}$ or nonlinear powers of $\lambda^{(2)}$. While 
such contractions appear in the \emph{products} of normal-ordered two-body operators, 
they cancel in the commutators (see \ref{app:NO}). 
This ensures that the MR-IMSRG only sums so-called \emph{connected} many-body 
diagrams (i.e., diagrams which do not contain intermediate insertions of the reference state) 
\cite{Brandow:1967tg,Shavitt:2009}. 

Let us conclude this section by briefly considering the numerical implementation 
of the MR-IMSRG(2) scheme. The computational effort is dominated by the 
two-body flow equation \eqref{eq:mr_flow_2b_tens}, which naively requires $\OC(N^6)$ 
operations, where $N$ denotes the single-particle basis size. This puts the 
MR-IMSRG(2) in the same category as its aforementioned ``cousins''
CCSD \cite{Shavitt:2009,Hagen:2014ve}, CTSD \cite{White:2002fk,Yanai:2006uq,Yanai:2007kx},
and ICSE(2) \cite{Nakatsuji:1976yq,Valdemoro:1987zl,Mukherjee:2001uq,Kutzelnigg:2002kx,Kutzelnigg:2004vn,Kutzelnigg:2004ys,Mazziotti:2006fk},
as well as the Self-Consistent Green's Function Approaches (SCGF) 
\cite{Dickhoff:2004fk,Barbieri:2007fk,Cipollone:2013uq,Soma:2013ys,Soma:2014fu}.
Fortunately, the flow equations can be expressed in terms of matrix products
and traces, allowing us to use optimized linear algebra libraries
provided by high-performance computing vendors. 

Moreover, we can reduce the computational cost in the single-reference case
by distinguishing particle and hole states, because the number of hole states 
$N_h$ is typically much smaller than the number of particle states $N_p\sim N$. 
The best scaling we can achieve in the IMSRG(2) depends on the choice of 
generator (see section \ref{sec:generators}). If the one- and two-body parts of the 
generator only consist of ph and pphhh type matrix elements (and their 
Hermitian conjugates), respectively, the scaling is reduced to $\OC(N_h^2N_p^4)$, 
which matches the cost of solving the CCSD amplitude equations.

\subsection{\label{sec:decoupling}Decoupling Strategy}
\begin{figure}[t]
  \begin{center}
    \includegraphics[width=0.9\textwidth]{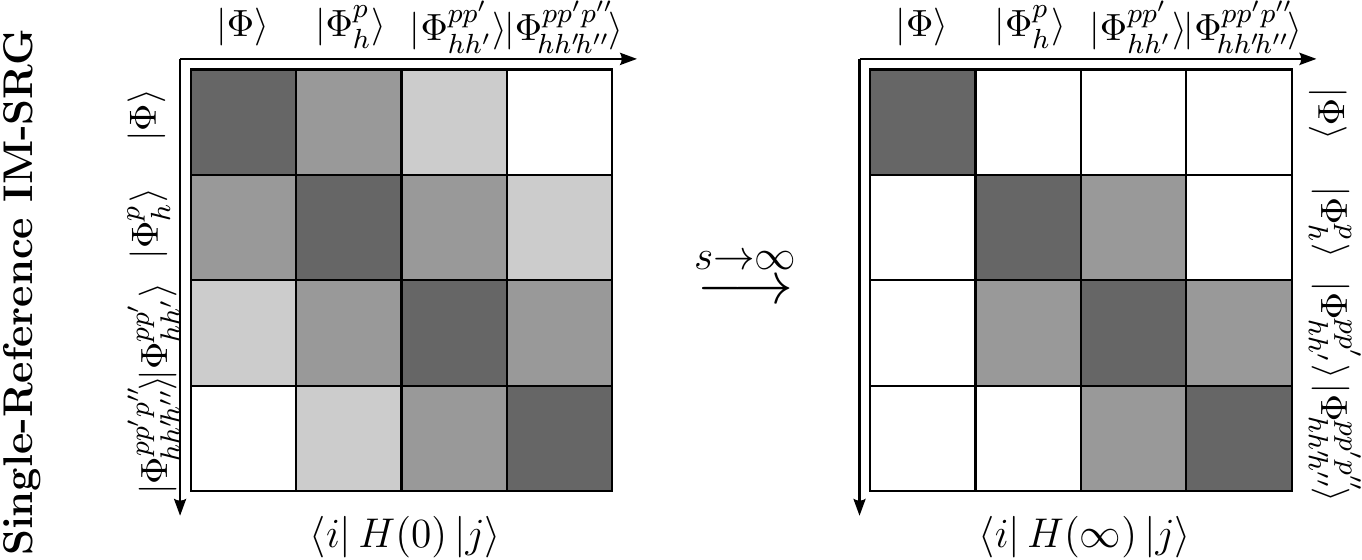}
  \end{center}
  \caption{\label{fig:imsrg}
  Schematic view of single-reference IMSRG decoupling in a many-body Hilbert space spanned by
  a Slater determinant reference $\ket{\Phi}$ and its particle-hole excitations $\ket{\Phi^{p\ldots}_{h\ldots}}$.
  }
\end{figure}

Having set up the flow equations in the previous section, we now need to
specify our decoupling strategy, i.e., how we split the Hamiltonian into
diagonal parts that we want to keep, and offdiagonal parts that we want
to suppress through the MR-IMSRG evolution (cf.~section \ref{sec:srg}). 
To do this, we refer to the
matrix representation of the Hamiltonian in a given $A$-body basis, which
is shown schematically for single- and multireference cases in
figures \ref{fig:imsrg} and \ref{fig:mrimsrg}, respectively. We stress that 
we do not actually need to construct the Hamiltonian matrix in this 
representation.

\subsubsection{IMSRG Decoupling in the Single-Reference Case}
Let us consider the simpler single-reference case first. We choose a Slater
determinant reference and construct a basis by considering all possible
particle-hole excitations (cf.~section \ref{sec:no}):
\begin{equation}
  \ket{\Phi},\,\nord{\AO^{p}_{h}}\ket{\Phi},\,\nord{\AO^{pp'}_{hh'}}\ket{\Phi},\ldots\,.
\end{equation}
Note that $\nord{\AO^{p_1\ldots p_i}_{h_1\ldots h_i}}=\AO^{p_1\ldots p_i}_{h_1\ldots h_i}$
because contractions of particle and hole indices vanish by construction. 
Using Wick's theorem, it is easy to see that the particle-hole excited Slater 
determinants are orthogonal to the reference state as well as each other.
In the Hilbert space spanned by this basis, the matrix representation of
our initial Hamiltonian in the NO2B approximation (or any two-body operator) has 
the structure shown in the left panel of figure \ref{fig:imsrg}, i.e., it is 
band-diagonal, and can at most couple $npnh$ and $(n\pm2)p(n\pm2)h$ 
excitations. 

We now have to split the Hamiltonian into appropriate diagonal and offdiagonal 
parts on the operator level, which is a non-trivial task (see, e.g., 
the extensive discussion in Refs.~\cite{Kutzelnigg:1982ly,Kutzelnigg:1983ve,Kutzelnigg:1984qf}
). Using a broad definition of diagonality is ill-advised because we must 
avoid to induce strong in-medium $3N, \ldots$ interactions to maintain the
validity of the IMSRG(2) truncation. We choose what we call a 
\emph{minimal decoupling scheme} that aims to decouple the one-dimensional block 
spanned by the reference state from all particle-hole excitations, as
shown in the right panel of figure \ref{fig:imsrg}. 

If we could implement this decoupling without approximations, 
we would extract a single eigenvalue and eigenstate of the many-body Hamiltonian
for the nucleus of interest in the limit $s\to\infty$. The eigenvalue
would simply be given by the zero-body piece of $H(\infty)$, while the eigenstate
is obtained by applying the unitary IMSRG transformation to the reference
state, $\UUO(\infty)\ket{\Phi}$. In practice, we end up with an approximate 
eigenvalue and mapping.

An important caveat is that we cannot guarantee a priori that we will 
target the true interacting ground state and its energy eigenvalue in 
this way. Empirically, the IMSRG flow is found to connect the reference
state to the eigenstate with which it has the highest overlap. In 
single-reference scenarios, a HF Slater determinant will typically
have the highest overlap with the exact ground state because it 
minimizes both the absolute energy and the correlation energy, the
latter being due to admixtures from particle-hole excitations. In
the multireference case, we have found examples where the MR-IMSRG
flow targets excited states, as discussed in sections \ref{sec:groundstate_Ne} 
and \ref{sec:sm_deformation}.

Analyzing the matrix elements between the reference state and its
excitations with the help of Wick's theorem, we first see that the 
Hamiltonian couples the $0$p$0$h block to $1$p$1$h excitations through 
the matrix elements
\begin{align}
\matrixe{\Phi}{H\nord{\AO^{p}_{h}}}{\Phi}
  &=E\matrixe{\Phi}{\nord{\AO^{p}_{h}}}{\Phi}
    +\sum_{ij}f^{i}_{j}\matrixe{\Phi}{\nord{\AO^{i}_{j}}\nord{\AO^{p}_{h}}}{\Phi}
    +\frac{1}{4}\sum_{ijkl}\Gamma^{ij}_{kl}\matrixe{\Phi}{\nord{\AO^{ij}_{kl}}\nord{\AO^{p}_{h}}}{\Phi}\notag\\
  &=\sum_{ij}f^{i}_{j}\delta^{i}_{h}\delta^{p}_{j}n_i\nn_j = f^{h}_{p} \label{eq:coupl_1p1h}
\end{align}
and their Hermitian conjugates. The contributions from the zero-body and 
two-body pieces of the Hamiltonian vanish because they are expectation
values of normal-ordered operators in the reference state (cf.~equation \eqref{eq:nord_ex}).
Likewise, the $0p0h$ and $2p2h$ blocks are coupled by the matrix elements
\begin{equation}
\matrixe{\Phi}{H\nord{\AO^{pp'}_{hh'}}}{\Phi} = \Gamma^{hh'}_{pp'}
\end{equation}
and their conjugates. It is precisely these two-body matrix elements 
that couple $npnh$ and $(n\pm2)p(n\pm2)h$ states and generate the
outermost side diagonals of the Hamiltonian matrix. This suggests that 
we can transform the Hamiltonian to the shape shown in the top right panel
of figure \ref{fig:imsrg} by defining its offdiagonal part as
\begin{equation}
  \HO_{od} \equiv \sum_{ph}f^{p}_{h}\nord{\AO^{p}_{h}} + 
              \frac{1}{4}\sum_{pp'hh'}\Gamma^{pp'}_{hh'}\nord{\AO^{pp'}_{hh'}} + \text{H.c.}\,.
\end{equation}
In section \ref{sec:features}, we will show that the IMSRG flow does indeed
exponentially suppress the matrix elements of $\HO_{od}$ and 
achieve the desired decoupling in the limit $s\to\infty$. 

\subsubsection{Variational Derivation of Minimal Decoupling}
Our minimal decoupling scheme is of course very reminiscent of the strategy 
followed in Coupled Cluster approaches \cite{Shavitt:2009,Hagen:2014ve}, except 
that we specifically use a unitary transformation instead of a general 
similarity transformation. It is also appealing for a different reason: As 
we will discuss now, it can be derived from a variational approach, tying
the seemingly unrelated ideas of energy minimization and renormalization 
in the many-body system together. To this end, we consider the energy
expectation value of the final IMSRG evolved Hamiltonian, 
\begin{equation}
  \Hfinal \equiv \HO(\infty)\,,
\end{equation}
in the reference state (which is assumed to be normalized):
\begin{equation}
  E = \dmatrixe{\Phi}{\Hfinal}\,.
\end{equation}
Next, we introduce a \emph{unitary} variation, which we can choose to apply either 
to the reference state\,,
\begin{equation}
  \ket{\Phi}\rightarrow e^\ZO\ket{\Phi}\,,\quad\ZZO=-\ZO\,,
\end{equation}
or, equivalently, to the Hamiltonian:
\begin{equation}
  e^{\ZZO}\Hfinal e^\ZO = e^{-\ZO}\Hfinal e^\ZO\,.
\end{equation}
The variation of the energy is
\begin{align}
  \delta E = \dmatrixe{\Phi}{e^{-\ZO}(\Hfinal - E) e^{\ZO}} 
           = \dmatrixe{\Phi}{\Hfinal-E} + \dmatrixe{\Phi}{\comm{\Hfinal-E}{\ZO}} + O(||\ZO||^2)\,,
\end{align}
with a suitable operator norm $||\cdot||$. The first term obviously vanishes, as does the
commutator of $\ZO$ with the energy. Thus, the energy is stationary if
\begin{equation}
  \delta E = \dmatrixe{\Phi}{\comm{\Hfinal}{\ZO}} = 0\,. 
\end{equation}
Expanding 
\begin{equation}
  \ZO =\sum_{ph} Z^p_{h}\nord{\AO^{p}_{h}} + \frac{1}{4}\sum_{pp'hh'} Z^{pp'}_{hh'}\nord{\AO^{pp'}_{hh'}} 
        + \text{H.c.} + \ldots\,,
\end{equation}
and using the independence of the expansion coefficients (save for the unitarity
conditions), we obtain the system of equations
\begin{align}
  \dmatrixe{\Phi}{\comm{\Hfinal}{\nord{\AO^{p}_{h}}}} &= 0 \,,\\
  \dmatrixe{\Phi}{\comm{\Hfinal}{\nord{\AO^{h}_{p}}}} &= 0 \,,\\
  \dmatrixe{\Phi}{\comm{\Hfinal}{\nord{\AO^{pp'}_{hh'}}}} &= 0 \,,\\
  \dmatrixe{\Phi}{\comm{\Hfinal}{\nord{\AO^{hh'}_{pp'}}}} &= 0 \,,\\
  \ldots\notag
\end{align}
which are special cases of the so-called \emph{irreducible Brillouin conditions (IBCs)}
\cite{Mukherjee:2001uq,Kutzelnigg:2002kx,Kutzelnigg:2004vn,Kutzelnigg:2004ys}.
Writing out the commutator in the first equation, we obtain
\begin{equation}
  \dmatrixe{\Phi}{\comm{\Hfinal}{\nord{\AO^{p}_{h}}}} 
    = \dmatrixe{\Phi}{\Hfinal\nord{\AO^{p}_{h}}} - \dmatrixe{\Phi}{\nord{\AO^{p}_{h}}\Hfinal}
    = \dmatrixe{\Phi}{\Hfinal\nord{\AO^{p}_{h}}} 
    = 0\,,
\end{equation}
where the second term vanishes because it is proportional to $n_p \nn_h = 0$.
The remaining equations can be evaluated analogously, and we find that the
stationarity conditions are satisfied if the IMSRG evolved Hamiltonian $\Hfinal$
no longer couples the reference state and its particle-hole 
excitations, as discussed above.
This connection between the decoupling conditions and the stationarity conditions
of an energy functional will prove useful in the multireference case. 

\begin{figure}[t]
  \begin{center}
    \includegraphics[width=0.9\textwidth]{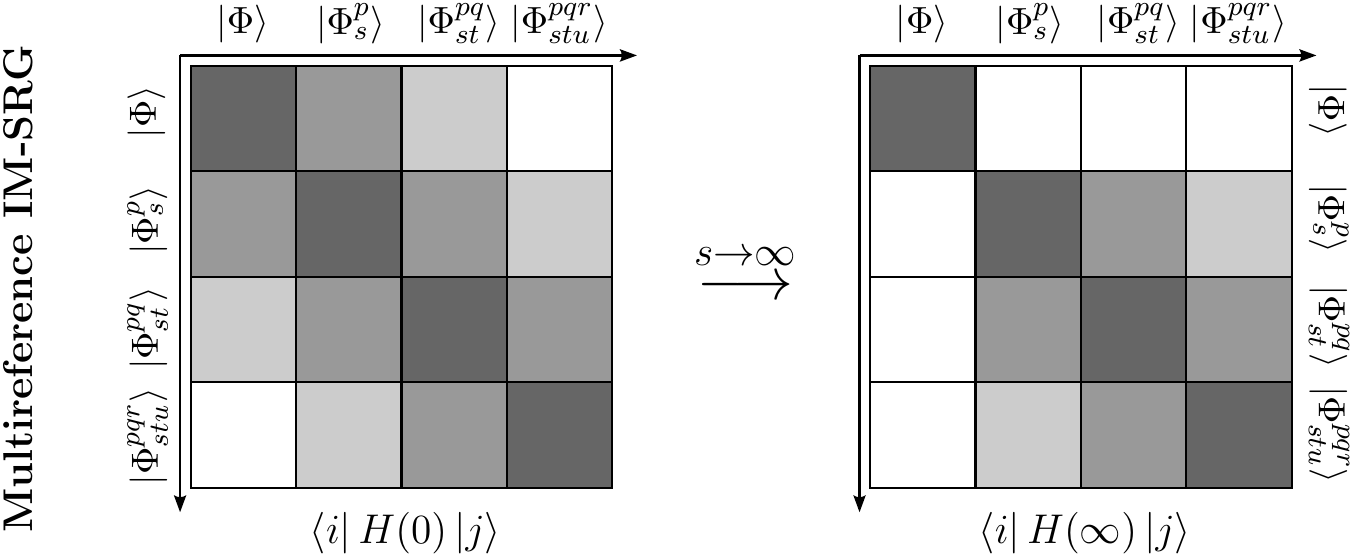}
  \end{center}
  \caption{\label{fig:mrimsrg}
  Schematic view of MR-IMSRG decoupling in the many-body Hilbert space. $\ket{\Phi}$ denotes an arbitrary 
  reference state, and $\ket{\Phi^{p\ldots}_{s\ldots}}$ are suitably defined quasi-particle excitations. }
\end{figure}

\subsubsection{\label{sec:mr_decoupling}MR-IMSRG Decoupling for Correlated Reference States}
In the multireference case, we choose a suitable correlated reference 
state, and construct its excitations by applying all possible one- 
and two-body operators:
\begin{equation}
  \ket{\Phi},\,\nord{\AO^{i}_{j}}\ket{\Phi},\,\nord{\AO^{ij}_{kl}}\ket{\Phi},\ldots\,.
\end{equation}
The properties of the normal ordering ensure that the excited states are
orthogonal to the reference state, but they are in general not orthogonal 
to each other: for instance,
\begin{equation}
  \dmatrixe{\Phi}{\nord{A^{i}_{j}}\nord{A^{k}_{l}}}
  =-\lambda^{i}_{l}\xi^{k}_{j} + \lambda^{ij}_{kl}
  =n_i\nn_j \delta^{i}_{l}\delta^{k}_{j} + \lambda^{ij}_{kl}\,,
\end{equation}
where $0\leq n_i, \nn_i \leq 1$. Moreover, there can be linear dependencies
between the excitations of the correlated reference state, so the matrix
representations of the Hamiltonian and other operators in this basis can 
be rank deficient. This poses a major challenge for multireference CC and
related schemes that obtain solutions of the many-body Schr\"odinger equation
by iterating a system of nonlinear equations. Numerical algorithms for solving
systems of nonlinear equations fail if the Jacobian of the system is singular,
and therefore one first must construct linearly independent excitations, e.g.,
through a costly diagonalization of the overlap matrix. For the MR-IMSRG,
in contrast, the linear dependencies merely imply that the flow is implicitly 
operating on a rank-deficient matrix that has additional spurious
zero eigenvalues. These eigenvalues are usually far removed from the low-lying 
part of the spectrum that interests us most. 

As shown in figure \ref{fig:mrimsrg}, the matrix representation of the 
initial NO2B Hamiltonian in the chosen excited states is again band-diagonal,
just like in the single-reference case. Following the minimal decoupling
approach discussed before, we want to transform the Hamiltonian to the
shape that is shown in the right panel of figure \ref{fig:mrimsrg}, with
\begin{align}
  \matrixe{\Phi}{\Hfinal\nord{\AO^{i}_{j}}}{\Phi}&=0\,,\label{eq:mr_decoupling1B}\\
  \matrixe{\Phi}{\Hfinal\nord{\AO^{ij}_{kl}}}{\Phi}&=0\,,\label{eq:mr_decoupling2B}\\
  \ldots& \notag
\end{align}
and corresponding conditions for the conjugate matrix elements. The matrix 
elements can be evaluated with the generalized Wick's theorem (see
section \ref{sec:no}), e.g.,
\begin{align}
  \dmatrixe{\Phi}{H\nord{\AO^{i}_{j}}}
  =\nn_i n_j f^{j}_{i}+\sum_{ab}f^{a}_{b}\lambda^{ai}_{bj}
   +\frac{1}{2}\sum_{abc} 
    \left( \nn_i \lambda^{bc}_{ja}\Gamma^{bc}_{ia} - n_j \Gamma^{ja}_{bc} \lambda^{ia}_{bc} \right)
   +\frac{1}{4}\sum_{abcd}\Gamma^{ab}_{cd}\lambda^{iab}_{jcd}\,.\label{eq:mr_coupling_1b}
\end{align}
Note that the first term is merely the generalization of the one-body
particle-hole matrix element, equation \eqref{eq:coupl_1p1h}: In the single-reference 
limit, the occupation number prefactor is nonzero if $i$ and
$j$ are particle and hole indices, respectively. In addition, the matrix
element depends on the irreducible densities $\lambda^{(2)}$ and $\lambda^{(3)}$
in a nontrivial manner due to the coupling of the Hamiltonian to 
correlated pairs and triples of nucleons in the reference state. The matrix 
element between the reference state and the two-body excitation is even more 
complicated:
\begin{align}
  &\matrixe{\Phi}{\HO\nord{\AO^{ij}_{kl}}}{\Phi}\notag\\
  &=\nn_i\nn_j n_k n_l \Gamma^{kl}_{ij}
    +(1-P_{kl})n_l\sum_{b}f^{l}_{b}\lambda^{ij}_{bk}
    -(1-P_{ij})\nn_{j}\sum_{a}f^{a}_{j}\lambda^{ai}_{kl}
    \notag\\[3pt]
  &\hphantom{=}
    +\frac{1}{2}\nn_i\nn_j \sum_{ab}\Gamma^{ab}_{ij}\lambda^{ab}_{kl}
    +\frac{1}{2}n_k n_l    \sum_{cd}\Gamma^{kl}_{cd}\lambda^{ij}_{cd} 
    \notag\\[3pt]
  &\hphantom{=}
    -(1-P_{ij})(1-P_{kl})n_k\nn_i \sum_{ad}\Gamma^{ak}_{id}\lambda^{aj}_{dl}
    \notag\\[3pt]
  &\hphantom{=}
    +\frac{1}{4}\sum_{abcd}\Gamma^{ab}_{cd}\lambda^{ab}_{kl}\lambda^{ij}_{cd}
    +\frac{1}{4}(1-P_{ij})\sum_{abcd}\Gamma^{ab}_{cd}
        \left(\lambda^{aj}_{kl}\lambda^{bi}_{cd}-\lambda^{ai}_{cd}\lambda^{bj}_{kl}\right)
    \notag\\[3pt]
  &\hphantom{=}
    -\frac{1}{2}(1-P_{kl})\sum_{abcd}\Gamma^{ab}_{cd}\lambda^{ab}_{ck}\lambda^{ij}_{dl}
    +\frac{1}{2}(1-P_{ij})(1-P_{kl})\sum_{abcd}\Gamma^{ab}_{cd}\lambda^{ai}_{ck}\lambda^{bj}_{dl}
    \notag\\[3pt]
  &\hphantom{=}
    +\sum_{ab}f^{a}_{b}\lambda^{aij}_{bkl}
    +\frac{1}{2}(1-P_{kl})n_k   \sum_{bcd}\Gamma^{kb}_{cd}\lambda^{bij}_{cdl} 
    -\frac{1}{2}(1-P_{ij})\nn_i \sum_{abd}\Gamma^{ab}_{id}\lambda^{abj}_{dkl}
    \notag\\[3pt]
  &\hphantom{=}
    +\frac{1}{4}\sum_{abcd}\Gamma^{ab}_{cd}\lambda^{abij}_{cdkl}\,,\label{eq:mr_coupling_2b}
\end{align}
with the permutation symbol $P_{ab}$ defined by (cf.~\ref{app:PNP})
\begin{equation}
  P_{ij} g(\ldots,i,\ldots,j) \equiv g(\ldots,j,\ldots,i)\,.
\end{equation}
Again, only the first term appears in the single-reference limit. While
the storage and manipulation of $\lambda^{(3)}$ is feasible for certain
types of reference states, treating $\lambda^{(4)}$ is essentially out
of the question. Thus, we are forced to introduce truncations in
equations \eqref{eq:mr_coupling_1b} and \eqref{eq:mr_coupling_2b}, and any
decoupling we can achieve will no longer be exact, in general. 

This is where the variational perspective introduced before becomes useful.
Formally, we can write, 
\begin{align}
  \dmatrixe{\Phi}{H\nord{\AO^{i}_{j}}} 
    &= \frac{1}{2}\dmatrixe{\Phi}{\acomm{H}{\nord{\AO^{i}_{j}}}} 
    + \frac{1}{2}\dmatrixe{\Phi}{\comm{H}{\nord{\AO^{i}_{j}}}}\,,\\
  \dmatrixe{\Phi}{H\nord{\AO^{ij}_{kl}}} 
    &= \frac{1}{2}\dmatrixe{\Phi}{\acomm{H}{\nord{\AO^{ij}_{kl}}}} 
    + \frac{1}{2}\dmatrixe{\Phi}{\comm{H}{\nord{\AO^{ij}_{kl}}}}\,,
\end{align}
and suppress the second term in a clean and controlled manner through 
what amounts to a minimization of the ground-state energy through unitary 
variation. As discussed above, the energy is stationary if the 
IBCs for the multireference case are satisfied \cite{Mukherjee:2001uq,Kutzelnigg:2002kx,Kutzelnigg:2004vn,Kutzelnigg:2004ys}:
\begin{align}
  \dmatrixe{\Phi}{\comm{\Hfinal}{\nord{\AO^{i}_{j}}}}&=0\,,\label{eq:ibc1}\\
  \dmatrixe{\Phi}{\comm{\Hfinal}{\nord{\AO^{ij}_{kl}}}}&=0\,.\label{eq:ibc2}
\end{align}
Evaluating the commutators with the generalized Wick's theorem of
section \ref{sec:no}, we obtain
\begin{align}
  \dmatrixe{\Phi}{\comm{H}{:\AO^{i}_{j}:}}
    &=(n_j-n_i)f^{j}_{i}-\frac{1}{2}\sum_{abc}\left(  
     \Gamma^{ja}_{bc}\lambda^{ia}_{bc}
    -\Gamma^{ab}_{ic}\lambda^{ab}_{jc}
    \right)\,,
\end{align}
\begin{align}
  &\dmatrixe{\Phi}{\comm{H}{:\AO^{ij}_{kl}:}}\notag\\
  &=
    \Gamma^{kl}_{ij}(\nn_i\nn_jn_kn_l - n_in_j\nn_k\nn_l)
  \notag\\[3pt]
  &\hphantom{=}
  +\sum_{a}
     \left(
     (1-P_{ij})f^{a}_{i}\lambda^{aj}_{kl}
    -(1-P_{kl})f^{k}_{a}\lambda^{ij}_{al}
    \right)
    \notag\\[3pt]
  &\hphantom{=}
    +\frac{1}{2}\left(
      (\lambda\Gamma)^{kl}_{ij}\left(1-n_i-n_j\right)
      -(\Gamma\lambda)^{kl}_{ij}\left(1-n_k-n_l\right)
    \right)
  \notag\\[3pt]
  &\hphantom{=}
    +(1-P_{ij})(1-P_{kl})\sum_{ac}
       \left(n_j-n_k\right)\Gamma^{ak}_{cj}\lambda^{ai}_{cl}
    \notag\\[3pt]
  &\hphantom{=}
    +\frac{1}{2}\sum_{abc}\left(
       (1-P_{kl})\Gamma^{ka}_{bc}\lambda^{aij}_{bcl} 
      -(1-P_{ij})\Gamma^{ab}_{ic}\lambda^{abj}_{ckl} 
    \right)\,.
\end{align}
Like the MR-IMSRG(2) flow equations \eqref{eq:mr_flow_0b_tens}--\eqref{eq:mr_flow_2b_tens},
these expressions only depend linearly on $\lambda^{(2)}$ and $\lambda^{(3)}$,
which makes untruncated implementations feasible.

By driving the Hamiltonian to a shape that satisfies the IBCs,
we will achieve at least some reduction of the coupling between
the reference state and excitations. The same also holds for the
side diagonals to some extent, although they have a much richer
structure than in the single-reference case. We also note that 
this is achieved by evolving the \emph{individual} one-body and two-body pieces of
the Hamiltonian, not just the specific linear combinations that 
enter into the decoupling conditions and IBCs. Thus, 
$\dmatrixe{\Phi}{\acomm{\HO}{\nord{\AO^{i}_{j}}}}$
and $\dmatrixe{\Phi}{\acomm{\HO}{\nord{\AO^{ij}_{kl}}}}$
will also be altered by the flow, and there is empirical evidence
for their reduction. A detailed investigation will be presented elsewhere.

\subsection{\label{sec:generators}Generators}
In the previous subsection we have specified the decoupling we want to
achieve, leading us to a definition of the offdiagonal part of the 
Hamiltonian that must be suppressed by the MR-IMSRG evolution. However,
we still have enormous freedom in choosing generators that implement
this decoupling, especially if we are only interested in the limit
$s\to\infty$ \cite{Hergert:2016jk}. Here, we restrict ourselves to those 
generators that we found to be most useful in practical applications. Let 
us discuss the single-reference case first.

\subsubsection{Construction of Generators for Single-Reference Applications}
A wide range of suitable generators for the single-reference case
is covered by the ansatz
\begin{equation}
  \eta= \sum_{ph} \eta^{p}_{h}\nord{\AO^{p}_{h}}
        +\frac{1}{4}\sum_{pp'hh'} \eta^{pp'}_{hh'}\nord{\AO^{pp'}_{hh'}} - \text{H.c.}\,,
\end{equation}
constructing the one- and two-body matrix elements directly from those
of the offdiagonal Hamiltonian and an object $G$ that ensures the 
anti-Hermiticity of $\eta$:
\begin{align}
  \eta^{p}_{h}     &\equiv G^{p}_{h} f^{p}_{h}\,,\\
  \eta^{pp'}_{hh'} &\equiv G^{pp'}_{hh'} \Gamma^{pp'}_{hh'}\,.
\end{align}
To see possible options for $G$, we consider the single-reference flow
equations in perturbation theory (see Ref.~\cite{Hergert:2016jk}
for a detailed discussion). We assume a Hartree-Fock reference state,
and partition the Hamiltonian as
\begin{equation}
  \HO = \HO_0 + g \HO_I\,,
\end{equation}
with 
\begin{align}
  \HO_0 &\equiv E + \sum_{i} f^{i}_{i}\nord{\AO^{i}_{i}}
                 + \frac{1}{4} \sum_{ij} \Gamma^{ij}_{ij}\nord{\AO^{ij}_{ij}}\,,\\
  \HO_I &\equiv \sum_{ij}^{i\neq j} f^{i}_{j}\nord{\AO^{i}_{j}} + \frac{1}{4} \sum_{ijkl}^{ij\neq kl} \Gamma^{ij}_{kl}\nord{\AO^{ij}_{kl}}\,.
\end{align}
In the space of up to $2$p$2$h excitations, this corresponds to 
a second-quantized form of Epstein-Nesbet partitioning 
\cite{Epstein:1926fp,Nesbet:1955lq}, and treats the proper diagonal
matrix elements in the aforementioned blocks of the Hamiltonian
matrix as unperturbed. Note that the one-body piece of the initial 
Hamiltonian is diagonal in the HF orbitals, which implies 
$f^{p}_{h}, \eta^{p}_{h}=0$. Inspecting the one-body flow equation, we see 
that corrections to $f$ that are induced during the flow are
at least of order $\OC(g^2)$, because no diagonal matrix elements 
of $\Gamma$ appear:
\begin{align}
  \left.\totd{}{s}f^{i}_{j}\right|_{s=0} &=
    \frac{1}{2}\sum_{abc}
    \left(\eta^{ia}_{bc}\Gamma^{bc}_{ja}-\Gamma^{ia}_{bc}\eta^{bc}_{ja}\right)\left(n_{a}\bar{n}_{b}\bar{n}_{c}+\bar{n}_{a}n_{b}n_{c}\right)=\OC(g^2)\,.
\end{align}
Using this knowledge, the two-body flow equation for the pphh matrix elements 
of the offdiagonal Hamiltonian reads
\begin{align}
  \totd{}{s}\Gamma^{pp'}_{hh'}&=  
  -\left(f^{p}_{p}+f^{p'}_{p'}-f^{h}_{h}-f^{h'}_{h'}\right)\eta^{pp'}_{hh'}
  -\left(\Gamma^{hh'}_{hh'}+\Gamma^{pp'}_{pp'}\right)\eta^{pp'}_{hh'}
  \notag\\
  &\hphantom{=}
    +\left(\Gamma^{p'h'}_{p'h'}+\Gamma^{ph}_{ph}+\Gamma^{ph'}_{ph'}+\Gamma^{p'h}_{p'h}
     \right)\eta^{pp'}_{hh'} + \OC(g^2)
  \notag\\
  &=-\Delta^{pp'}_{hh'}\eta^{pp'}_{hh'} + \OC(g^2)\,.\label{eq:flow2b_pert}
\end{align}
Note that the factors $\tfrac{1}{2}$ in the particle-particle and
hole-hole ladder summation (line 2 of equation \eqref{eq:mr_flow_2b_tens}) are 
canceled by factors $2$ from the unrestricted summation over indices, e.g.,
\begin{equation}
  \frac{1}{2}\sum_{h_1h_2}\eta^{pp'}_{h_1h_2}\Gamma^{h_1h_2}_{hh'}(1-n_{h_1}-n_{h_2})
  =-\frac{1}{2}\eta^{pp'}_{hh'}\Gamma^{hh'}_{hh'}-\frac{1}{2}\eta^{pp'}_{h'h}\Gamma^{h'h}_{hh'}
  =-\eta^{pp'}_{hh'}\Gamma^{hh'}_{hh'}\,.
\end{equation}
In equation \eqref{eq:flow2b_pert}, we have introduced the quantity 
\begin{align}
  \Delta^{pp'}_{hh'}&\equiv
    f^{p}_{p}+f^{p'}_{p'}-f^{h}_{h}-f^{h'}_{h'} + \Gamma^{hh'}_{hh'}+\Gamma^{pp'}_{pp'}
    -\Gamma^{ph}_{ph}-\Gamma^{p'h'}_{p'h'}-\Gamma^{ph'}_{ph'}-\Gamma^{p'h}_{p'h} \notag\\
    &=\dmatrixe{\Phi}{\nord{\AO^{hh'}_{pp'}}\HO\nord{\AO^{pp'}_{hh'}}}-\dmatrixe{\Phi}{\HO}\notag\\
    &=\dmatrixe{\Phi}{\nord{\AO^{hh'}_{pp'}}\HO_0\nord{\AO^{pp'}_{hh'}}}-\dmatrixe{\Phi}{\HO_0}\,,
    \label{eq:def_epsteinnesbet_2b}
\end{align}
i.e., the unperturbed energy difference between the two states
that are coupled by the matrix element $\Gamma^{pp'}_{hh'}$, namely
the reference state $\ket{\Phi}$ and the excited state $\nord{\AO^{pp'}_{hh'}}\ket{\Phi}$.
Since it is expressed in terms of diagonal matrix elements, $\Delta^{pp'}_{hh'}$
would appear in precisely this form in appropriate energy denominators
of Epstein-Nesbet perturbation theory. 

Plugging our ansatz for $\eta$ into equation \eqref{eq:flow2b_pert}, we obtain
\begin{equation}
  \totd{}{s}\Gamma^{pp'}_{hh'} 
  =-\Delta^{pp'}_{hh'}G^{pp'}_{hh'}\Gamma^{pp'}_{hh'} + \OC(g^2)\,,
\end{equation}
Neglecting $\OC(g^2)$ terms in the flow equations, the one-body part of $\HO$ 
remains unchanged, and assuming that $G$ itself is independent of $s$
at order $\OC(g)$, we can integrate equation \eqref{eq:flow2b_pert}:
\begin{equation}\label{eq:Gamma_pert}
  \Gamma^{pp'}_{hh'}(s) = \Gamma^{pp'}_{hh'}(0) e^{-\Delta^{pp'}_{hh'}G^{pp'}_{hh'} s}\,.
\end{equation}
Clearly, the offdiagonal matrix elements of the Hamiltonian will be
suppressed for $s\to\infty$, provided the product $\Delta^{pp'}_{hh'} G^{pp'}_{hh'}$ 
is positive. $G^{pp'}_{hh'}$ also allows us
to control the details of this suppression, e.g., the decay scales.
To avoid misconceptions, we stress that we do not impose perturbative
truncations in practical applications, and treat all matrix elements 
and derived quantities, including the $\Delta^{pp'}_{hh'}$, as 
$s$-dependent.

\subsubsection{The Imaginary-Time Generator}
Using $G^{pp'}_{hh'}$ to ensure that the energy denominator is always
positive, we obtain the so-called \emph{imaginary-time generator}
\cite{Morris:2015ve,Hergert:2014vn,Hergert:2016jk},
which is inspired by imaginary-time evolution techniques that are 
frequently used in Quantum Monte Carlo methods, for instance (see, 
e.g., \cite{Carlson:2015lq} and references therein). Explicitly 
indicating the flow parameter dependence of all quantities, we
define
\begin{align}
  \etaO^\text{IT}(s)&\equiv\sum_{ph} \sgn\!\left(\Delta^{p}_{h}(s)\right) f^{p}_{h}(s)\nord{\AO^{p}_{h}}\notag\\
  &\hphantom{=}
   +\frac{1}{4}\sum_{pp'hh'}\sgn\!\left(\Delta^{pp'}_{hh'}(s)\right)\Gamma^{pp'}_{hh'}(s)
    \nord{\AO^{pp'}_{hh'}}-\text{H.c.}\,,\label{eq:eta_imtime}
\end{align}
where
\begin{align}
  \Delta^{p}_{h}&\equiv
    f^{p}_{p}-f^{h}_{h} + \Gamma^{ph}_{ph} 
    =\dmatrixe{\Phi}{\nord{\AO^{h}_{p}}\HO\nord{\AO^{p}_{h}}}-\dmatrixe{\Phi}{\HO}\,.\label{eq:def_epsteinnesbet_1b}
\end{align}

For this generator, the perturbative analysis of the offdiagonal 
two-body matrix elements yields
\begin{equation}
  \Gamma^{pp'}_{hh'}(s) = \Gamma^{pp'}_{hh'}(0) e^{-|\Delta^{pp'}_{hh'}|s}\,,
\end{equation}
ensuring that they are driven to zero by the evolution. We also
note that the energy difference $\Delta^{pp'}_{hh'}$ controls
the scales of the decay. Matrix elements between states with large 
energy differences are suppressed more rapidly than those which
couple states that are close in energy. This means that $\eta^\text{IT}$ 
generates a proper renormalization group flow \cite{Kehrein:2006kx,Hergert:2016jk}.

\subsubsection{\label{sec:white}The White Generator}
A generator that is particularly powerful in numerical applications goes back
to the work of White on canonical transformation theory in quantum chemistry 
\cite{White:2002fk,Tsukiyama:2011uq,Hergert:2016jk}. In the language we have
set up above, it uses $G^{pp'}_{hh'}$ to \emph{remove} the scale dependence
of the IMSRG flow. The White generator is defined as
\begin{align}
  \etaO^\text{W}(s)
  &\equiv\sum_{ph}\frac{f^{p}_{h}(s)}{\Delta^{p}_{h}(s)}\nord{\AO^{p}_{h}}
  +\frac{1}{4}\sum_{pp'hh'}\frac{\Gamma^{pp'}_{hh'}(s)}{\Delta^{pp'}_{hh'}(s)}
    \nord{\AO^{pp'}_{hh'}}-\;\text{H.c.}\label{eq:eta_white}\,,
\end{align}
where the Epstein-Nesbet denominators use the energy differences defined in 
equations \eqref{eq:def_epsteinnesbet_2b} and \eqref{eq:def_epsteinnesbet_1b}.

Referring again to our perturbative analysis of the offdiagonal 
two-body matrix elements, we find
\begin{equation}
  \Gamma^{pp'}_{hh'}(s) = \Gamma^{pp'}_{hh'}(0) e^{-s}\,,
\end{equation}
i.e., the White generator suppresses \emph{all} offdiagonal matrix 
elements simultaneously with a decay scale identical (or close to) 1 
\cite{Hergert:2016jk}. 
While this means that $\eta^\text{W}$ does \emph{not} generate a proper RG
flow, this is inconsequential if we are only interested in the final Hamiltonian 
$\HO(\infty)$, because all unitary transformations which suppress $\HO_{od}$ 
must be equivalent up to truncation effects \cite{Hergert:2016jk}.

A benefit of the White generator is that its matrix elements are defined 
as ratios of energies, and therefore the Hamiltonian only contributes linearly 
to the magnitude of the right-hand side of the flow equations \eqref{eq:mr_flow_0b_tens}--
\eqref{eq:mr_flow_2b_tens}. This leads to a significant reduction of the 
ODE system's stiffness compared to the other generators discussed here or in
Ref.~\cite{Hergert:2016jk}, and greatly reduces the numerical effort for the
ODE solver. However, the dependence of $\etaO^\text{W}$ on energy denominators
can also be a drawback if $\Delta^{p}_{h}$ and/or $\Delta^{pp'}_{hh'}$ become
small, and cause some of its matrix elements to diverge. This can be
mitigated by using an alternative ansatz that is also inspired by White's work 
\cite{White:2002fk}:
\begin{align}
  \etaO^\text{W'}(s)
  &\equiv\frac{1}{2}\sum_{ph}\arctan\frac{2 f^{p}_{h}(s)}{\Delta^{p}_{h}(s)}\nord{\AO^{p}_{h}}
  +\frac{1}{8}\sum_{pp'hh'}\arctan\frac{2\Gamma^{pp'}_{hh'}(s)}{\Delta^{pp'}_{hh'}(s)}
    \nord{\AO^{pp'}_{hh'}}-\;\text{H.c.}\label{eq:eta_whiteatan}\,.
\end{align}
This form emphasizes that the unitary transformation can be thought of as an
abstract rotation of the Hamiltonian. The matrix elements of $\eta^\text{W'}$
are regularized by the $\arctan$ function, and explicitly limited to the 
interval $]-\tfrac{\pi}{4},\tfrac{\pi}{4}[$. Expanding the function for 
small arguments, we recover our initial ansatz for the White generator, 
equation \eqref{eq:eta_white}.

\subsubsection{Generators for the Multireference Case}
The imaginary-time and White generators introduced in the previous
subsections can be generalized to the multireference case by evaluating
$\dmatrixe{\Phi}{\HO\nord{\AO^{i}_{j}}}$, $\dmatrixe{\Phi}{\HO\nord{\AO^{ij}_{kl}}}$
and the diagonal matrix elements $\dmatrixe{\Phi}{\nord{\AO^{j}_{i}}\HO\nord{\AO^{i}_{j}}}$
$\dmatrixe{\Phi}{\nord{\AO^{kl}_{ij}}\HO\nord{\AO^{ij}_{kl}}}$ that enter
equations \eqref{eq:def_epsteinnesbet_1b} and \eqref{eq:def_epsteinnesbet_2b} with 
the Wick's theorem for correlated reference states. As we have seen in 
section \ref{sec:decoupling}, the offdiagonal matrix element, 
equation \eqref{eq:mr_coupling_2b}, depends on $\lambda^{(4)}$, and 
the diagonal matrix elements contain terms that are proportional to
$\lambda^{(5)}$ and $\lambda^{(6)}$, in general. This forces us to 
introduce approximations that may adversely impact the behavior of 
the MR-IMSRG flow for the ground-state energy, e.g., by causing 
oscillations (see section \ref{sec:flow_decoupling} for an example).

In section \ref{sec:mr_decoupling}, we argued that a formally cleaner
approach can be devised where we forego the decoupling
conditions \eqref{eq:mr_decoupling1B} and \eqref{eq:mr_decoupling2B}
in favor of the IBCs \eqref{eq:ibc1} and
\eqref{eq:ibc2}, tying the MR-IMSRG flow to a variational minimization
of the energy under unitary transformations. We define our so-called 
\emph{Brillouin} generator as
\begin{equation}
  \eta^\text{B}\equiv\sum_{ij}\eta^{i}_{j}\nord{\AO^{i}_{j}} 
    + \frac{1}{4}\sum_{ijkl}\eta^{ij}_{kl}\nord{\AO^{ij}_{kl}}\,,\label{eq:eta_brillouin}
\end{equation}
with the one- and two-body matrix elements given by 
\begin{align}
  \eta^{i}_{j}&\equiv
    \dmatrixe{\Phi}{\comm{H}{:\AO^{i}_{j}:}}
    =(n_j-n_i)f^{j}_{i}-\frac{1}{2}\sum_{abc}\left(  
     \Gamma^{ja}_{bc}\lambda^{ia}_{bc}
    -\Gamma^{ab}_{ic}\lambda^{ab}_{jc}
    \right)\,,
    \\[10pt]
%
  \eta^{ij}_{kl}&\equiv\dmatrixe{\Phi}{\comm{H}{:\AO^{ij}_{kl}:}}\notag\\
  &=
    \Gamma^{kl}_{ij}(\nn_i\nn_jn_kn_l - n_in_j\nn_k\nn_l)
  \notag\\[3pt]
  &\hphantom{=}
  +\sum_{a}
     \left(
     (1-P_{ij})f^{a}_{i}\lambda^{aj}_{kl}
    -(1-P_{kl})f^{k}_{a}\lambda^{ij}_{al}
    \right)
    \notag\\[3pt]
  &\hphantom{=}
    +\frac{1}{2}\left(
      (\lambda\Gamma)^{kl}_{ij}\left(1-n_i-n_j\right)
      -(\Gamma\lambda)^{kl}_{ij}\left(1-n_k-n_l\right)
    \right)
  \notag\\[3pt]
  &\hphantom{=}
    +(1-P_{ij})(1-P_{kl})\sum_{ac}
       \left(n_j-n_k\right)\Gamma^{ak}_{cj}\lambda^{ai}_{cl}
    \notag\\[3pt]
  &\hphantom{=}
    +\frac{1}{2}\sum_{abc}\left(
       (1-P_{kl})\Gamma^{ka}_{bc}\lambda^{aij}_{bcl} 
      -(1-P_{ij})\Gamma^{ab}_{ic}\lambda^{abj}_{ckl} 
    \right)\,.
\end{align}
Like the MR-IMSRG(2) flow equations \eqref{eq:mr_flow_0b_tens}--\eqref{eq:mr_flow_2b_tens},
$\eta^\text{B}$ only depends on $\lambda^{(2)}$ and $\lambda^{(3)}$,
and higher-rank irreducible density matrices appear only linearly.

Because the matrix elements of $\eta^\text{B}$ are directly given
by the residuals of the IBCs, it can be interpreted as the \emph{gradient}
of the energy with respect to the parameters of the unitary transformation 
at each step of the flow. At the fixed point of the flow, $\eta^\text{B}=0$,
and the flowing zero-body part of the Hamiltonian, $E(\infty)$, will be an extremum 
of the energy. Indeed, $\eta^\text{B}$ has behaved in this manner in all 
numerical applications to date, generating a monotonic flow of the  
energy towards the converged results (see section \ref{sec:flow_gs}).

We conclude this section by elucidating the relation between the
Brillouin and imaginary-time generators. The multireference version of
the latter has the matrix elements
\begin{align}
  \left(\eta^\text{IT}\right)^{i}_{j}
   &=\sgn\left(\Delta^{i}_{j}\right)\dmatrixe{\Phi}{\HO\nord{\AO^{i}_{j}}}
     -\sgn\left(\Delta^{j}_{i}\right)\dmatrixe{\Phi}{\HO\nord{\AO^{j}_{i}}}^*\,,\\
  \left(\eta^\text{IT}\right)^{ij}_{kl}
   &=\sgn\left(\Delta^{ij}_{kl}\right)\dmatrixe{\Phi}{\HO\nord{\AO^{ij}_{kl}}}
     -\sgn\left(\Delta^{kl}_{ij}\right)\dmatrixe{\Phi}{\HO\nord{\AO^{kl}_{ij}}}^*\,.
\end{align}
Note that it is not possible to express $\Delta^{i}_{j}$ in terms of $\Delta^{j}_{i}$, 
because they are matrix elements of the Hamiltonian in the 
unrelated states $\nord{\AO^{i}_{j}}\ket{\Phi}$ and 
$\nord{\AO^{j}_{i}}\ket{\Phi}$. The same is the
case for the two-body excitation. However, if \emph{all} basic excitations
have higher energies than the reference state for all values of $s$, i.e.,
$\Delta^{i}_{j},\Delta^{ij}_{kl}>0$, the matrix elements read
\begin{align}
  \left(\eta^\text{IT}\right)^{i}_{j}
   &=\dmatrixe{\Phi}{\HO\nord{\AO^{i}_{j}}}
     -\dmatrixe{\Phi}{\HO\nord{\AO^{j}_{i}}}^*\notag\\
   & 
    =\dmatrixe{\Phi}{\HO\nord{\AO^{i}_{j}}}
     -\dmatrixe{\Phi}{\nord{\AO^{i}_{j}}\HO}
    =\dmatrixe{\Phi}{\comm{\HO}{\nord{\AO^{i}_{j}}}}\,,\\
  \left(\eta^\text{IT}\right)^{ij}_{kl}
   &=\dmatrixe{\Phi}{\comm{\HO}{\nord{\AO^{ij}_{kl}}}}\,,\label{eq:eta_it_brillouin}
\end{align}
and the Brillouin and imaginary-time generators are identical. Of course, is it difficult to
ascertain in general that the condition $\Delta^{i}_{j},\Delta^{ij}_{kl}>0$ 
is satisfied in the multireference case. In the single-reference limit, 
on the other hand, the condition reduces to $\Delta^{p}_{h},\Delta^{pp'}_{hh'}>0$, 
which is typically satisfied if we start from HF reference states
for nuclei with strong shell closures.

\section{\label{sec:features}Features of IMSRG and MR-IMSRG Flows}
Before we launch into the discussion of prior and new MR-IMSRG
ground state results in section \ref{sec:groundstate}, we want to illustrate 
some of the features of MR-IMSRG flows with concrete numerical examples.
More details can also be found in Ref.~\cite{Hergert:2016jk}.

\subsection{\label{sec:flow_implementation}Interactions and Implementation}
Let us start by providing some details on the implementation and typical
interactions, for use both here and in later sections. 

As made evident in section \ref{sec:no}, we use the intrinsic nuclear Hamiltonian
in our calculations, and employ interactions from chiral EFT both with
and without free-space SRG evolution (section \ref{sec:srg}). Our
primary choice for the $NN$ sector is the \NNNLO{} interaction by Entem
and Machleidt, with cutoff $\Lambda_{NN}=500\,\MeV/c$ \cite{Entem:2003th,Machleidt:2011bh}.
Unless specifically stated otherwise, this interaction will be supplemented
by a local \NNLO{} $3N$ interaction with cutoff $\Lambda_{3N}=400\,\MeV/c$
\cite{Roth:2011kx,Gazit:2009qf}. This Hamiltonian, referred to as $NN\!+\!3N(400)$
in the following, has been used widely in the \emph{ab initio} nuclear structure 
literature in recent years, serving as the ``parent'' for families of
interactions that are generated by varying $\Lambda_{NN}$, $\Lambda_{3N}$
and the SRG resolution scale $\lambdaSRG$. 

The shortcomings of the $NN\!+\!3N(400)$ Hamiltonian, e.g., the underestimation
of nuclear charge radii or the overbinding of $pf$-shell nuclei (see section 
\ref{sec:groundstate} and 
\cite{Binder:2013zr,Binder:2014fk,Soma:2014eu,Hergert:2013mi,Hergert:2014vn}), 
have sparked efforts to derive and optimize next-generation chiral forces
\cite{Ekstrom:2013uq,Ekstrom:2015fk,Hebeler:2015qr,Epelbaum:2015fb,Binder:2016la,Entem:2015qf,Entem:2015hl,Gezerlis:2014zr,Lynn:2016ec}. In section \ref{sec:groundstate}, we will present MR-IMSRG results
with one of the first new $NN\!+\!3N$ interactions that resulted from these efforts,
\NNLOsat{} \cite{Ekstrom:2015fk}. By taking select many-body data into
account in the optimization procedure, the creators of \NNLOsat{} were able
to improve the interaction's saturation properties, allowing an accurate 
description of the ground-state energies and radii of $\nuc{Ca}{40,48}$ 
\cite{Hagen:2015ve,Garcia-Ruiz:2016fk}.

We perform our calculations in a spherical harmonic oscillator (SHO) configuration space,
with a truncation in the energy quantum number:
\begin{equation}
  e = (2n+l) \leq \eMax\,.
\end{equation}
While mature techniques to extrapolate results to infinite HO bases
are available \cite{Coon:2012uq,Furnstahl:2012ys,More:2013bh,Furnstahl:2014vn,Furnstahl:2015sf,Wendt:2015zl,Odell:2016qq},
we limit ourselves to finite bases here, using sufficiently large $\eMax$
values to eliminate the single-particle basis truncation as a relevant
source of uncertainty, typically up to $\eMax=14$ (15 major HO shells). 
An additional truncation is necessary to manage the enormous memory 
requirements of $3N$ interaction matrix elements. We only keep matrix 
elements involving three-body HO states that satisfy
\begin{equation}
  e_1 + e_2 + e_3 \leq \EMax\,.
\end{equation}
For nuclei up to the calcium and nickel region, careful analyses have 
shown that it is sufficient to use $\EMax=14$ or $16$ \cite{Binder:2014fk,Hergert:2014vn}
for soft interactions like $NN\!+\!3N(400)$. These $\EMax$ require $\sim5$GB 
and $\sim25$ GB of memory, respectively, to store matrix elements in single 
precision. This exponential growth makes it challenging to push calculations
to heavier nuclei, and it is clearly not feasible to store the entirety of the $3N$ 
interaction for a given $\eMax$, which would require $\EMax=3\eMax.$

Reference states for closed- and open-shell nuclei are obtained by solving 
spherical Hartree-Fock and Hartree-Fock-Bogoliubov equations, respectively, 
using the code described in Ref.~\cite{Hergert:2009zn}. In this calculation
step, $3N$ interactions can be included exactly (up to $\eMax$ and $\EMax$
truncations). The HFB solutions are projected on good proton and neutron 
numbers, yielding a correlated state that must be treated in the 
multireference formalism. Details 
on the calculation of the irreducible density matrices of particle-number 
projected HFB states, referred to as \emph{PNP reference states} in the following, 
can be found in \ref{app:PNP}.

With the reference state and its density matrices at our disposal,
we normal-order the Hamiltonian using the techniques discussed in 
section \ref{sec:no}, discard the residual $3N$ interaction, and eventually 
perform the IMSRG(2) or MR-IMSRG(2) evolution.

\subsection{\label{sec:flow_gs}Ground-State Calculations}

\begin{figure}[t]
  \begin{center}
    \includegraphics[width=\textwidth]{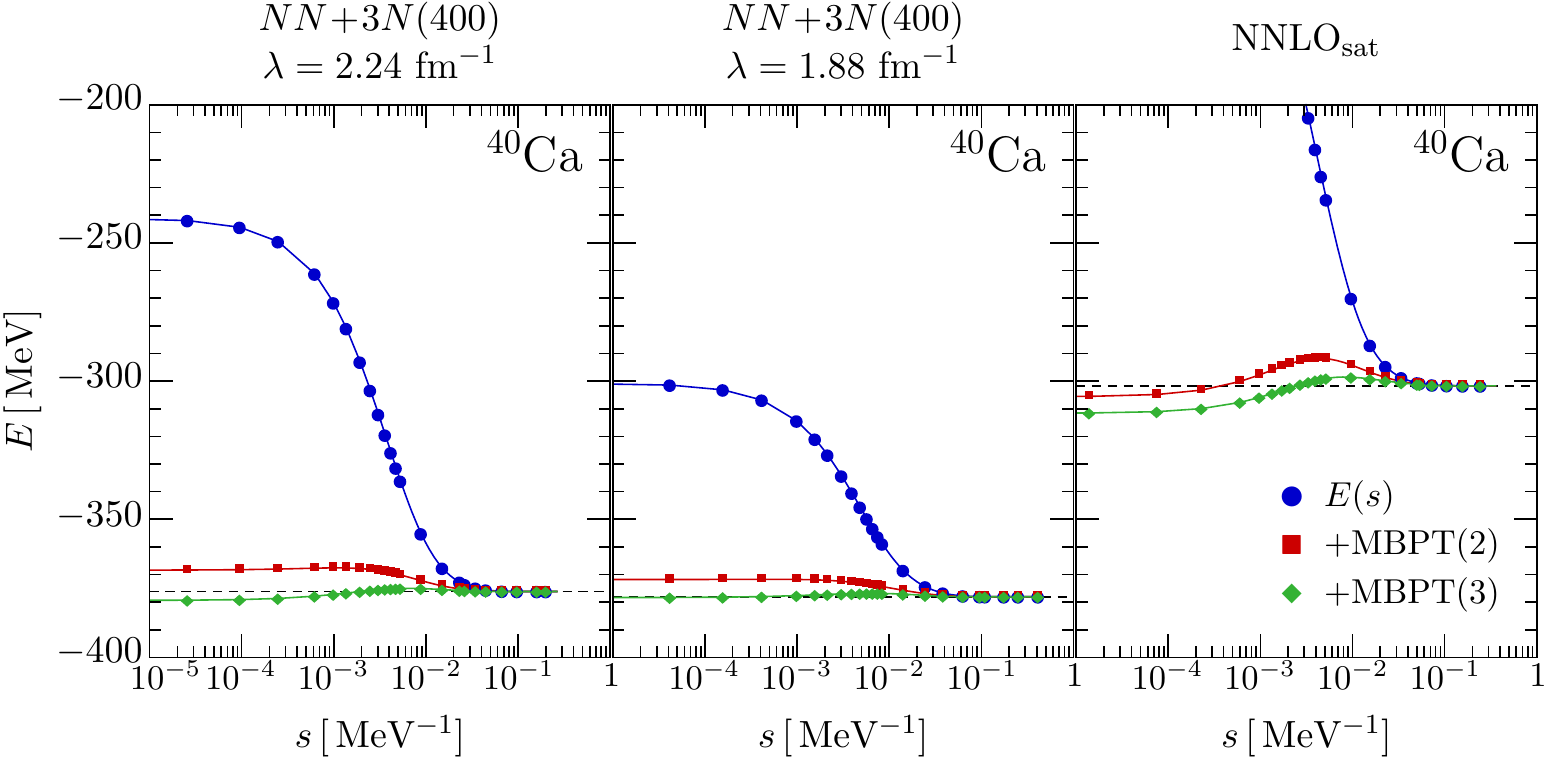}
  \end{center}
  \caption{\label{fig:mbpt}IMSRG(2) flow for $\nuc{Ca}{40}$ using different chiral $NN\!+\!3N$ interactions,
    obtained with the single-reference version of the Brillouin generator, equation \eqref{eq:eta_brillouin} 
    ($\eMax=14, \EMax=14,$ optimal $\hw$).
    We show the flowing ground-state energy $E(s)$, and the sum of $E(s)$ and perturbative energy
    corrections evaluated with the flowing Hamiltonian $\HO(s)$, to illustrate the re-shuffling
    of correlations into the Hamiltonian (see text). The dashed lines indicate the final IMSRG(2)
    energies.
  }
\end{figure}

As a first example, we consider IMSRG(2) ground-state calculations 
for the magic nucleus $\nuc{Ca}{40}$, using the single-reference version 
of the Brillouin $\eta^\text{B}$ generator, equation \eqref{eq:eta_brillouin}, 
and different chiral $NN\!+\!3N$ interactions (figure \ref{fig:mbpt}). Globally, 
sizable amounts of correlation energy are re-shuffled into the zero-body piece 
of the Hamiltonian. We note that the specific size of these contributions
changes significantly with the resolution scale. For $NN\!+\!3N(400)$ with
$\lambdaSRG=2.24\,\fmi$, we gain about $130\,\MeV$ of binding. For the
Hamiltonian with the lower resolution $\lambdaSRG=1.88\,\fmi$, the HF
reference state is already significantly lower in energy, so the energy
gains from many-body correlations are less pronounced. This behavior is
expected as interactions become increasingly soft, and thereby more 
perturbative (see, e.g., \cite{Bogner:2010pq}). Note that the final
ground-state energies for $\lambdaSRG=2.24\,\fmi$ and $1.88\,\fmi$ are 
almost identical, namely $-376.1\,\MeV$ and $-378.0\,\MeV$.
As discussed in section \ref{sec:srg}, in ideal implementations, all results
should be invariant under arbitrary changes of $\lambdaSRG$, which 
appears to be satisfied to a high degree here. However, we caution that 
the $NN\!+\!3N(400)$ 
is tuned to minimize induced $4N,\ldots$ forces
\cite{Roth:2011kx,Roth:2014fk}, so we have to reconsider the uncertainties 
due to these omitted terms for other interactions and observables
(see, e.g., \cite{Calci:2016hb}).

In the rightmost panel of figure \ref{fig:mbpt}, we show the result
of a calculation with \NNLOsat{}, which is considerably different from
the $NN\!+\!3N(400)$ Hamiltonian. For instance, the HF ground-state energy
of $\nuc{Ca}{40}$ is merely $-96.4\,\MeV$, and the binding energy gain due 
to the IMSRG(2) evolution is about $200\,\MeV$, which is a first indicator
that its resolution scale is higher than that of the other two
interactions used in the figure. The softened $NN\!+\!3N(400)$ Hamiltonians
overestimate the binding energy compared to the experimental value of 
$-342\,\MeV$ \cite{Wang:2012uq} (cf.~section \ref{sec:groundstate_CaNi}), and 
yield a charge radius of $3.0\fm$, which is about 15\% smaller than the 
experimental one \cite{Angeli:2013rz}. In contrast, \NNLOsat{} is underbound 
at the IMSRG(2) level, and the charge radius is about $0.1\,\fm$ too 
large. In CC calculations with \NNLOsat{}, the binding energy at the
CCSD(T) level is $-326\,\MeV$, \cite{Ekstrom:2015fk}, and we expect 
a comparable result from a similar approximate treatment of the  
next-level IMSRG truncation, denoted IMSRG(3). Work in this direction
is in progress.

The mechanism by which the flowing ground-state energy is absorbing
correlation energy can be understood if we consider the zero-body flow equation
\eqref{eq:mr_flow_0b_tens} in the perturbative approach we introduced 
in section \ref{sec:generators}. In the single-reference case, we have
\begin{align}
  \totd{E}{s} &=     
    \underbrace{\sum_{ab}(n_{a}-n_{b})\eta^{a}_{b}f^{b}_{a}}_{\OC(g^4)}
    +\underbrace{\frac{1}{4}\sum_{abcd}
        \left(\eta^{ab}_{cd}\Gamma^{cd}_{ab}-\Gamma^{ab}_{cd}\eta^{cd}_{ab}\right)
        n_{a}n_{b}\bar{n}_{c}\bar{n}_{d}}_{\OC(g^2)}\,.
\end{align}
Assuming an imaginary-time (or Brillouin) generator, and recalling
\begin{equation}
    \Gamma^{pp'}_{hh'}(s)=\Gamma^{pp'}_{hh'}(0)e^{-|\Delta^{pp'}_{hh'}|s}\,,\quad 
    \Gamma^{hh'}_{pp'}(s)=\Gamma^{hh'}_{pp'}(0)e^{-|\Delta^{pp'}_{hh'}|s}\,,
\end{equation}
we have to $\OC(g^2)$
\begin{equation}
  \totd{E}{s}=\frac{1}{2}\sum_{pp'hh'}|\Gamma^{pp'}_{hh'}(0)|^2e^{-2|\Delta^{pp'}_{hh'}|s}\,.
\end{equation}
Integrating over the flow parameter, we obtain
\begin{equation}
  E(s) = E(0) - \frac{1}{4}\sum_{pp'hh'}\frac{|\Gamma^{pp'}_{hh'}(0)|^2}{|\Delta^{pp'}_{hh'}|}
      \left(1 - e^{-2|\Delta^{pp'}_{hh'}|s}\right)\,.\label{eq:flow_E_pert}
\end{equation}
We recognize the second-order energy correction, evaluated with the
initial Hamiltonian, and see that $E(s)$ will decrease 
with $s$ (i.e., the binding energy increases). In the limit $s\to\infty$
the entire correction is shuffled into the zero-body piece of the evolved 
Hamiltonian. The complete IMSRG(2) obviously performs a more complex 
resummation of correlations, but we can see from figure \ref{fig:mbpt} 
that it encompasses the complete second order. In fact, we see that 
the third-order correction is completely absorbed into the final $E(\infty)$ 
as well. An extensive discussion of how higher-order corrections are
resummed can be found in Ref.~\cite{Hergert:2016jk} (also see 
\cite{Morris:2015ve}).

The perturbative analysis also gives us a rough understanding of the three
characteristic regions in the flowing energy shown in figure \ref{fig:mbpt}. 
From $s=0\,\MeVi$ to about $s=10^{-3}\,\MeVi$, the energy is renormalized 
only weakly, followed by a rapid drop from $10^{-3}$ to $10^{-2}\,\MeVi$, 
and an eventual slow decay from $10^{-2}\,\MeVi$ onward. At a given value
of $s$, the offdiagonal matrix elements that couple states with energy 
differences $\Delta^{pp'}_{hh'}=1/s$ have been suppressed by a factor $1/e$. 
Thus the transitions in the energy flow occur when $2$p$2$h excitations
up to $~1\,\GeV$ and $~100\,\MeV$, respectively, have been suppressed.
For the softer $NN\!+\!3N(400)$ interaction with $\lambdaSRG=1.88\,\fmi$,
the coupling between the reference state and such excitations is weaker 
than for $\lambdaSRG=2.24\,\fmi$ (or \NNLOsat{}), and less 
correlation energy is gained by evolving.

\begin{figure}[t]
  \begin{center}
    \includegraphics[width=0.66\textwidth]{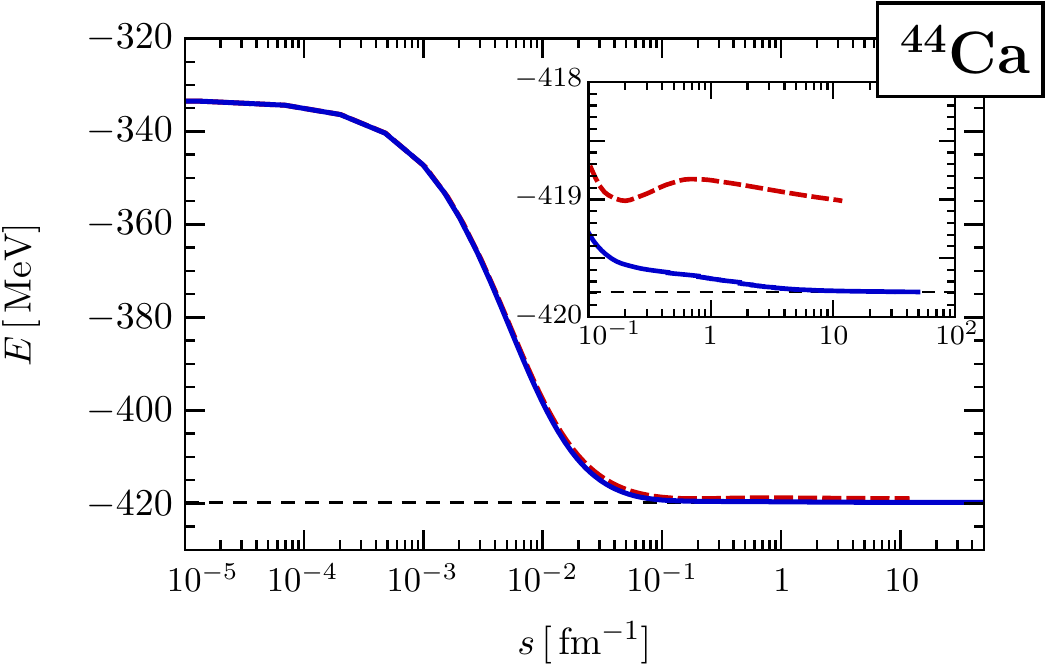}
  \end{center}
  \caption{\label{fig:flow_Ca44}Flow of the MR-IMSRG(2) ground-state energy for $\nuc{Ca}{44}$, generated by
  the multireference Brillouin (equation \eqref{eq:eta_brillouin}, solid lines) and 
  imaginary-time generators (equation \eqref{eq:eta_imtime}, dashed lines). In the latter,
  $\lambda^{(k\geq2)}$ terms have been truncated (see text). 
  Both calculations use the chiral $NN\!+\!3N(400)$ Hamiltonian, SRG-evolved to $\lambdaSRG=1.88\fmi$
  (cf.~section \ref{sec:flow_implementation}), a PNP reference state, and a single-particle 
  basis of 15 major HO shells ($\eMax=14,\hw=24\,\MeV$).
  }
\end{figure}

As another example, we consider an MR-IMSRG(2) calculation for
the semi-magic open-shell nucleus $\nuc{Ca}{44}$, using a PNP reference
state and the $NN\!+\!3N(400)$ Hamiltonian with $\lambdaSRG=1.88\,\fmi$. In 
the top panel of figure \ref{fig:flow_Ca44}, we show the flow of the 
ground-state energy for the multireference Brillouin 
generator $\eta^\text{B}$, and an approximate version of the multireference 
imaginary-time generator where terms involving $\lambda^{(k\geq2)}$ have
been truncated. Superficially, the flow of the energy is
similar to the $\nuc{Ca}{40}$ single-reference examples discussed above,
and the two generators seem to perform equally well. However, the inset
in the panel reveals differences for $s>0.1\,\MeVi$. The flow generated
by $\eta^\text{B}$ is causing a monotonic decrease of the energy, while 
the approximate $\eta^\text{IT}$ exhibits oscillatory behavior. These 
oscillations die out if we evolve to sufficiently large $s$, and a
converged energy of $-419.0\,\MeV$ is obtained, compared to $-419.8\,\MeV$
for the Brillouin generator. The inclusion of the terms 
that are linear in $\lambda^{(2)}$ and $\lambda^{(3)}$ removes the oscillation 
and brings the $\eta^\text{IT}$ flow in agreement with $\eta^\text{B}$, 
suggesting that the sign factors are consistently positive and the relation 
\eqref{eq:eta_it_brillouin} holds. Terms that are quadratic in $\lambda^{(2)}$
or linear in $\lambda^{(4)}$ cancel.

\subsection{\label{sec:flow_decoupling}Decoupling}

In section \ref{sec:decoupling}, we discussed in detail how we have to define 
the offdiagonal Hamiltonian to ensure that the reference state is decoupled
from excitations by the (MR-)IMSRG flow. Let us now demonstrate that the
matrix elements in question are indeed suppressed as intended. Since only
approximate decoupling can be achieved in the multireference case if we 
want to avoid the extremely costly inclusion of irreducible density matrices
$\lambda^{(k\geq4)}$, we use a single-reference calculation for $\nuc{Ca}{40}$
as our example. For this doubly magic nucleus, we can use the White generator,
equation \eqref{eq:eta_white}. Based on our perturbative analysis in 
section \ref{sec:generators}, we expect offdiagonal matrix elements to decay
according to
\begin{equation}\label{eq:decay_white}
  \Gamma^{pp'}_{hh'}(s) = \Gamma^{pp'}_{hh'}(0)e^{-s}\,
\end{equation}
(note that the flow parameter is dimensionless for the White generator).

\begin{figure}[t]
  \begin{center}
    \includegraphics[width=\textwidth]{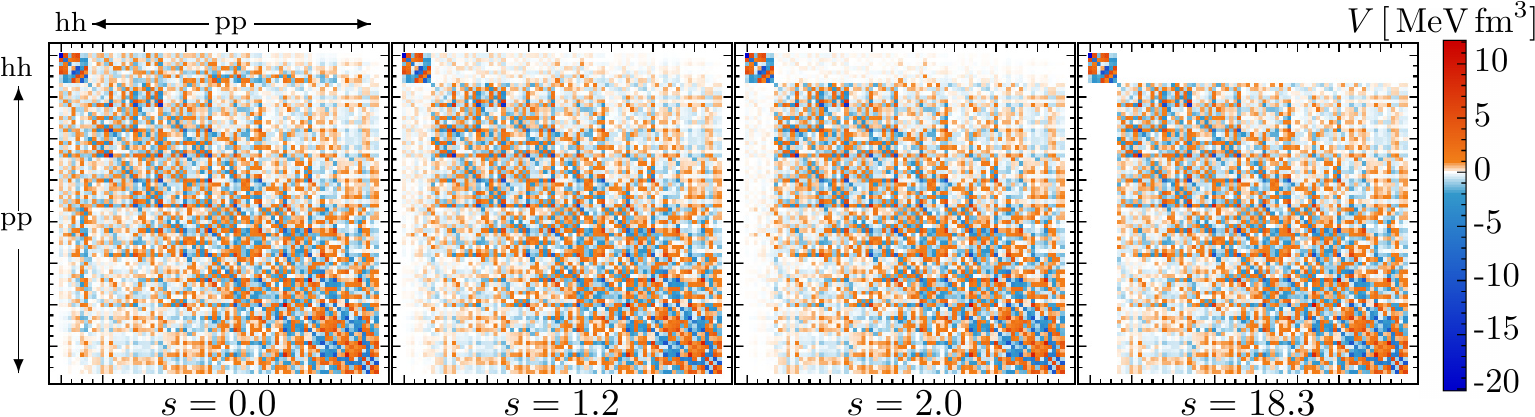}
  \end{center}
  \caption{\label{fig:decoupling}Decoupling for the White generator, 
  equation \eqref{eq:eta_white}, in the $J^\pi=0^+$ neutron-neutron interaction 
  matrix elements of $\nuc{Ca}{40}$ ($\eMax=8, \hbar\omega=20\,\MeV$, Entem-Machleidt N${}^{3}$LO(500) 
  evolved to $\lambda=2.0\,\fm^{-1}$, no induced or initial $3N$ forces). 
  Only hhhh, hhpp, pphh, and pppp blocks of the matrix are shown.}
\end{figure}

In figure \ref{fig:decoupling}, we show the pppp, hhhh, pphh and hhpp
matrix elements of the normal-ordered Hamiltonian in the $J^\pi=0^+$ 
neutron-neutron partial wave (the phph and hphp matrix elements are omitted 
to avoid clutter). As we integrate the IMSRG(2) flow equations, the offdiagonal 
matrix elements ($\Gamma^{pp'}_{hh'}$ and $\Gamma^{hh'}_{pp'}$) are suppressed 
rapidly, as suggested by equation \eqref{eq:decay_white}. We stop the evolution at 
$s=18.3$, where the second-order energy correction calculated with $\HO(s)$, 
falls below $10^{-6}\,\MeV$. 

\subsection{\label{sec:flow_heff}Effective Hamiltonians}

With the suppression of pphh and hhpp matrix elements, the matrix representation
of the Hamiltonian in our many-body Hilbert space is driven to the simplified 
form shown in figure \ref{fig:imsrg}, eliminating the outermost diagonals that 
are coupling $n$p$n$h and $(n\pm2)$p$(n\pm2)$h excitations. In the MR-IMSRG,
we can at least reduce the strength of the couplings between generalized 
excitations. At any finite value of $s$, correlations due to such couplings have
been reshuffled into the \emph{diagonal} Hamiltonian. Thus, we can also
view the MR-IMSRG as a tool to generate RG-improved effective Hamiltonians,
analogous to the free-space SRG discussed in section \ref{sec:srg}.

Formally, all quantum many-body methods either implicitly or explicitly
approximate the eigenvalues and eigenvectors of an initial Hamiltonian 
that are given by
\begin{equation}
  \HO(0) \ket{\Psi_n} = E_n \ket{\Psi_n}\,.
\end{equation}
For an ``exact'' method like FCI, the only approximation is the use of
a finite basis to represent the Hamiltonian matrix, while approaches like 
CI, CC or MR-IMSRG also introduce systematic truncations. Let us write
the approximate eigenvalues and eigenstates of a real many-body calculation
as
\begin{equation}
  \HO(0) \ket{\Phi_n} = E'_n \ket{\Phi_n}\,,\quad \ket{\Phi_n}\approx\ket{\Psi_n}\,, E'_n \approx E_n\,.
\end{equation}
The eigenvalues are invariant under a unitary transformation, e.g.,
an untruncated MR-IMSRG evolution,
\begin{equation}
  \HO(s)\UO(s)\ket{\Psi_n} \equiv \UO(s) \HO \UUO(s) \UO(s)\ket{\Psi_n} = E_n \UO(s) \ket{\Psi_n}\,.
\end{equation}
Thus, it may be fruitful to use an effective Hamiltonian that has absorbed 
many-body correlations, e.g., through IMSRG or MR-IMSRG improvement,
as input for a quantum many-body calculation. In this case, the many-body
method would need to approximate the transformed eigenstate,
\begin{equation}\label{eq:approx_eigenstate}
  \ket{\Phi_n}\approx\UO(s)\ket{\Psi_n}\,
\end{equation}
instead of $\ket{\Psi_n}$, which may be a less demanding task. For example, the 
momentum-space decoupling achieved by the free-space SRG allows us to use low-energy 
(low-momentum) model spaces to accurately describe low-lying nuclear spectra 
(cf.~section \ref{sec:srg}). Similarly, the IMSRG can be used to build correlations 
into the Hamiltonian that are not accessible by a truncated many-body method, 
improving the quality of the approximation \eqref{eq:approx_eigenstate}.
Examples are a CI method with up to $npnh$ excitations that can probe $(n+2)p(n+2)h$
excitations if an IMSRG(2) Hamiltonian is used (cf.~section \ref{sec:decoupling}),
the physics of the inert core and excluded space in the interacting Shell
model / valence-space CI (see section \ref{sec:sm}), or excitations 
beyond a specific $\Nmax$ model space used in the NCSM \cite{Gebrerufael:2016rp}.

\begin{figure}[t]
  \begin{center}
    \includegraphics[width=0.66\textwidth]{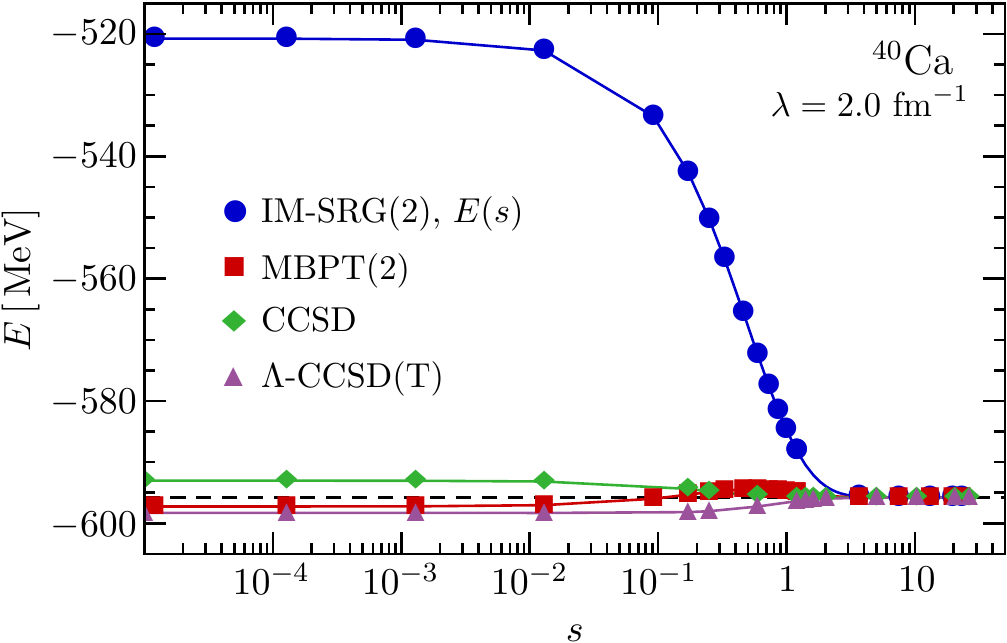}
  \end{center}
  \caption{\label{fig:flow_methods} IMSRG(2) ground-state energy of $\nuc{Ca}{40}$ 
  as a function of the flow parameter $s$, compared to MBPT(2), CCSD, and $\Lambda$-CCSD(T) 
  energies with the IMSRG-evolved Hamiltonian $\HO(s)$. We only show part of the data 
  points to avoid clutter. Calculations were done for $\eMax=10$ and optimal 
  $\hbar\omega=24\,\MeV$, using our standard \NNNLO{} interaction (cf.~section \ref{sec:flow_implementation})
  at $\lambdaSRG=2.0\,\fmi$, without initial or induced $3N$ forces. The dashed 
  lines indicate the final IMSRG(2) energies. 
  }
\end{figure}

For numerical illustration, we use the IMSRG Hamiltonian $\HO(s)$ from a single-reference 
ground-state calculation of $\nuc{Ca}{40}$ as input for second-order MBPT
(denoted MBPT(2)) and CC with singles and doubles excitations (CCSD, \cite{Shavitt:2009}), 
as well as non-iterative triples corrections ($\Lambda$-CCSD(T), \cite{Taube:2008kx,Taube:2008vn}). 
The resulting ground-state 
energies are compared to the flowing IMSRG(2) energy in figure \ref{fig:flow_methods}. Once we
reach $s=2.0$, the offdiagonal matrix elements of the Hamiltonian have been
strongly suppressed (cf.~figure \ref{fig:decoupling}), and the ground-state
energies of all methods collapse to the same result, namely the IMSRG(2)
ground-state energy. The CC results can be viewed as an extension of our
discussion of figure \ref{fig:mbpt}, showing that for $s>2.0$ there is practically
no more correlation energy to be gained from MBPT corrections, not even
when terms are summed to infinite order.

While the FCI ground-state energy is independent of $s$ under exact IMSRG 
transformations of the Hamiltonian (also see section \ref{sec:srg}), we have to carefully 
assess the interplay of all practical truncations if we use IMSRG evolved 
Hamiltonians as an input for non-exact many-body
methods \cite{Hergert:2016jk}. If an approximate many-body method is less 
complete than the IMSRG in a specific truncation, we will obtain an RG 
improvement towards the exact result, as discussed above.
However, if the many-body method contains terms beyond the truncated IMSRG,
the final result can be an inferior approximation of the true eigenvalue and 
eigenstate than the one obtained with the unevolved Hamiltonian. Figure 
\ref{fig:flow_methods} contains examples for both cases: MBPT(2) is less 
complete than the IMSRG(2), so the MBPT(2) energy is improved 
towards the exact energy. Note that the improvement in the energy can amount
to an attractive or a repulsive correction, depending on the initial Hamiltonian.
For soft interactions like the one used here, MBPT(2) tends to provide too 
much binding 
\cite{Roth:2006lr,Guenther:2010ge,Roth:2010ys,Bogner:2010pq,Hebeler:2011dq,Tsukiyama:2011uq,Langhammer:2012uq,Tichai:2016vl}, 
hence the final IMSRG(2) ground-state energy increases (the binding energy 
decreases). On the other hand, the $\Lambda$-CCSD(T) energy contains 
fourth-order $3$p$3$h (triples) correlations that are missing in the 
IMSRG(2) \cite{Hergert:2016jk}, hence it is a better approximation to the true 
ground-state energy of the initial Hamiltonian than the result obtained with
the IMSRG(2) Hamiltonian for large $s$.

In general, the capability to ``split'' correlations between the wave function 
and the effective Hamiltonian can be used to greatest effect if complementary 
types of correlations are handled better by each ingredient. For instance, reference 
states that are projected from symmetry-broken mean fields offer a very efficient 
way to capture \emph{static correlations} that would require an explicit 
treatment of up to $A$p$A$h (or generalized $A$-body) excitations in the 
many-body bases we have discussed so far. Conversely, these latter bases are 
well-suited for the description of \emph{dynamical correlations} (i.e., the dynamics of 
nucleon pairs, triples, \ldots) inside the nucleus. The MR-IMSRG offers us a 
framework that can harness both types of correlations, by building dynamical 
correlations on top of statically correlated reference states. A concrete 
example are the calculations based on PNP reference states discussed above 
and in section \ref{sec:groundstate}. 

\section{\label{sec:groundstate}Ground-State Calculations for Closed- and Open-Shell Nuclei}

In recent years, we have applied the MR-IMSRG to study the ground-state
energies of semi-magic isotopic chains and gauge the quality of chiral 
interactions like $NN\!+\!3N(400)$ through their confrontation with experimental 
data \cite{Hergert:2013mi,Hergert:2013ij,Hergert:2014vn}. We will review 
the salient findings of these investigations in the following, using the 
opportunity to repeat the MR-IMSRG(2) calculations 
with the Brillouin generator \eqref{eq:eta_brillouin} instead of our earlier
choices. Thus, the present work serves as a benchmark for the new generator. 
The outcome of this benchmark process can be anticipated based on our discussion in 
sections \ref{sec:generators} and \ref{sec:flow_gs}: Results obtained with
$\eta^\text{B}$ agree with those for (approximate) multireference 
imaginary-time and White generators on the level of $0.1-0.2\%$, which
is currently a negligible contribution to the uncertainties of our 
calculations. On the many-body side, these uncertainties are due to
truncation effects (i.e., the omission of three- and higher-body
terms that are induced by the MR-IMSRG flow) and the NO2B approximation.
As discussed in Refs.~\cite{Hergert:2013mi,Hergert:2013ij,Hergert:2014vn}
and the remainder of this section, 
these uncertainties can be quantified and controlled reasonably well, so 
that the main source of uncertainty are the input Hamiltonian 
and the impact of changing the resolution scale $\lambdaSRG$.

\subsection{\label{sec:groundstate_O}Oxygen Isotopes}
\begin{figure}[t]
  \begin{center}
    \includegraphics[width=0.66\textwidth]{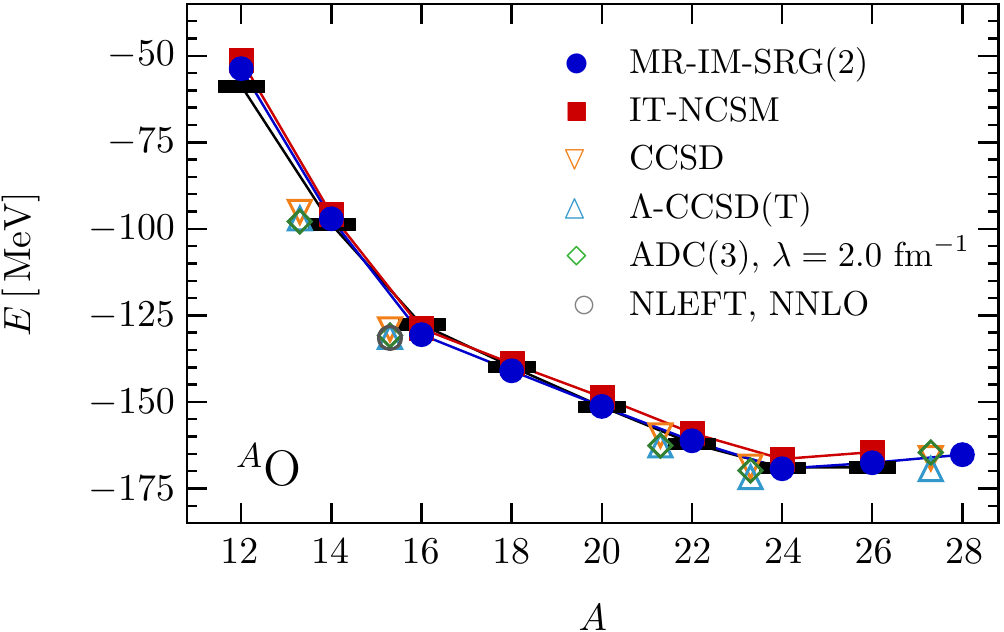}
  \end{center}
  \caption{\label{fig:OXX_methods} Ground-state energies of the oxygen isotopes from 
    MR-IMSRG(2) and other many-body approaches, using the $NN\!+\!3N(400)$ interaction
    at $\lambdaSRG=1.88\fmi$. Some data points were offset horizontally
    to enhance the readability of the figure. MR-IMSRG(2) calculations were performed using 
    the Brillouin generator ($\eMax=14,\EMax=14$, and optimal $\hw$), updating 
    earlier results shown in Refs.~\cite{Hergert:2013ij,Hergert:2016jk}. Note that 
    the ADC(3) Self-Consistent Green's Function results \cite{Cipollone:2013uq,Cipollone:2015fk}
    were obtained for $\lambdaSRG=2.0\,\fmi$, but the dependence of energies
    on $\lambdaSRG$ is very weak. Black bars indicate experimental data \cite{Wang:2012uq}.
    }
\end{figure}

The oxygen isotopic chain has become a testing ground for \emph{ab initio}
nuclear structure methods in recent years 
\cite{Otsuka:2010cr,Roth:2011kx,Hagen:2012oq,Hergert:2013ij,Cipollone:2013uq,Epelbaum:2014kx,Cipollone:2015fk,Holt:2013fk,Bogner:2014tg,Stroberg:2016fk},
mainly for two reasons: First, Otsuka \emph{et al.}~\cite{Otsuka:2010cr} showed
the impact of $3N$ forces on the location of the neutron drip line, and
second, exact results from the importance-truncated NCSM (IT-NCSM)
\cite{Roth:2007fk,Roth:2009eu,Roth:2011kx,Roth:2014fk} are available for
the ground states and low-lying excitations, allowing us to assess the
many-body uncertainties of our calculations. 

The semi-magicity of the oxygen isotopes also allows us to enforce spherical
symmetry in our calculations to boost the numerical efficiency. For instance, 
an IMSRG(2) calculation for a closed-shell oxygen isotope, using a
spherical HF reference state and 15 major HO shells, requires about 20 core hours on current 
high-performance computing hardware. An MR-IMSRG(2) calculation for an open-shell
isotope, based on a spherical PNP reference state, takes about 500-1000
core hours to converge. In contrast, an IT-NCSM calculation requires on 
the order of 100,000 core hours for neutron-rich oxygen nuclei \cite{Hergert:2013ij}.

\begin{figure}[t]
  \begin{center}
    \includegraphics[width=\textwidth]{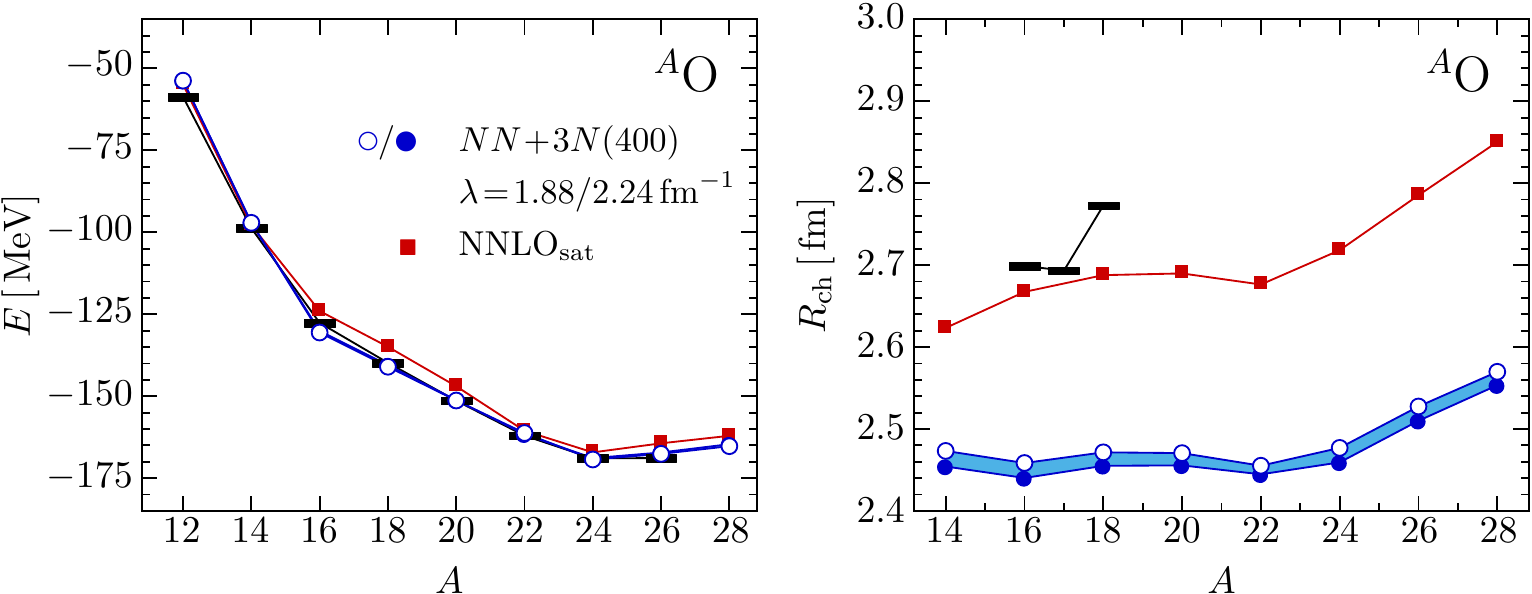}
  \end{center}
  \caption{\label{fig:OXX} MR-IMSRG(2) ground-state energies and charge radii of the oxygen 
    isotopes for \NNLOsat{} and $NN\!+\!3N(400)$ at $\lambdaSRG=1.88,\ldots,2.24\fmi$ 
    ($\eMax=14,\EMax=14$, and optimal $\hw$). Black bars indicate experimental 
    data \cite{Wang:2012uq,Angeli:2013rz}.
  }
\end{figure}

In figure \ref{fig:OXX_methods}, we compare MR-IMSRG(2) results for the oxygen
ground-state energies with a variety of configuration-space many-body
methods. In addition to IT-NCSM, CCSD, and $\Lambda$-CCSD(T) (cf.~\ref{sec:flow_heff}), 
we also include results from the Self-Consistent Green's Function (SCGF)
approach in the ADC(3) scheme \cite{Cipollone:2013uq,Cipollone:2015fk}. 
The latter are obtained at a slightly different resolution scale $\lambdaSRG=2.0\,\fmi$, 
but the dependence of the ground-state energies on $\lambdaSRG$ is very 
weak, at least in the range $\lambdaSRG=1.88,\ldots,2.24\,\fmi$. For example,
the ground-state energy of $\nuc{O}{24}$ changes by 0.2\% under this variation
(cf.~figure \ref{fig:OXX}). The insensitivity of the ground-state energies to
variations of $\lambdaSRG$ is due to a cancellation of  $4N$ forces that 
are induced by lowering the resolution scale of the initial $NN$ and $3N$ 
forces, respectively (see section \ref{sec:flow_gs} and \cite{Binder:2013zr,Hergert:2013ij}). 
We will illustrate below that this tuning does not hold for general observables.

For the same $NN\!+\!3N(400)$ input Hamiltonian, all used methods give 
consistent results that agree within a few percent with experimental ground
state energies. 
The systematically truncated methods, i.e., MR-IMSRG(2), CCSD, $\Lambda$-CCSD(T)
and ADC(3), agree very well with the exact IT-NCSM results, on the
level of 1\%--2\%. Since the IT-NCSM includes the complete $3N$ interaction,
in accordance with its model space truncation, this deviation is due
to the combined effects of the NO2B approximation \cite{Roth:2012qf,Binder:2013zr,Binder:2013fk}, 
as well as truncated many-body correlations. The $\Lambda$-CCSD(T) method 
gains about 2\% of additional binding energy compared to CCSD through the 
inclusion of triples correlations, giving us an in-method measure of the
scheme's many-body uncertainty, and indicating the rapid convergence of the 
many-body expansion for low-momentum Hamiltonians \cite{Bogner:2006qf,Bogner:2010pq,Tichai:2016vl}.
However, this particular CC method is known to over-predict ground-state
energies in quantum chemistry compared to exact diagonalization methods
like FCI, which is the counterpart to the IT-NCSM in our case. For this
reason, improved triples approaches like the completely renormalized CR-CC(2,3) 
scheme have been introduced in the literature \cite{Piecuch:2005dp,Binder:2013fk},
which we will use for comparison in section \ref{sec:groundstate_CaNi}. 
For an approximate or complete implementation of MR-IMSRG(3), we can expect 
binding energy gains of comparable size because it will probe dynamical 
correlations due to nucleon triples in a similar fashion. For more details,
we refer the reader to the in-depth analysis published in Ref.~\cite{Hergert:2016jk}. 

The MR-IMSRG(2) ground-state energy of $\nuc{O}{16}$, $-130.1\,\MeV$, 
also agrees well with the result of a recent Nuclear Lattice EFT (NLEFT) 
calculation, which is $-131.3(5)\,\MeV$. This ground-state 
energy is obtained with an \NNLO{} Lagrangian, including $NN$ and $3N$ 
interactions, as well as a tuned regularized $4N$ contact force 
\cite{Epelbaum:2014kx}. The net effect of the $4N$ term is repulsive; without it, 
the ground-state energy is $-138.8(5)\,\MeV$. Since the treatment of
the nuclear many-body problem in NLEFT is completely different from all the 
other approaches compared here \cite{Lee:2009bh}, the consistency of the results 
for comparable inputs is very encouraging.

The \emph{ab initio} calculations clearly predict the neutron drip line at 
$\nuc{O}{24}$, matching experimental findings \cite{Hoffman:2008ly}. While
absolute ground-state energies can change significantly under variations
of the $3N$ cutoffs or other modifications of the initial Hamiltonian, the 
drip line signal turns out to be rather robust \cite{Hergert:2013ij}. All methods
predict the $\nuc{O}{26}$ resonance at an energy $E_x\approx1-2\,\MeV$ above 
the $\nuc{O}{24}$ ground state, 
which is considerably higher than the current experimental limits 
$E_x\approx 50\,\keV$ \cite{Kohley:2015ay,Caesar:2013qf,Lunderberg:2012cr}.
In part, this is due to the omission of continuum effects in all
calculations that are shown here. However, we also see indications
that while the $NN\!+\!3N(400)$ Hamiltonian reproduces the ground-state 
energy trends along the isotopic chain quite well, it enhances certain shell 
closures compared to experiment. This causes an overestimation of
the experimental $\nuc{O}{16}$ binding energy, for instance. We will 
find further examples of enhanced shell closures in other 
isotopic chains. The interplay of nuclear interactions, many-body
and continuum effects that causes the flat trend of the experimental 
ground-state resonance energies beyond $\nuc{O}{24}$ suggests that the 
oxygen isotopes will remain an important 
testing ground for nuclear Hamiltonians and many-body methods for the 
foreseeable future.

While $NN\!+\!3N(400)$ gives a good reproduction of the oxygen ground-state
energies, an issue with the Hamiltonian's saturation properties is revealed
by inspecting the oxygen charge radii (see figure \ref{fig:OXX}). 
The theoretical charge radii are about 10\% smaller
than the experimental charge radius of $\nuc{O}{16}$, $\Rch=2.70\,\fm$ 
\cite{Angeli:2013rz}, and the sharp increase for $\nuc{O}{18}$ is missing 
entirely. 

The underestimation of nuclear radii was one of the deficiencies
that inspired efforts by multiple groups to improve the construction and 
optimization of chiral interactions. One of the first new interactions to
come out of these efforts is \NNLOsat{} by Ekstr\"{o}m and collaborators 
\cite{Ekstrom:2015fk}. As the name suggests, it is defined at chiral order
\NNLO{}, and contains $NN$ and $3N$ interactions. The creators of this interaction
chose to include select many-body data in the optimization protocol for the 
interaction's low-energy constants (LECs), including the $\nuc{O}{16}$ 
ground-state energy and charge radius. Note that perfect agreement with
experimental data is not enforced, because the optimization procedure
aims to account for uncertainties due to truncation at chiral order \NNLO{},
and the many-body approach used for accessing the medium-mass nuclei (CCSD).

In figure \ref{fig:OXX}, we compare MR-IMSRG(2) results for \NNLOsat{} and
$NN\!+\!3N(400)$. For the latter, we indicate the effects of varying $\lambdaSRG$
from $1.88\,\fmi$ to $2.24\,\fmi$ through a shaded band. As discussed above,
the ground-state energies only vary by 0.2\% due to a fine-tuned cancellation,
but the change in the radii is as large as 1\%. Interestingly, $\Rch$
grows larger as $\lambdaSRG$ decreases. This is consistent with a 
recent study in light nuclei \cite{Schuster:2014oq}, which found that 
two- and three-body terms that are induced by consistently evolving
the charge radius operator to lower $\lambdaSRG$ have the opposite
effect and \emph{reduce} 
its expectation value. These terms have not been included here.

The MR-IMSRG(2) ground-state energies obtained with \NNLOsat{} are 
slightly \emph{lower} than those for $NN\!+\!3N(400)$ in the proton-rich
isotopes $\nuc{O}{12,14}$, and above the $NN\!+\!3N(400)$ energies in 
$\nuc{O}{16-28}$. From $\nuc{O}{16-22}$, the \NNLOsat{} ground-state energies
exhibit a parabolic behavior as opposed to the essentially linear trend
we find for $NN\!+\!3N(400)$. A possible cause is the inclusion of the $\nuc{O}{22,24}$
ground-state energies in the optimization protocol, which constrains
the possible energy deviation in these nuclei. \NNLOsat{} predicts the
drip line at $\nuc{O}{24}$, and the trend for the $\nuc{O}{26,28}$
resonance energies is similar to the $NN\!+\!3N(400)$ case.

For \NNLOsat{}, the charge radii for the bound oxygen isotopes are 
about 10\% larger than for $NN\!+\!3N(400)$, which is expected given 
the use of the $\nuc{O}{16}$ charge radius in the optimization of 
the LECs (also see Ref.~\cite{Lapoux:2016xu}). For the resonant states, 
the increase is even larger, but continuum effects must be considered to 
make a meaningful comparison. We note that \NNLOsat{} also fails to describe 
the sharp jump in $\Rch$ at $\nuc{O}{18}$.

\subsection{\label{sec:groundstate_CaNi}Calcium and Nickel Isotopes}

In Ref.~\cite{Hergert:2014vn}, we applied the MR-IMSRG(2) to study the ground-state
energies of calcium and nickel isotopes. As for the oxygen isotopes reviewed in the
previous subsection, we exploited the semi-magicity of these nuclei and enforced
spherical symmetry in our calculations. 

\begin{figure}[t]
  \begin{center}
    \includegraphics[width=0.66\textwidth]{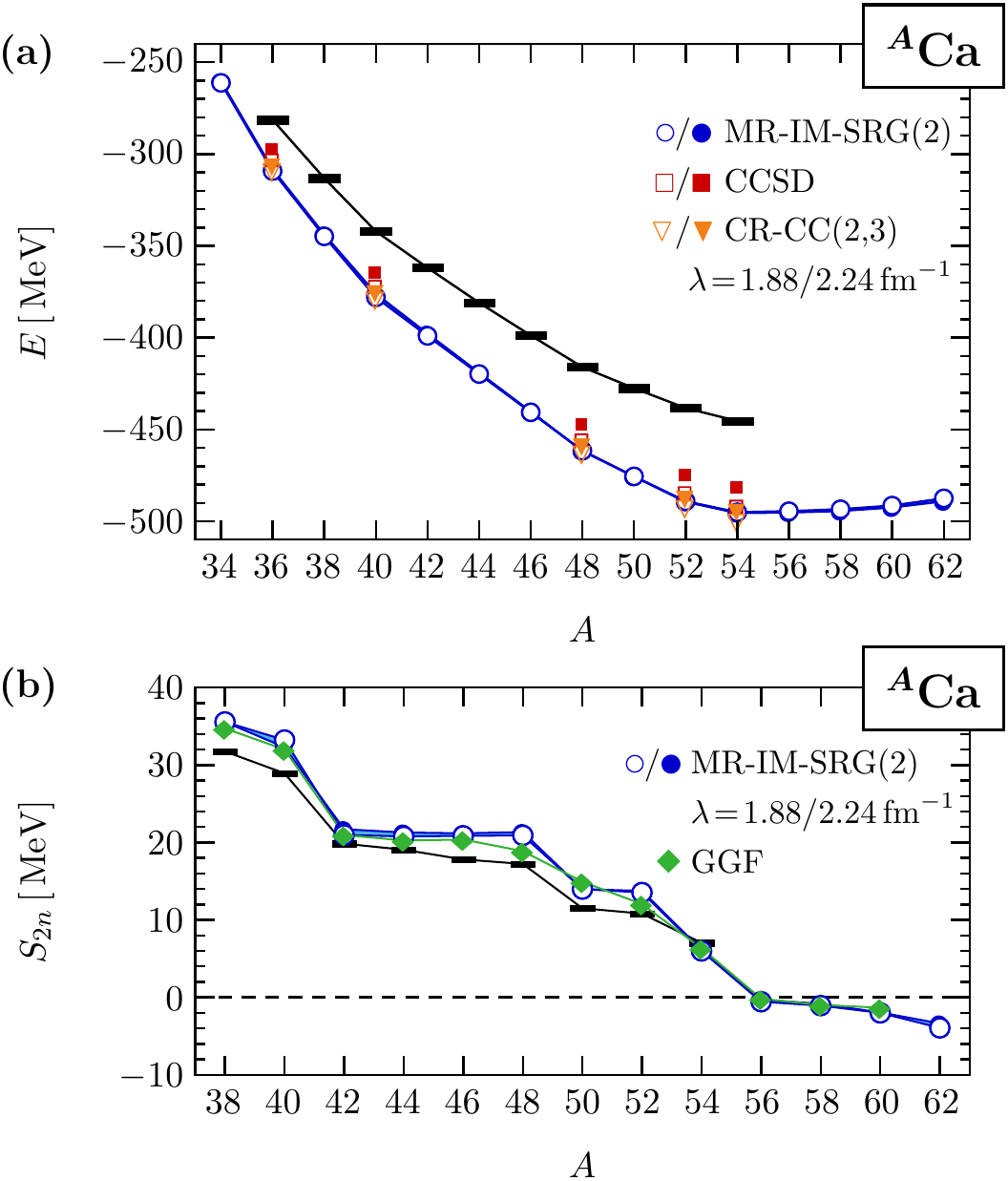}
  \end{center}
  \caption{\label{fig:CaXX} 
    MR-IMSRG(2) ground-state energies (top) and two-neutron separation energies
    (bottom) of the calcium isotopes, for the $NN\!+\!3N(400)$ Hamiltonian with 
    $\lambdaSRG=1.88,\ldots,2.24\fmi$ ($\eMax=14,\EMax=14$, optimal $\hw$). 
    All calculations were performed with the Brillouin generator, updating 
    the previous work \cite{Hergert:2014vn}. For nuclei with neutron sub-shell 
    closures, we show ground-state energies from CCSD and CR-CC(2,3) 
    calculations for comparison (see text and Refs.~\cite{Binder:2014fk,Binder:2013fk}). 
    Two-neutron separation energies are compared to results from self-consistent 
    second-order Gor'kov Green's Function (GGF) calculations with the $NN\!+\!3N(400)$ 
    Hamiltonian at $\lambdaSRG=2.0\,\fmi$ \cite{Soma:2014eu,Hergert:2014vn}. 
    Black bars indicate experimental data \cite{Wang:2012uq,Wienholtz:2013bh}.    
  }
\end{figure}

Figure~\ref{fig:CaXX}(a) shows the MR-IMSRG(2) ground-state energies for the
calcium isotopic chain. The deficient saturation properties of the $NN\!+\!3N(400)$
interaction are now fully apparent, causing an overbinding compared to experiment
that increases from 8\% to 12\% along the known isotopes $\nuc{Ca}{36,54}$.
In nuclei with sub-shell closures, the MR-IMSRG(2) energies are consistent 
with results from CCSD and CR-CC(2,3) calculations with the same Hamiltonian,
just as in the oxygen case. The ground-state energy gains from the inclusion 
of triples are on the order of 2\% for the Hamiltonian used here, which can
serve as an indicator of the uncertainty due to the many-body truncation.
The energies are insensitive to variations of the resolution scale $\lambdaSRG$ 
in a window around $2.0\fmi$, which suggests that the cancellation of induced 
$4N$ interactions works as in the oxygen chain. The residual changes are about 
0.2\% for MR-IMSRG(2), 2\% for CCSD, and 1\% for CR-CC(2,3). 

The presence of the $3N$ force in our Hamiltonian causes the appearance of a 
flat trend in the ground-state energies beyond $\nuc{Ca}{54}$ \cite{Hergert:2014vn}.
Similar behavior was found in CC calculations that used a more phenomenological
treatment of the $3N$ force, normal ordering it in symmetric nuclear matter to derive 
the in-medium contributions to lower-rank parts of the Hamiltonian, and readjusting
the LECs \cite{Hagen:2012nx}. It will be interesting to see if this trend will be 
confirmed experimentally in the coming years, since it would have strong implications
for the location of the neutron drip line in calcium. 

In figure \ref{fig:CaXX}(b), we show the MR-IMSRG(2) results for the two-neutron 
separation energies, defined as
\begin{equation}
  S_{2n}(Z,N) \equiv E(Z,N-2)-E(Z,N)\,.
\end{equation}
Despite the overestimation of the calcium binding energies, the $NN\!+\!3N(400)$ 
Hamiltonian gives a reasonable reproduction of the experimental trends. Most
notably, the major shell closure at the magic neutron number $N=20$ is too
pronounced, continuing behavior we saw in the oxygen isotopes. The drops in the
theoretical $S_{2n}$ in $\nuc{Ca}{48,52,54}$, corresponding to the hypothetical
shell closures $N=28,32,34$, suggests that these nuclei are magic for 
the used interaction, matching predictions from Shell model calculations with
interactions derived from chiral $NN\!+\!3N$ forces in MBPT \cite{Holt:2012fk,Holt:2014vn}. 
While this view was supported by precision mass measurements 
\cite{Gallant:2012kx,Wienholtz:2013bh,Steppenbeck:2013dq}, a recent experiment
found an unexpectedly large charge radius of $\nuc{Ca}{52}$, which puts
the magicity of the neutron number $N=32$ in question \cite{Garcia-Ruiz:2016fk}
(also see \cite{Kreim:2014cr}). 

The flat trend of the calcium ground-state energies is reflected by the small
values of the separation energies in isotopes beyond $\nuc{Ca}{54}$. In fact, 
our calculations predict the $S_{2n}$ to be negative, rendering these isotopes 
unbound with respect to $\nuc{Ca}{54}$. Taking into account the uncertainties
of our calculation, including the missing continuum effects, the $S_{2n}$ may
well be positive in a more refined treatment. Thus, we cannot presently 
identify the drip line location for the $NN\!+\!3N(400)$ Hamiltonian.

Since CCSD and CR-CC(2,3) are single-reference methods, they can only be applied
in nuclei that have good sub-shell closures. Thus, we cannot obtain $S_{2n}$
values from these methods for comparison with our MR-IMSRG(2) results. However, 
in recent years, Som\`{a} \emph{et al.} have extended the SCGF approach to open-shell 
nuclei by using the Gor'kov formalism for systems with broken particle number
symmetry \cite{Soma:2011vn,Soma:2013ys,Soma:2014fu,Soma:2014eu}. In figure \ref{fig:CaXX}(b),
we compare $S_{2n}$ results from this self-consistent second-order 
Gor'kov Green's Function (GGF) method with those from the MR-IMSRG(2). While
the GGF scheme contains less many-body correlations than the MR-IMSRG(2), the 
differences primarily affect absolute energies, as is evident from the agreement
we see in figure \ref{fig:CaXX}(b). The main discrepancy between the two
methods are seen near the sub-shell closures, where the $S_{2n}$ from the GGF
approach behave more smoothly. This is a consequence of the broken particle
number symmetry, which causes a mixing of neighboring even-even nuclei if the
sub-shell closures are sufficiently weak (note that there is no smooth transition
in the GGF results at the major shell-closure $N=20$). Thus, the $S_{2n}$ from 
both methods are consistent when the same input Hamiltonian is used. 

\begin{figure}[t]
  \begin{center}
    \includegraphics[width=0.66\textwidth]{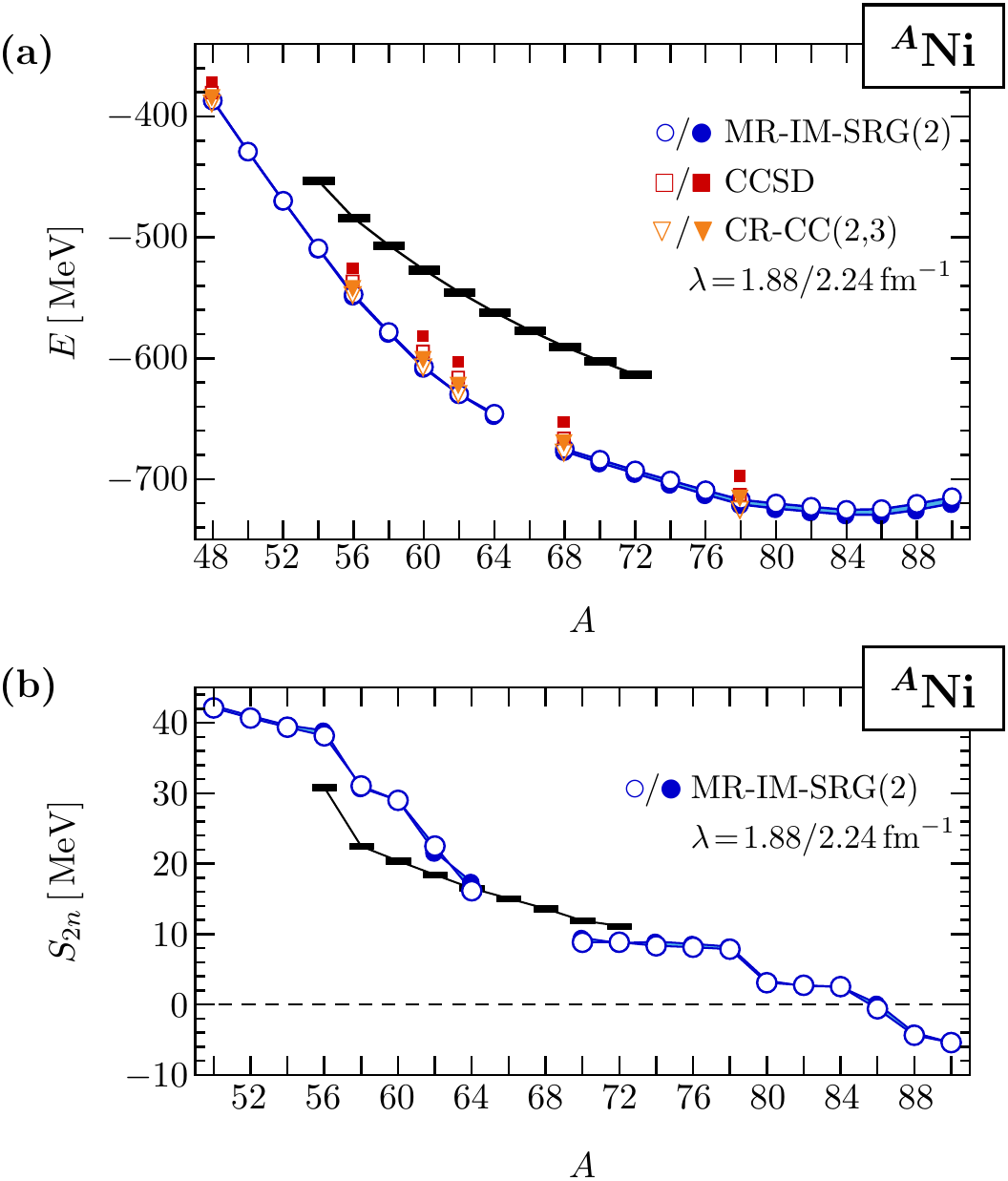}
  \end{center}
  \caption{\label{fig:NiXX} 
    MR-IMSRG(2) ground-state energies (top) and two-neutron separation energies
    (bottom) of the nickel isotopes, for the $NN\!+\!3N(400)$ Hamiltonian with 
    $\lambdaSRG=1.88,\ldots,2.24\fmi$ ($\eMax=14,\EMax=14$, optimal $\hw$). 
    All calculations were performed with the Brillouin generator, updating 
    the previous work \cite{Hergert:2014vn}. For nuclei with neutron sub-shell 
    closures, we show ground-state energies from CCSD and CR-CC(2,3) 
    calculations for comparison (see text and Refs.~\cite{Binder:2014fk,Binder:2013fk}). 
    Black bars indicate experimental data \cite{Wang:2012uq}.    
  }
\end{figure}

Moving on to the nickel isotopes, we show the MR-IMSRG(2) ground-state
energies and two-neutron separation energies in figure \ref{fig:NiXX}. 
The binding energies of the known nickel isotopes are overestimated by
about 13\%. The variation of the $\nuc{Ni}{48-78}$ ground-state energies 
with $\lambdaSRG$ is again very weak. For CCSD and CR-CC(2,3), it is
comparable to the variation in the calcium energies, while the variation
of the MR-IMSRG(2) results grows to 0.5\% in $\nuc{Ni}{78}$, and eventually
to 0.7\% in $\nuc{Ni}{86}$. The nickel isotopes also exhibit a flat trend 
in the neutron-rich region, although it is not quite as pronounced as in the 
calcium isotopic chain. Consequently, the $S_{2n}$ are quite small. Figure 
\ref{fig:NiXX}(b) shows that they become negative in $\nuc{Ni}{86}$, 
but the uncertainties of our calculations are too large for a conclusive 
identification of the neutron drip line for the $NN\!+\!3N(400)$ Hamiltonian.

The $S_{2n}$ also indicate the presence of sub-shell closures in $\nuc{Ni}{60}$ 
and $\nuc{Ni}{62}$ that are not seen in the experimental data. These isotopes 
have neutron numbers $N=32$ and $34$, respectively, placing them in the same
isotonic chains as $\nuc{Ca}{52,54}$, where we first saw these sub-shell 
closures. This is another example of enhanced shell closures with 
$NN\!+\!3N(400)$. The mounting evidence suggests deficiencies in the tensor
and spin-orbit structures of the Hamiltonian, which are the main drivers
for the details of the shell evolution. 

In our original study in Ref.~\cite{Hergert:2014vn}, we experienced numerical
problems with the MR-IMSRG(2) flow of the $\nuc{Ni}{64,66}$ ground-state
energies. Using an approximate imaginary-time generator, the $\nuc{Ni}{64}$ 
energy exhibited energy oscillations in the several-percent range that did 
not fall off over large ranges of the flow parameter (cf.~section \ref{sec:flow_gs}), 
while the energy of $\nuc{Ni}{66}$ diverges around $s\approx1\,\MeVi$. The
complete Brillouin generator we use here fixes the issue in $\nuc{Ni}{64}$,
but the divergence in $\nuc{Ni}{66}$ remains. A possible reason could be the 
enforcement of spherical symmetry in our calculations. 
Recent experiments have revealed the coexistence of spherical ground states 
and axially deformed states with excitation energies below $3\,\MeV$
in $\nuc{Ni}{68}$ and its vicinity \cite{Chiara:2012ys,Recchia:2013kx,Suchyta:2014vn}. 
In the next subsection, we will discuss examples in which the MR-IMSRG(2)
successfully deals with the presence of both spherical and deformed states in the 
spectrum of neon isotopes, but we note that the states in question have 
much larger energetic separations of $7-8\,\MeV$.

\subsection{\label{sec:groundstate_Ne}Towards Doubly Open-Shell Nuclei: Neon Isotopes}

We want to conclude our discussion by addressing applications of the 
MR-IMSRG(2) away from the semi-magic chains. The biggest obstacle is the 
tendency of doubly open-shell nuclei to undergo transitions in their intrinsic 
shapes. In our calculation for semi-magic isotopic chains, we enforce spherical
symmetry when we calculate HF and PNP reference states, which greatly
facilitates the handling of the full $3N$ interaction at that stage.
These reference states have $J^\pi=0^+$, which implies that their
density matrices are scalars under rotations and block-diagonal in angular 
momentum in the individual one-body, two-body, \ldots sectors. As a consequence,
the normal-ordered Hamiltonian will be represented by block-diagonal
matrices in each sector as well, and $\eta$ and $\totd{H}{s}$ inherit
this structure through their relation with the Hamiltonian \footnote{For 
reference states with $J^\pi\neq0^+$, the density
matrices have non-scalar components that ultimately cause angular-momentum
changing blocks in the $f$ and $\Gamma$ to be non-zero. Of course, these
are coupled with the tensorial densities and creation / annihilation operators
so that the $\HO$ is a scalar overall. In the MR-IMSRG flow equations,
we would have to couple terms consisting of three spherical tensors, namely 
$\eta^{(1)}$ or $\eta^{(2)}$, $f$ or $\Gamma$, and a density matrix. Because
of the complicated angular momentum algebra that results, it is most 
likely an easier option to solve the flow equations in an $M$-scheme 
approach instead.}. The block-diagonal form of the MR-IMSRG flow equations
holds for general $J^\pi=0^+$ reference states, not just intrinsically 
spherical ones. We can just as well start from an intrinsically deformed state, 
e.g., from a symmetry-broken HF or HFB calculation, and project it on good 
angular momentum --- the strategy is the same as in the PNP case. The actual
computational challenge is the implementation of a converged deformed
HFB calculation with $3N$ forces, which we defer for now.

\begin{figure}[t]
  \begin{center}
    \includegraphics[width=0.66\textwidth]{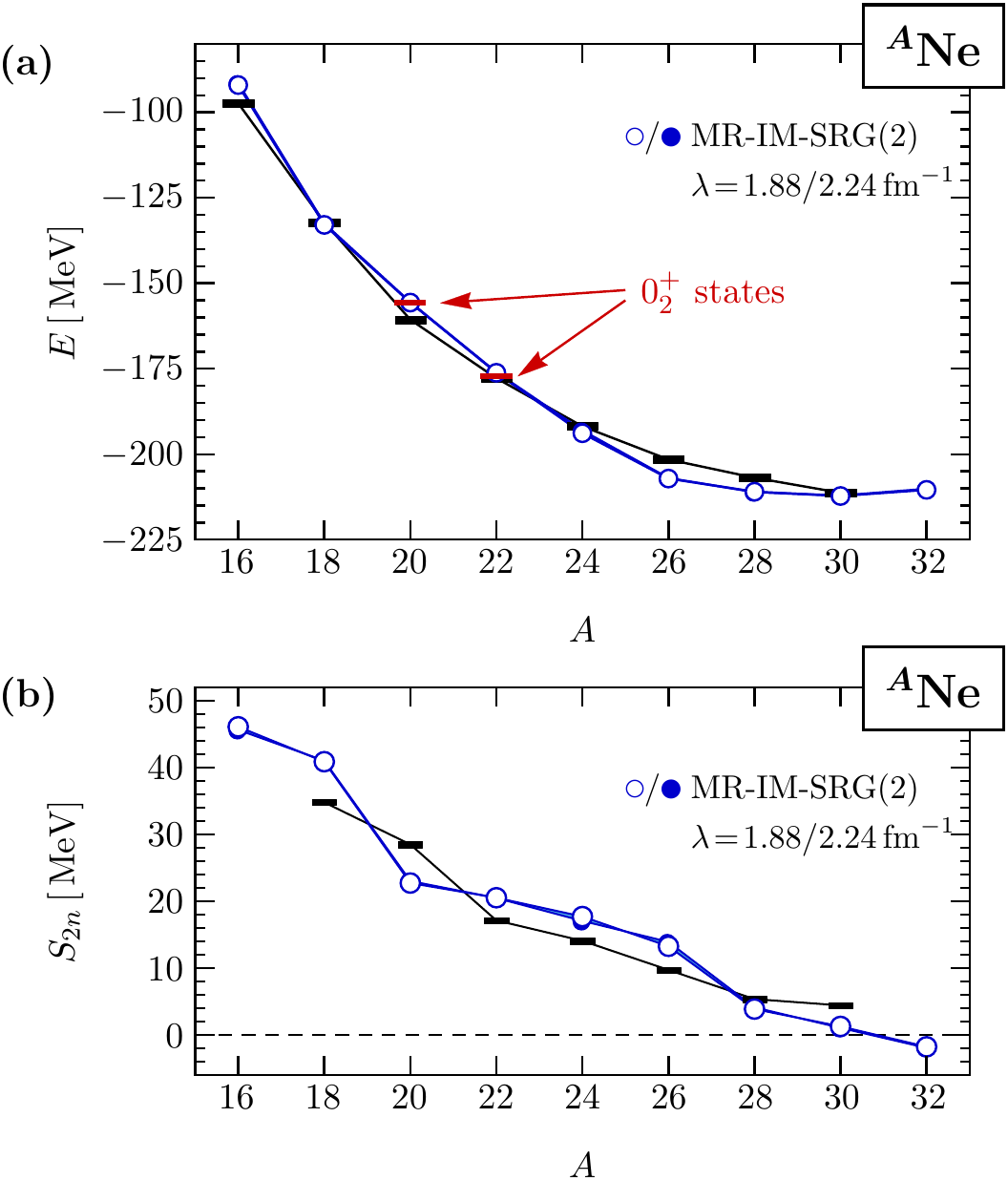}
  \end{center}
  \caption{\label{fig:NeXX} 
    MR-IMSRG(2) ground-state energies (top) and two-neutron separation energies
    (bottom) of the neon isotopes, for the $NN\!+\!3N(400)$ Hamiltonian with 
    $\lambdaSRG=1.88,\ldots,2.24\fmi$ ($\eMax=14,\EMax=14$, optimal $\hw$). 
    Spherical symmetry is enforced for the reference states in the calculation. Red bars indicate the
    absolute energies of $0^+_{2}$ excited states from Shell model calculations
    with IMSRG-derived effective interactions that are based on the same $NN\!+\!3N(400)$
    initial Hamiltonian. Black bars indicate experimental data \cite{Wang:2012uq}.    
  }
\end{figure}

Let us consider the neon isotopic chain as an example. As we can see in 
figure \ref{fig:NeXX}(a), the theoretical ground-state energies lie within a 
few $\MeV$ of experimental data, and they are insensitive to variations of 
$\lambdaSRG$. There are significant deviations between the theoretical and 
experimental energy trends along the isotopic chain, with the MR-IMSRG(2) 
energies alternating between lying above and below the experimental data. 
The reason for this behavior is the explicit spherical symmetry of 
the PNP reference states that we use in these calculations. As discussed in 
section \ref{sec:decoupling}, it is not guaranteed that the
MR-IMSRG flow will extract the ground state, and in practice, we find that 
the overlap between our chosen reference state and the targeted eigenstate 
plays an important role. This is illustrated beautifully in the examples of 
$\nuc{Ne}{20}$ and $\nuc{Ne}{22}$, which both have intrinsically deformed 
ground states \cite{Marinova:2011lo}. In section \ref{sec:sm}, we will show that
Shell model calculations with IMSRG effective interactions based on the
$NN\!+\!3N(400)$ Hamiltonian \cite{Bogner:2014tg,Stroberg:2016fk} yield 
deformed ground states for $\nuc{Ne}{20,22}$. In these calculations, 
we also find intrinsically spherical $0^+_2$ excited states whose absolute 
energies are in excellent agreement with the MR-IMSRG(2) energies. We indicate 
these states by red bars in figure \ref{fig:NeXX}(a). Thus, the MR-IMSRG(2) 
appears to target the eigenstate whose overlap with the spherical reference 
state is largest.

Finally, we show the neon $S_{2n}$ in figure \ref{fig:NeXX}(b). In 
$\nuc{Ne}{22-26}$, the trend actually follows experimental data rather well,
although the $NN\!+\!3N(400)$ results are offset by an almost constant shift.
The drops of the theoretical $S_{2n}$ at $\nuc{Ne}{18}$ and $\nuc{Ne}{26}$ 
are further examples of over-pronounced sub-shell closures, namely for $N=8$ and
$16$ (also see \cite{Gibelin:2007jw,Lepailleur:2013kh}. This supports our argument 
that the discrepancy between the experimental 
and theoretical energies for the $\nuc{O}{26}$ resonance is not entirely due 
to absence of continuum effects in our calculations.

\section{\label{sec:sm}Non-empirical Shell Model Interactions from the IMSRG}

In the previous section, we have reviewed MR-IMSRG results for ground
state energies of semi-magic isotopic chains, i.e., singly open-shell
nuclei, and presented a first look at doubly open-shell nuclei. The neon
chain which served as an illustrative example is actually one of the
harder cases we could have investigated, because it contains several
isotopes with considerable intrinsic deformation, including clustering
in the $N=Z$ nucleus $\nuc{Ne}{20}$. As discussed in Sec.~\ref{sec:groundstate_Ne}, 
the MR-IMSRG is formally capable of dealing with intrinsic deformation, 
but a practical implementation is very challenging and computationally
demanding, and ultimately, we are still only considering ground states
(or individual excited states selected by the MR-IMSRG flow).

The IMSRG framework provides us with another route for attacking the
nuclear many-body problem, building on our considerations of the effective
Hamiltonian in section \ref{sec:flow_heff}. As we have shown in a series of
publications \cite{Tsukiyama:2012fk,Bogner:2014tg,Stroberg:2016fk}, we
can use IMSRG flows to construct nonempirical interactions for use in
valence-space CI approaches like the interacting Shell model. In this way, 
we can systematically link 
Shell model phenomenology to the underlying nuclear interactions in the
vacuum, and through them to QCD if we start from chiral $NN$ and $3N$
interactions. The Shell model gives us immediate access to intrinsically 
deformed nuclei, excited states, transitions, etc., with the added
benefits that we can systematically study the mechanisms by which 
many-body correlations are absorbed into the valence-space interactions,
and maintain control over the input and many-body uncertainties. The
drawback of such a combined IMSRG+Shell model (IMSRG+SM) approach
is that we remain bound to the factorial computational scaling of the 
exact diagonalization in the valence space.

In the following, we will discuss the implementation of valence-space
decoupling via a straightforward modification of the ground-state
decoupling, and review results from recent applications
\cite{Bogner:2014tg,Stroberg:2016fk,Caceres:2015fk}.

\subsection{\label{sec:sm_decoupling}Valence-Space Decoupling}

\begin{figure}[t]
  \begin{center}
    \includegraphics[width=0.36\textwidth]{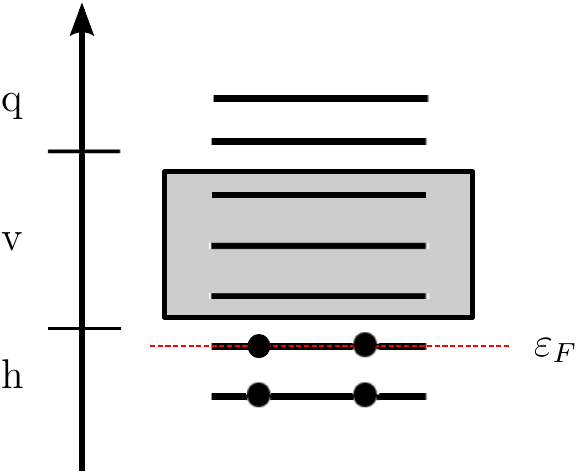}
  \end{center}
  \caption{\label{fig:valence_space}
    Separation of the single-particle basis into hole (h), valence particle (v)
    and non-valence particle (q) states. The Fermi energy of the fully occupied
    core, $\varepsilon_F$, is indicated by the red dashed line.
  }
\end{figure}

In section \ref{sec:decoupling}, we discussed in depth how we can use
the (MR-)IMSRG evolution to decouple a suitable reference state from
$npnh$ or general $n$-body excitations. From a more general point
of view, we actually decouple different \emph{sectors} of the many-body
Hilbert space by driving the matrix elements of the Hamiltonian that 
couple these sectors to zero. To decouple multiple states in a valence
space, we only need to extend our previous definition of the offdiagonal 
Hamiltonian from section \ref{sec:decoupling} in a suitable manner!

Let us follow established conventions and split the 
single-particle basis in our calculation into core or hole (h), valence 
particle (v) and non-valence particle (q) orbitals (see 
figure \ref{fig:valence_space}). The actual Shell model calculation for a nucleus
with $A$ nucleons is an exact diagonalization of the Hamiltonian matrix in a 
subspace of the Hilbert space that is spanned by configurations 
of the form
\begin{equation}\label{eq:def_configurations}
  \ket{\aaO_{v_1}\ldots\aaO_{v_{A_v}}} \equiv \aaO_{v_1}\ldots\aaO_{v_{A_v}}\ket{\Phi}\,,
\end{equation}
where $\ket{\Phi}$ is the wave function for a suitable core with $A_c$
nucleons, and the $A_v$ valence nucleons are distributed over the valence
orbitals $v_i$ in all allowed ways. Since the Shell model assumes the core 
to be inert, it can be viewed as a vacuum state for the valence configurations. 
The matrix representation of the Hamiltonian in the space spanned by these
configurations is
\begin{equation}
  \matrixe{v'_{1}\ldots v'_{A_v}}{\HO}{v_{1}\ldots v_{A_v}}
  = \matrixe{\Phi}{\aO_{v'_{A_v}}\ldots\aO_{v'_1}\HO\aaO_{v_1}\ldots\aaO_{v_{A_v}}}{\Phi}\,.
\end{equation}
For our purposes, this suggests that we normal order the Hamiltonian and other 
operators with respect to the core wave function $\ket{\Phi}$, which will
take on the role of the reference state for the IMSRG flow.
We obtain $\ket{\Phi}$ by solving the HF equations for the core, but use
the mass number $A$ of the target nucleus (instead of $A_c$) in the 
intrinsic Hamiltonian \eqref{eq:def_Hint}. This is appropriate, because
the IMSRG+SM calculation is supposed to replicate the results of an FCI 
calculation for the target nucleus.

\begin{figure}[t]
  \begin{center}
    \includegraphics[width=0.78\textwidth]{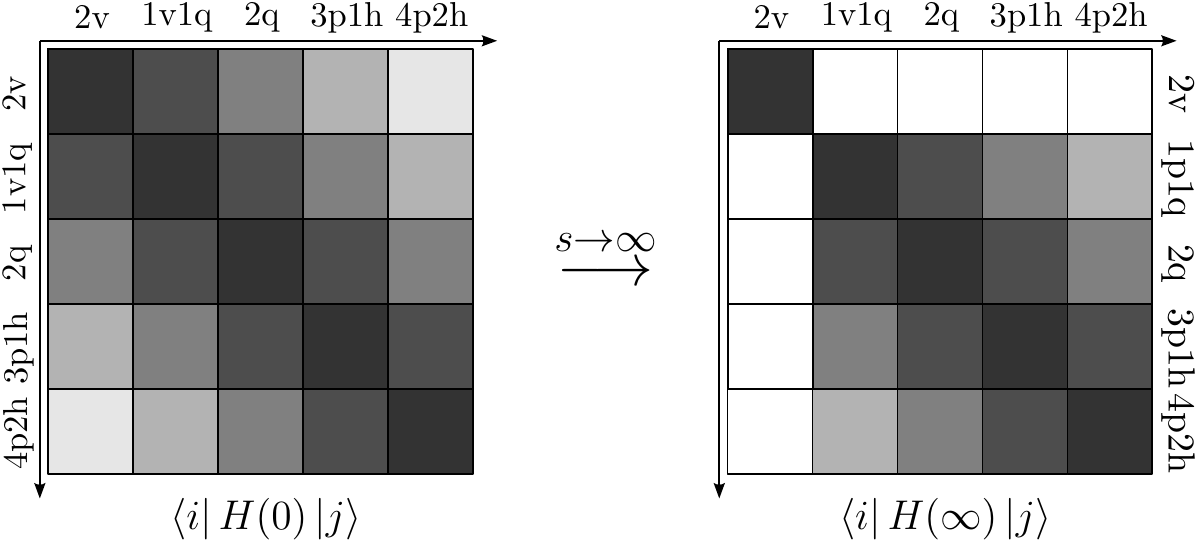}
  \end{center}
  \caption{\label{fig:valence_decoupling}
    Schematic view of IMSRG valence-space decoupling for two valence nucleons (p$\,=\,$v,\,q).
  }
\end{figure}

We want to use the IMSRG evolution to decouple the configurations \eqref{eq:def_configurations}
from states that involve excitations of the core, just as in the
ground-state calculations. In addition, we need to decouple them 
from states containing nucleons in non-valence particle states.
To illustrate the decoupling and identify the offdiagonal matrix 
elements, we consider two particles in the valence space, and show 
a schematic representation of the many-body Hamiltonian in a basis 
spanned by ($n$+2)p$n$h excitations of the reference state $\ket{\Phi}$ 
in figure \ref{fig:valence_decoupling}. In table \ref{tab:diagrams}, we classify 
the matrix elements of $\HO$ which couple $2$v to $1$q$1$v, $2$q, $3$p$1$h, 
and $4$p$2$h excitations, respectively, where p=v,q. For each type 
of matrix element, we show antisymmetrized Goldstone diagrams (see, 
e.g., \cite{Shavitt:2009}) involving the one- and two-body vertices 
$f$ and $\Gamma$ (three-body vertices are omitted because of the NO2B
approximation). Additional diagrams due to permutations of the 
nucleons or taking Hermitian adjoints are suppressed for brevity.

Diagrams (I) and (II) are eliminated if matrix elements of $f$ and 
$\Gamma$ that contain at least one q index are chosen to be offdiagonal. Diagrams 
(III) and (V) are eliminated by the reference state decoupling condition, 
which defines $f^{p}_{h}$ and $\Gamma^{pp'}_{hh'}$ as offdiagonal 
(cf.~section \ref{sec:decoupling}). This only leaves diagram (IV), 
which vanishes if matrix elements of the type $\Gamma^{pp'}_{vh}$ 
vanish. Thus, we define \cite{Tsukiyama:2012fk}
\begin{equation}\label{eq:def_Hod}
  H_{od} \equiv \sum_{i\neq i'}f^{i}_{i'}\nord{\AO^{i}_{i'}}
           + \frac{1}{4}\left(\sum_{pp'hh'}\Gamma^{pp'}_{hh'}\nord{\AO^{pp'}_{hh'}}
           + \sum_{pp'vh}\Gamma^{pp'}_{vh}\nord{\AO^{pp'}_{vh}}
           + \sum_{pqvv'}\Gamma^{pq}_{vv'}\nord{\AO^{pq}_{vv'}}\right)
           + \text{H.c.}\,.
\end{equation}
This definition of the offdiagonal Hamiltonian holds for an arbitrary
number of valence particles $A_v$. For $A_v=1$, diagram (II) vanishes, 
while diagrams (I) and (III)-(V) have the same topology, but one 
less spectator nucleons. Analogously, diagrams (I)-(V) merely contain 
additional spectator nucleons for $A_v>2$.

\begin{table*}[t]
  \setlength{\unitlength}{0.1\textwidth}
  \begin{tabular*}{\textwidth}{lm{0.16\textwidth}m{0.2\textwidth}m{0.45\textwidth}}
    \hline\hline
     no. & type & diagram & energy difference $\Delta$ \\
    \hline\\[-5pt]
     I & $\matrixe{2p}{\HO}{2p}$ & 
    \begin{picture}(2.0000,1.3000)
      \put(0.0000,0.1500){\includegraphics[height=\unitlength]{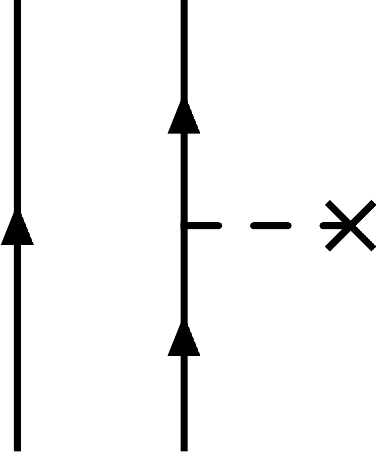}}
      \put(0.3750,1.2000){\footnotesize$p$}
      \put(0.3750,0.0000){\footnotesize$p'$}
    \end{picture}
    & 
    $f^{p}_{p} - f^{p'}_{p'}$\\
    II & $\matrixe{2p}{\HO}{2p}$ & 
    \begin{picture}(2.0000,1.4000)
      \put(0.0000,0.1500){\includegraphics[height=\unitlength]{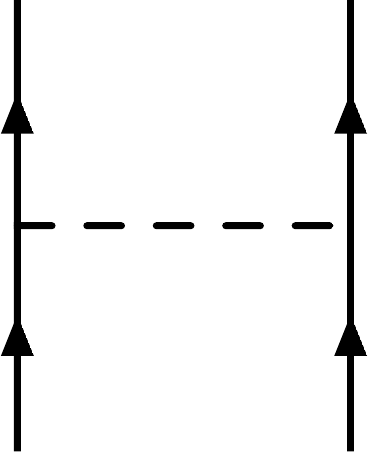}}
      \put(0.0150,1.2000){\footnotesize$p$}
      \put(0.7400,1.2000){\footnotesize$p'$}
      \put(0.0150,0.0000){\footnotesize$p''$}
      \put(0.7400,0.0000){\footnotesize$p'''$}
    \end{picture}
     & 
    $f^{p}_{p} + f^{p'}_{p'} - f^{p''}_{p''} - f^{p'''}_{p'''} + \Gamma^{pp'}_{pp'} - \Gamma^{p''p'''}_{p''p'''}$\\
    III & $\matrixe{3p1h}{\HO}{2p}$ &
    \begin{picture}(2.0000,1.4000)
      \put(0.0000,0.1500){\includegraphics[height=\unitlength]{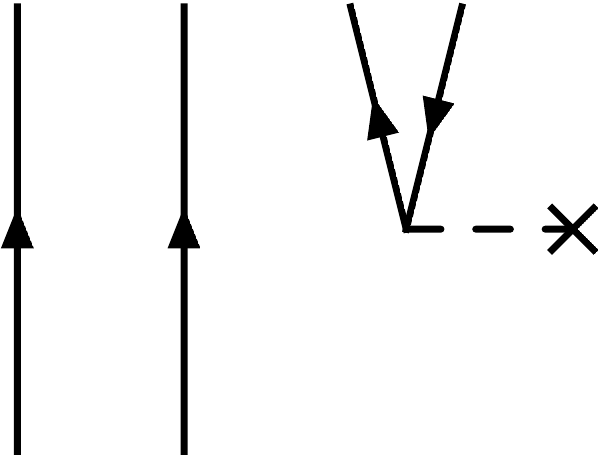}}
      \put(0.7400,1.2000){\footnotesize$p$}
      \put(0.9600,1.2000){\footnotesize$h$}
  \end{picture}
    & 
    $f^{p}_{p} - f^{h}_{h} - \Gamma^{ph}_{ph}$
    \\
    IV & $\matrixe{3p1h}{\HO}{2p}$ &
    \begin{picture}(2.000,1.4000)
      \put(0.0000,0.1500){\includegraphics[height=\unitlength]{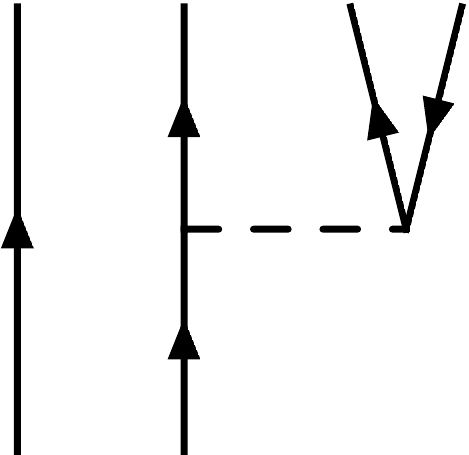}}
      \put(0.3750,1.2000){\footnotesize$p$}
      \put(0.7400,1.2000){\footnotesize$p'$}
      \put(0.9600,1.2000){\footnotesize$h$}
      \put(0.3750,0.0000){\footnotesize$p''$}
    \end{picture}
    & 
    $f^{p}_{p} + f^{p'}_{p'} - f^{p''}_{p''}-f^{h}_{h} + \Gamma^{pp'}_{pp'}-\Gamma^{ph}_{ph}-\Gamma^{p'h}_{p'h}$ 
    \\
    V & $\matrixe{4p2h}{\HO}{2p}$ &
    \begin{picture}(2.000,1.4000)
      \put(0.0000,0.1500){\includegraphics[height=\unitlength]{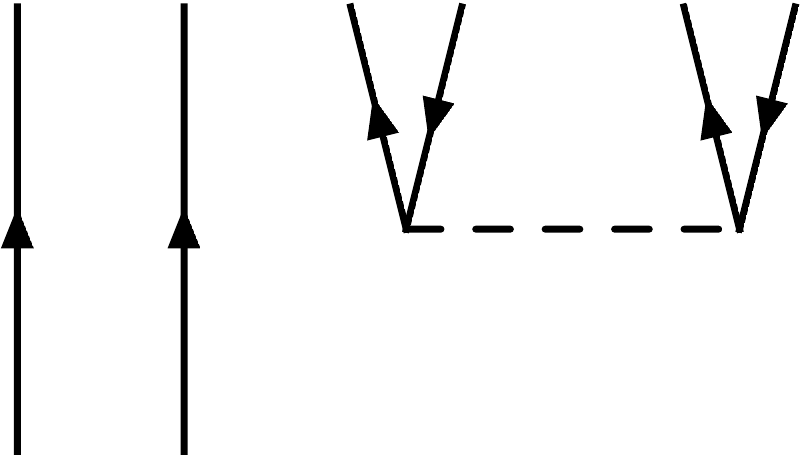}}
      \put(0.7400,1.2000){\footnotesize$p$}
      \put(0.9600,1.2000){\footnotesize$h$}
      \put(1.4800,1.2000){\footnotesize$p'$}
      \put(1.7000,1.2000){\footnotesize$h'$}
    \end{picture}
    &
    $f^{p}_{p} + f^{p'}_{p'} - f^{h}_{h} - f^{h'}_{h'} 
    + \Gamma^{pp'}_{pp'} + \Gamma^{hh'}_{hh'} - \Gamma^{ph}_{ph} 
    - \Gamma^{p'h'}_{p'h'} - \Gamma^{ph'}_{ph'}-\Gamma^{p'h}_{p'h}$
    \\ 
    \hline\hline
  \end{tabular*}
  \caption{\label{tab:diagrams} 
    Classification of matrix elements of the many-body Hamiltonian 
    in the many-body Hilbert space spanned by $(n+2)$p$n$h excitations 
    of the reference state (cf.~\ref{fig:valence_decoupling}). For 
    each matrix element, we show the corresponding antisymmetrized 
    Goldstone diagrams \cite{Shavitt:2009} involving the one- and 
    two-body parts of $\HO$ (permutations involving spectator particles 
    which are required by antisymmetry are implied), as well as the 
    energy differences appearing in the matrix elements for $\eta(s)$ 
    in each case (see text).
  }
\end{table*}

Using $H_{od}$ in the construction of generators, we evolve the 
Hamiltonian by solving the flow equations 
\eqref{eq:mr_flow_0b_tens}--\eqref{eq:mr_flow_2b_tens} in IMSRG(2) 
truncation. Since the core wave functions are HF Slater determinants, 
we can work in the single-reference limit. The final Hamiltonian is 
given by
\begin{equation}
 \Hfinal = \UO(\infty)\HO\UUO(\infty) = E + \sum_{v}f^{v}_{v}\nord{\AO^{v}_{v}} + \frac{1}{4}\sum_{v_i,v_j,v_k,v_l}
  \Gamma^{v_iv_j}_{v_kv_l}\nord{\AO^{v_iv_j}_{v_kv_l}} + \ldots\,,
\end{equation}
where the explicitly shown terms are the core energy, single-particle
energies, and two-body matrix elements that are used as input for 
a Shell model diagonalization. The solutions of that diagonalization 
are given by
\begin{equation}
  \ket{\overline{\Psi}_n} = \sum_{v_1,\ldots,v_{A_v}} C^{(n)}_{v_1\ldots v_{A_v}}\aaO_{v_1}\ldots\aaO_{v_{A_v}}\ket{\Phi}\,,
\end{equation}
and they are related to the eigenstates of the initial Hamiltonian (up to 
truncation errors) by
\begin{equation}
  \ket{\Psi_n} = \UUO(\infty)\ket{\overline{\Psi}_n}\,.
\end{equation}

The naive computational scaling for the valence-decoupling procedure 
described here is $\OC(N^6)$, just like that of MR-IMSRG(2) or IMSRG(2) 
ground-state calculations. In practice, individual calculations require
about 100-1000 core hours, putting the effort between that of  
single-reference and multireference ground-state calculations. 
Compared to other nonperturbative approaches for the construction of 
nonempirical Shell model interactions, this effort is low \cite{Jansen:2014qf,Lisetskiy:2008fk,Dikmen:2015fk}. 
Moreover, we note that we obtain consistent neutron-neutron, 
proton-neutron, and proton-proton interactions from the same IMSRG 
evolution.

\subsection{Ground-state Energies and Targeted Normal Ordering}

As a first application \cite{Bogner:2014tg}, we tested the IMSRG+SM 
approach in the oxygen isotopic chain, where results from 
large-scale MR-IMSRG ground state calculations and a variety of other exact 
and approximate \emph{ab initio} methods are available for comparison 
(cf.~section \ref{sec:groundstate}). 

\begin{figure}[t]
  \begin{center}
    \includegraphics[width=0.6\textwidth]{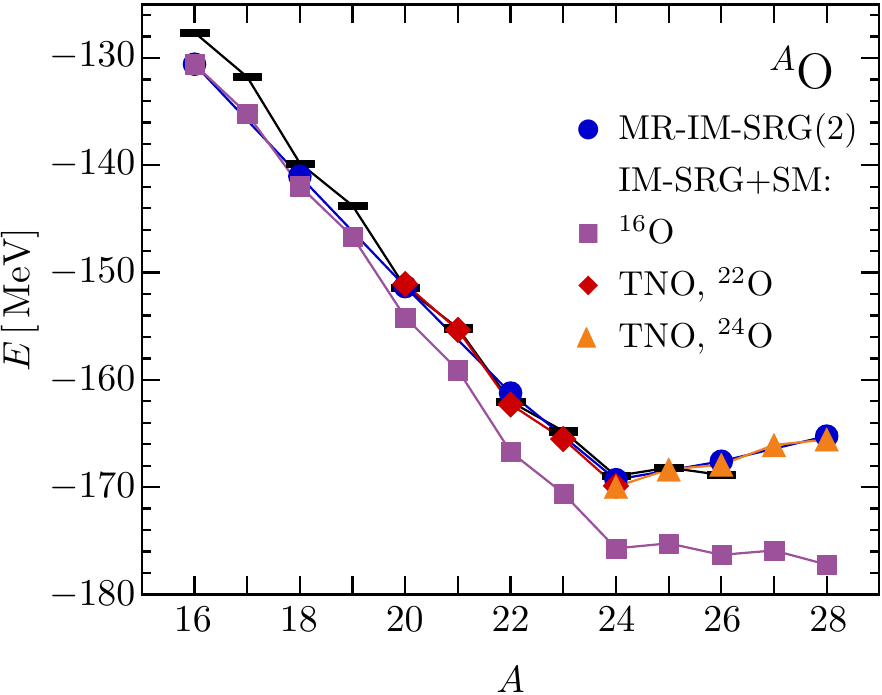}
  \end{center}
  \caption{\label{fig:OXX_sm_gs}
    Ground-state energies of the oxygen isotopes, calculated from nonempirical
    IMSRG Shell model interactions derived from $NN\!+\!3N(400)$ at a resolution scale
    $\lambdaSRG=1.88\,\fmi$ and $\hw=24\,\MeV$. We compare results from the original 
    IMSRG+SM approach discussed in \cite{Bogner:2014tg} and an improved version using
    the so-called targeted normal ordering (TNO, \cite{Stroberg:2016fk}), using $\nuc{O}{22}$
    and $\nuc{O}{24}$ as reference states.
    Black bars indicate experimental data \cite{Wang:2012uq}.
  }
\end{figure}

In figure \ref{fig:OXX_sm_gs} we show the oxygen ground-state energies, calculated 
with effective interactions derived from the $NN\!+\!3N(400)$ Hamiltonian (see 
section \ref{sec:flow_implementation}) at a resolution scale $\lambdaSRG=1.88\,\fmi$. 
Note that we include ground-state energies for the odd oxygen isotopes, which
are easily obtainable from a Shell model calculation. In the vicinity of 
$\nuc{O}{16}$, the ground-state energies obtained from IMSRG+SM and MR-IMSRG(2) 
calculations agree well with each other and experimental data, but for growing 
neutron number $N$, the IMSRG+SM solutions are increasingly overbound. 

The origin of this discrepancy can be traced back to the reference state 
we use for both the normal ordering and the IMSRG valence-space decoupling. 
Initially, we used HF solutions for the $\nuc{O}{16}$ core, only changing 
the mass number of the intrinsic Hamiltonian to that of the target nucleus, 
as explained in the previous subsection. When we normal order the Hamiltonian
and perform the IMSRG evolution, we miss contributions from the valence
nucleons that are taken into account fully in the MR-IMSRG ground-state 
calculations. In Ref.~\cite{Stroberg:2016fk}, we took a first step towards
remedying this deficiency by means of a so-called \emph{targeted normal-ordering}
(TNO) procedure, in which the reference state for the normal ordering and 
decoupling is a HF solution for a closed sub-shell nucleus in close 
proximity to the target nucleus.
Then, the Hamiltonian is re-normal ordered with respect to the $\nuc{O}{16}$
core that is assumed by the Shell model calculation. As shown in figure \ref{fig:OXX_sm_gs},
this procedure essentially eliminates the difference between the 
IMSRG+SM and MR-IMSRG(2) ground-state energies, reducing the 
overbinding of the neutron-rich oxygen isotopes by several MeV. For 
$\nuc{O}{24}$, we can use both $\nuc{O}{22}$ and $\nuc{O}{24}$ as 
reference states for the TNO and decoupling. The resulting ground-state 
energies are $-169.874\,\MeV$ and $-169.956\,\MeV$, respectively, in 
excellent agreement with each other and the MR-IMSRG(2) result for the 
same initial Hamiltonian, which is $-169.491\,\MeV$. 

\begin{figure}[t]
  \begin{center}
    \includegraphics[width=0.43\textwidth]{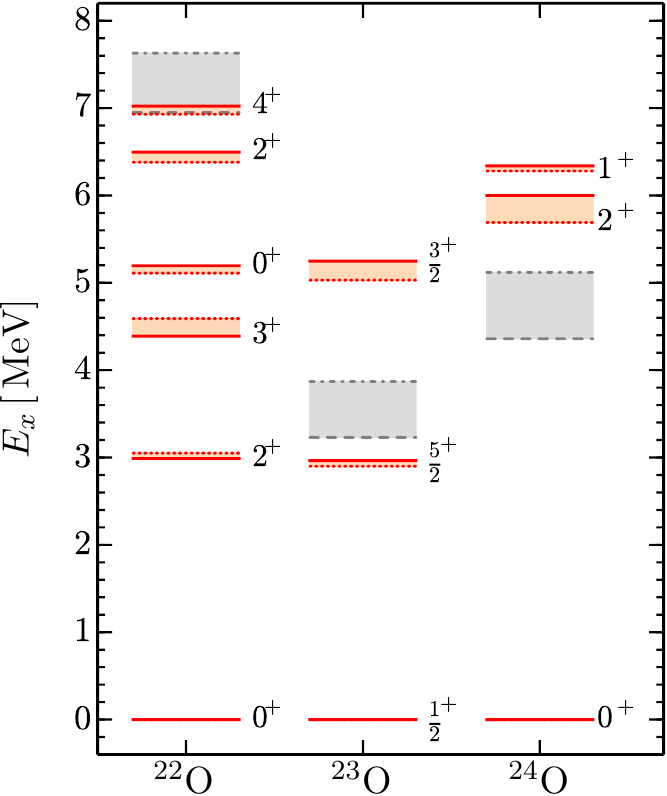}
  \end{center}
  \caption{
    \label{fig:OXX_tno}
    Excitation spectra of $\nuc{O}{22-24}$ from IMSRG+SM calculations with (solid
    lines)
    and without targeted normal ordering (dotted lines, see text). All effective interactions 
    are derived from the chiral $NN\!+\!3N(400)$ interaction at resolution scale 
    $\lambdaSRG=1.88\,\fmi$ ($\eMax=14, \EMax=14, \hw=24\MeV$). The gray dashed 
    and dash-dotted lines indicate the neutron separation energies.
  }
\end{figure}

In figure \ref{fig:OXX_tno}, we show the effect of the TNO on the low-lying
excitation spectra of $\nuc{O}{22-24}$ and the neutron separation energies
\begin{equation}
  S_n(Z,N)=E(Z,N-1)-E(Z,N)\,.
\end{equation}
Calculations were performed with an $\nuc{O}{22}$ reference state. If we use 
an $\nuc{O}{24}$ reference state instead, both the excitation energies and 
neutron separation energies change by $80\,\keV$ or less (not shown). Figure 
\ref{fig:OXX_tno} illustrates that the TNO improves the core energy, single-particle 
energies, and two-body matrix elements, in decreasing order of importance. For 
the nuclei shown here, the core energy is raised by $3.5-4\,\MeV$, which 
accounts for the bulk of the ground-state energy improvement. The input 
single-particle energies for protons and neutrons are increased by up to 
$1.3\,\MeV$ ($\pi0d_{3/2}$) and $200\,\keV$ ($\nu1s_{1/2}$), respectively.
The $S_{n}$ decrease rather uniformly by $600\,\keV$ for our sample nuclei. 
Finally, the effect of the TNO on the
two-body matrix elements is weak, so the orderings and level spacings of the 
excitation spectra are hardly affected. The largest change in excitation 
energy is about $250\,\keV$.

\subsection{ Spectroscopy of $sd$-shell nuclei }

\begin{figure}[t]
  \begin{center}
    \includegraphics[width=\textwidth]{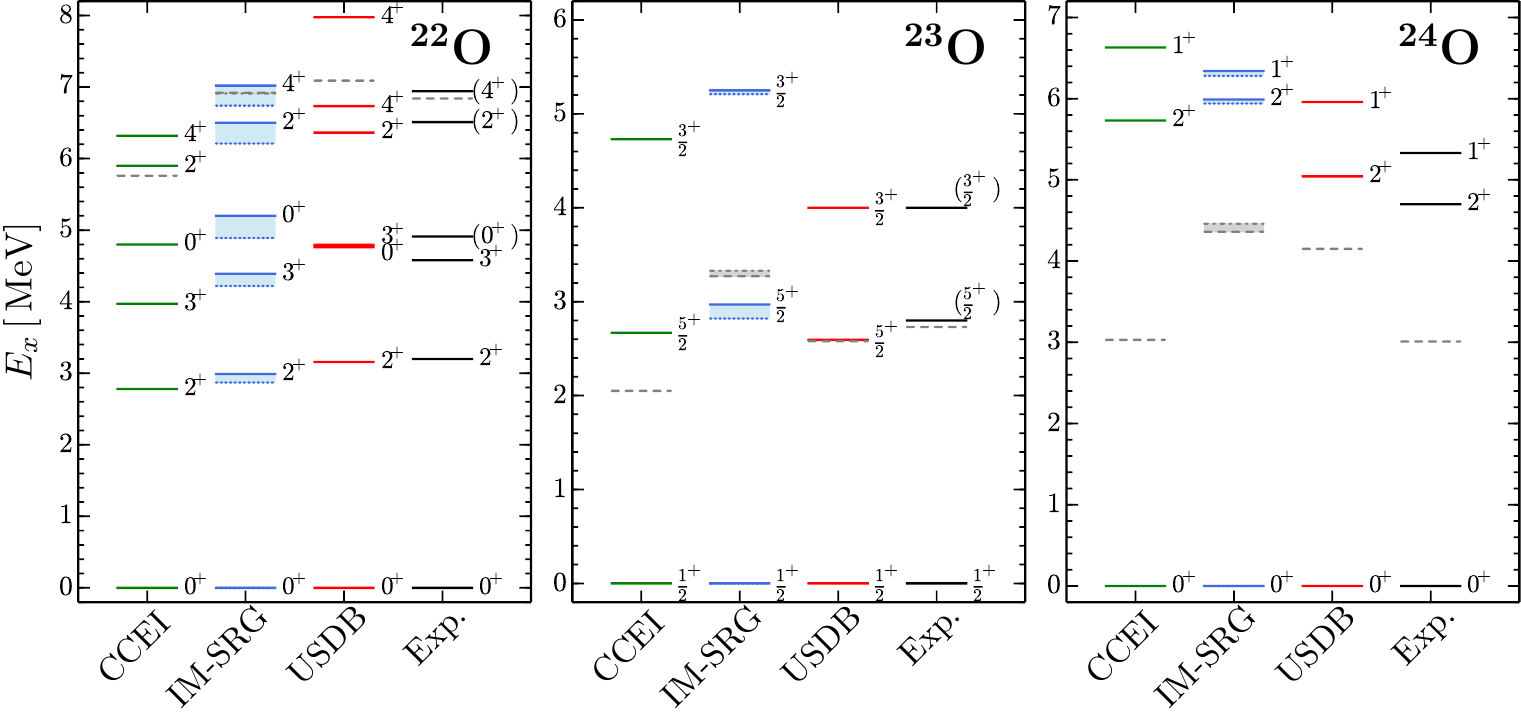}
  \end{center}
  \caption{\label{fig:OXX_spectra}
    Excitation spectra of $\nuc{O}{22-24}$ based on the chiral $NN\!+\!3N(400)$ interaction at 
    resolution scale $\lambdaSRG=1.88\,\fmi$. We compare results for effective interactions
    derived by IMSRG valence-space decoupling ($\eMax=14, \EMax=14, \hw=20\MeV$ (dashed 
    lines) and $24\,\MeV$ (solid lines)), the $A$-dependent CCEI approach of \cite{Jansen:2014qf} 
    ($\eMax=12, \EMax=14, \hw=20\,\MeV$), and the phenomenological USDB interaction 
    \cite{Brown:2006fk} to experimental data \cite{NuDat:2014}. The dashed lines represent 
    the neutron separation energies. 
  }
\end{figure}

Let us now discuss IMSRG+SM results for the spectra of selected $sd$-shell nuclei,
starting with $\nuc{O}{22-24}$. In figure \ref{fig:OXX_spectra}, we show results obtained with
the $NN\!+\!3N(400)$ interaction at a resolution scale $\lambdaSRG=1.88\,\fmi$. The
shaded band results from varying the oscillator basis parameter from $\hw=20\,\MeV$ 
to $24\,\MeV$, which serves as an indicator for the convergence of a specific
excited state. Factors that can affect the convergence are complex intrinsic 
shapes, the extension of the wave function's tail, etc. Overall, the convergence
is satisfactory for the nuclei shown here. The IMSRG+SM results agree impressively
well with experimental data, given that the parameters of our Hamiltonian have not 
been adjusted to the oxygen isotopes at all (see Ref.~\cite{Bogner:2014tg} for a 
more detailed analysis). The inclusion of $3N$ interactions in the initial 
Hamiltonian is crucial for achieving this good reproduction of the experimental level data:
It stabilizes the spacing between the neutron $0d_{3/2}$ orbital and the other levels
in the $sd$-shell as neutrons are added, which governs the energies of low-lying
excitations and the location of the neutron drip line 
\cite{Otsuka:2010cr,Holt:2013fk,Hergert:2013ij,Cipollone:2013uq}.
We note that the excited states in $\nuc{O}{23,24}$ are overestimated to
some degree, but this is expected because our Shell model calculations do not 
explicitly treat the continuum coupling at present, and the $NN\!+\!3N(400)$
Hamiltonian produces a too-pronounced shell closure at $N=16$
(see section \ref{sec:groundstate_O}).

In figure \ref{fig:OXX_spectra}, we also compare our spectra to results obtained 
with the phenomenological USDB interaction \cite{Brown:2006fk}, as well as 
nonempirical valence-space Hamiltonians obtained within the Coupled 
Cluster Effective Interaction (CCEI) approach \cite{Jansen:2014qf,Jansen:2016kq}. 
We note that the former depends on the mass number $A$ of the target nucleus 
through a scaling of the two-body matrix elements \cite{Brown:2006fk}.
In the latter, interactions for specific target masses were constructed for use 
in the oxygen isotopes, starting from the same Hamiltonian that we used for the
IMSRG+SM here. The CCEI and IMSRG results for $\hw=20\,\MeV$ are in very good 
agreement. Since 
CCEI is built from CCSD and its Equation-of-Motion extension to excited states 
\cite{Jansen:2014qf,Hagen:2010uq,Jansen:2013zr}, the reshuffling of correlations 
into the valence-space effective interaction should be similar to that of 
IMSRG(2) valence decoupling, and therefore reflect the similarity of CCSD and 
IMSRG(2) ground-state results (see sections \ref{sec:flow_heff}, \ref{sec:groundstate}).
The biggest discrepancy occurs for the neutron separation energies, which are 
lower for CCEI because neutron-rich oxygen isotopes are increasingly underbound
(see \cite{Jansen:2014qf}).

\begin{figure}[t]

  \begin{center}
    \includegraphics[width=0.425\textwidth]{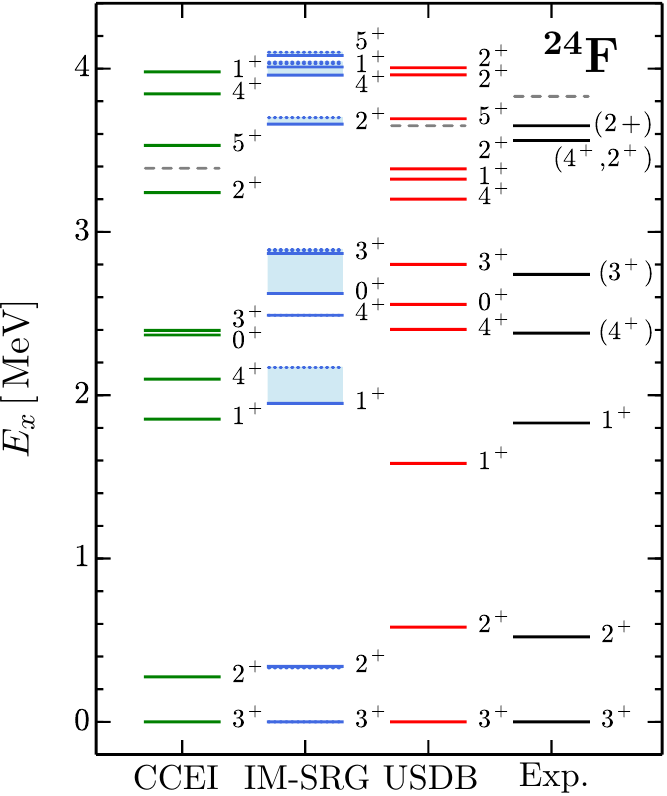}
  \end{center}
  \caption{\label{fig:F24_spectra}
    Excitation spectrum of $\nuc{F}{24}$ based on the chiral $NN\!+\!3N(400)$ interaction 
    at $\lambdaSRG=1.88\,\fmi$. 
    We compare results for effective interactions
    derived by IMSRG valence-space decoupling ($\eMax=14, \EMax=14, \hw=20\MeV$ (dashed 
    lines) and $24\,\MeV$ (solid lines)), the $A$-independent CCEI interaction \cite{Jansen:2016kq} 
    ($\eMax=12, \EMax=14, \hw=20\,\MeV$), and the phenomenological USDB interaction 
    \cite{Brown:2006fk} to experimental data \cite{Caceres:2015fk,NuDat:2014}. The 
    dashed lines represent the neutron separation energies. 
  }
\end{figure}

The USDB interaction is optimized to simultaneously describe more than 600
excited states in $sd$-shell nuclei \cite{Brown:2006fk}, hence it is not surprising
that the USDB spectra agree very well with experiment. We observe the most notable
deviation for the second $0^+$ and the $3^+$ state in $\nuc{O}{22}$, which are
nearly degenerate and whose ordering is inverted compared to experiment. The 
CCEI and IMSRG interactions describe the level ordering correctly. The latter even 
seems to give the correct level spacing for $\hw=24\,\MeV$, although it is necessary to
reduce the $\hw$ variation of the $3^+$ state to make a conclusive claim here.

As mentioned in section \ref{sec:sm_decoupling}, the IMSRG valence-space decoupling
provides us with a consistent set of proton-proton, proton-neutron, and 
neutron-neutron interaction matrix elements at the same time, so we can
easily extend our calculations into the $sd$-shell. For instance, we calculated
the excitation spectrum of $\nuc{F}{24}$ in support of a recent experiment at
GANIL \cite{Caceres:2015fk}. In figure \ref{fig:F24_spectra}, we again compare
IMSRG+SM (with $\hw$ variation) to experimental data and other theoretical 
results. States below $3\,\MeV$ are described well by the IMSRG, and aside
from the $0^+$ and $1^+$ states, very well converged. The IMSRG interaction 
exhibits a gap in the spectrum between the $3_2^+$ and $2_2^+$ 
states, which USDB fills with a group of states that has not been observed
at the corresponding energy in the experiment.

\begin{figure}[t]


  \begin{center}
    \includegraphics[width=0.78\textwidth]{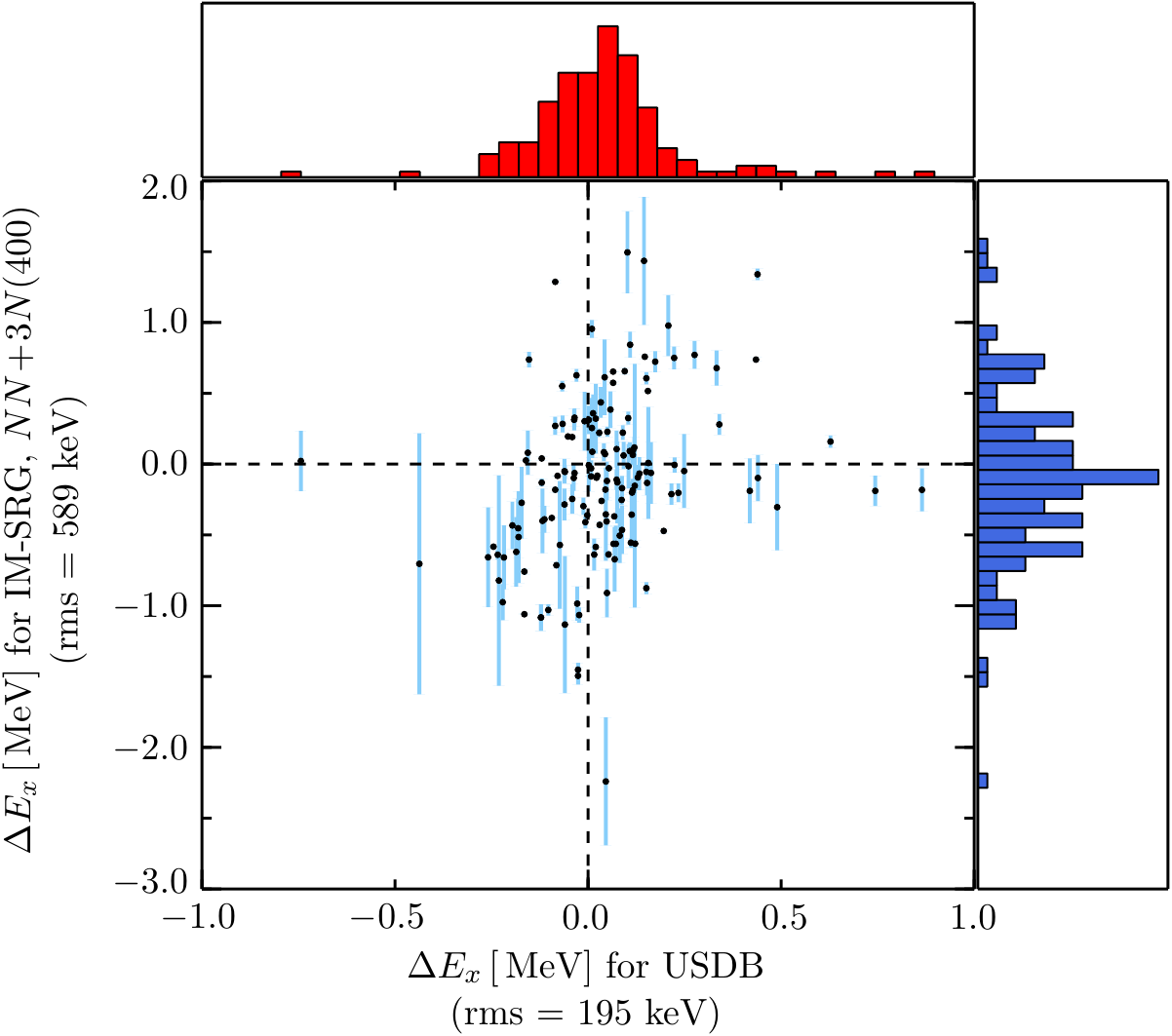}
  \end{center}
  \caption{\label{fig:rms}
    Deviation of theoretical excitation energies from experimental data for 144
    levels in $\mathrm{O,Fe,Ne,Na,Mg}$, calculated with USDB and IMSRG effective
    interactions derived from $NN\!+\!3N(400)$ at $\lambdaSRG=1.88\,\fmi$. The 
    error bars indicate the uncertainty of IMSRG excitation energies from varying
    $\hw$ of the underlying oscillator basis (see text).
  }
\end{figure}

The CCEI results we show in figure \ref{fig:F24_spectra} were obtained with
the \emph{$A$-independent} interaction that was recently published in 
Ref.~\cite{Jansen:2016kq}. While the ordering of the low-lying states is 
the same as for the IMSRG interactions, the positions and spacings of the
levels are notably different. We assume that this is caused by fixing $A$
to the mass number of the $\nuc{O}{16}$ core (and its vicinity) instead of
the target nucleus when the interaction is constructed \cite{Jansen:2014qf,Jansen:2016kq}.
This issue will be investigated further elsewhere.

Encouraged by the good agreement of our excitation energies with results
for the USDB interaction, we decided to broaden our perspective beyond
individual nuclei. In figure \ref{fig:rms}, we compare the deviations 
of theoretical and experimental excitation energies of 144 excited states 
in the $\mathrm{O, Fe, Ne, Na}$ and $\mathrm{Mg}$ isotopes, for both
USDB and the $A$-dependent IMSRG valence-space interactions discussed 
here. For USDB, the root-mean-square (rms) deviation is a mere $195\,\keV$,
with individual deviations ranging from $-1\,\MeV$ to $1\,\MeV$.
Aside from a few outliers, the bulk of the deviations for the IMSRG 
interactions fall in a similar range, but the distribution 
is wider, leading to an rms deviation of 589\,\keV. The ``error bars'' on the
IMSRG results indicate the uncertainty of the excitation energies due 
to a variation of $\hw$ from $20\,\MeV$ to $24\,\MeV$ --- we remind the
reader that the effect of such variations depends on the structure and 
convergence of individual states, as mentioned above. In summary, while
the description of the states in the lower $sd$-shell with IMSRG derived
interactions is not on the same level of accuracy as with USDB, it is
very encouraging that an rms of 589\,\keV can be achieved without adjusting
the parameters of the Hamiltonian to the nuclei in the region.

\subsection{\label{sec:sm_deformation} Deformation and Rotational Bands }

\begin{figure}[t]

  \begin{center}
      \includegraphics[width=0.73\textwidth]{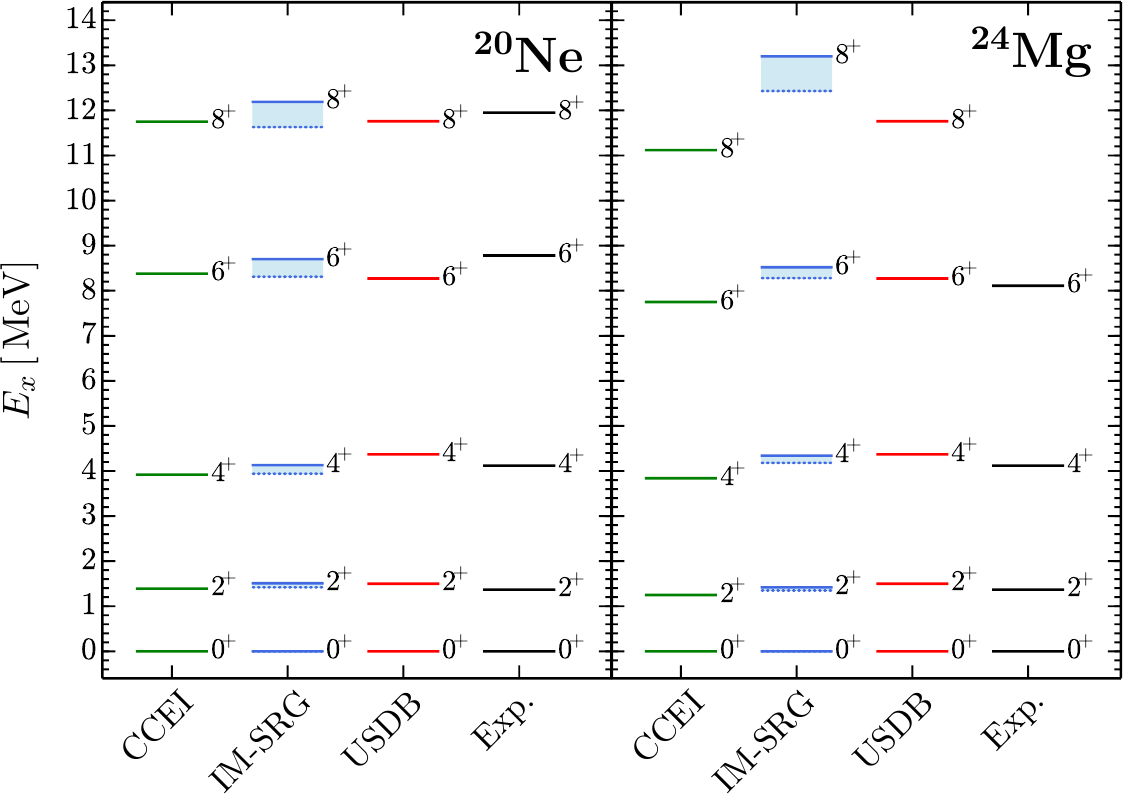}
  \end{center}
  \caption{\label{fig:rot_spectra}
    Ground-state rotational bands of $\nuc{Ne}{20}$ and $\nuc{Mg}{24}$, based on 
    the $NN\!+\!3N(400)$ interaction at $\lambdaSRG=1.88\,\fmi$. 
    We compare results for effective interactions
    derived by IMSRG valence-space decoupling ($\eMax=14, \EMax=14, \hw=20\MeV$ (dashed 
    lines) and $24\,\MeV$ (solid lines)), the $A$-independent CCEI interaction \cite{Jansen:2016kq} 
    ($\eMax=12, \EMax=14, \hw=20\,\MeV$), and the phenomenological USDB interaction 
    \cite{Brown:2006fk} to experimental data \cite{Caceres:2015fk,NuDat:2014}. The 
    dashed lines represent the neutron separation energies. 
  }
\end{figure}

As discussed above, the Shell model gives us access to nuclei with
intrinsic deformation. The ground-state rotational bands of $\nuc{Ne}{20}$
and $\nuc{Mg}{24}$ are shown in figure \ref{fig:rot_spectra}. The levels
obtained with the IMSRG and the $A$-independent CCEI interactions,
both based on the $NN\!+\!3N(400)$ Hamiltonian ($\lambdaSRG=1.88,\fmi$),
are in good agreement with each other as well as the USDB interaction.
While rotational bands emerge naturally in these nuclei even without an
initial chiral $3N$ force, its inclusion markedly improves the agreement
of the theoretical excitation energies and level spacings with experimental
data \cite{Stroberg:2016fk}.

\begin{figure}[t]





  \begin{center}
    \includegraphics[width=0.99\textwidth]{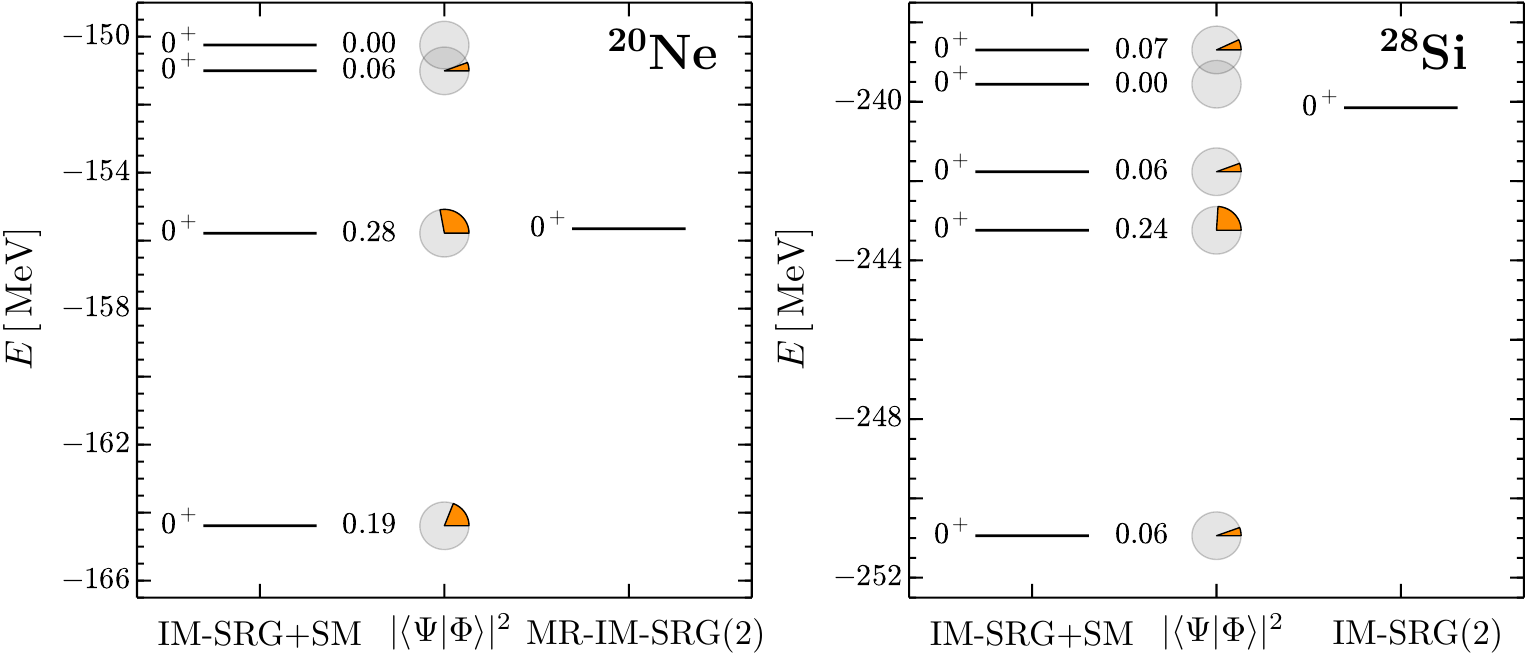}
  \end{center}
  \caption{
    \label{fig:rot}
    Absolute energies of the $0^+$ states in $\nuc{Ne}{20}$ and $\nuc{Si}{28}$ 
    from Shell model calculation with an IMSRG derived interaction, 
    compared to the result of MR-IMSRG(2) ground-state calculations.
    The $NN\!+\!3N(400)$ Hamiltonian at $\lambdaSRG=1.88\,\fmi$ 
    served as input in all cases. The middle column shows the overlap 
    of the Shell model solutions with intrinsically spherical Shell model 
    configurations:
    $\ket{[\pi 0d_{5/2}]^2 J_\pi=0, [\nu 0d_{5/2}]^2 J_\nu=0; J=0 }$
    for $\nuc{Ne}{20}$, and 
    $\ket{[\pi 0d_{5/2}]^6 J_\pi=0, [\nu 0d_{5/2}]^6 J_\nu=0; J=0 }$
    for $\nuc{Si}{28}$.
  }
\end{figure}

The Shell model's capability to describe intrinsically deformed nuclei allows
us to follow up on our discussion of $\nuc{Ne}{20}$ from section \ref{sec:groundstate_Ne}
now. There, we claimed that the MR-IMSRG(2) extracts an excited state with 
spherical intrinsic structure. In the left panel of figure \ref{fig:rot}, we 
show the absolute energies of the four lowest $0^+$ states in $\nuc{Ne}{20}$, 
and the MR-IMSRG(2) energy for the same Hamiltonian. The MR-IMSRG(2) energy is in excellent 
agreement with that of the $0_2^+$ state. The middle column of the panel 
shows the overlap of the Shell model eigenstate with the spherical configuration
(suppressing the core wave function)
$\ket{[\pi 0d_{5/2}]^2 J_\pi=0, [\nu 0d_{5/2}]^2 J_\nu=0; J=0 }$,
which is a fair approximation to the spherical PNP reference state we 
use for the MR-IMSRG(2). This overlap is indeed largest for the $0^+_2$
state, but also considerable for the $0_1^+$ state.

As another example, we consider $\nuc{Si}{28}$. For the Hamiltonian we
use here, there is a stable HF solution with closed proton and neutron 
subshells and spherical intrinsic structure. The MR-IMSRG(2) energy 
(or rather, IMSRG(2) energy because of the HF reference state) is within
$400\,\keV$ of the $0_4^+$ state, but that specific Shell model solution
has practically no overlap with the 
$\ket{[\pi 0d_{5/2}]^6 J_\pi=0, [\nu 0d_{5/2}]^6 J_\nu=0; J=0 }$
configuration that is identical to the reference state used for both 
the IMSRG(2) calculation and the (TNO) valence decoupling. The overlap 
is largest for the $0^+_2$ state (24\%), whose absolute energy is 
$-243.2\,\MeV$, compared to the IMSRG(2) energy of $-240.2\,\MeV$.

An absolute energy difference of $3\,\MeV$ between the IMSRG+SM 
and IMSRG(2) results is well within the realm of possibility, given
the estimated uncertainties due to the many-body truncation. This is
especially relevant because $\nuc{Si}{28}$ is an $N=Z$ nucleus, and
therefore exhibits $\alpha$-cluster correlations in excited states
\cite{Oertzen:2006ec}. The IMSRG(2) energy contains only limited 
contributions from $4p4h$ excitations, which appear first as immediate 
excitations in fourth-order MBPT diagrams (see \cite{Hergert:2016jk}).
A complete treatment of $4p4h$ excitations would make it necessary
to work in IMSRG(4) truncation (or use a reference state with $\alpha$
correlations in the MR-IMSRG). In contrast, the Shell model 
diagonalization can readily access $4p4h$ excitations in the 
valence space. From this perspective, it is perhaps more surprising
that the IMSRG+SM  and MR-IMSRG(2) energies for the $0_2^+$
state in the $\nuc{Ne}{20}$ are practically identical, because
that nucleus should exhibit $\alpha$ correlations as well. We
conclude our discussion here, and defer further investigations
of this issue to a future publication.

\section{\label{sec:epilogue}Conclusions and Outlook}
Over the course of this article, we have strived to give a pedagogical
introduction to the description of nuclear many-body physics in the 
In-Medium SRG framework. The IMSRG belongs to a family of 
efficient, systematically extendable many-body approaches that have 
extended the reach of \emph{ab initio} nuclear structure theory well
into the medium-mass region of the nuclear chart in recent years 
(see, e.g., \cite{Binder:2014fk,Hergert:2016jk}). The MR-IMSRG and
IMSRG+SM, specifically, are ideally suited to investigate 
the properties of open-shell nuclei with a systematic assessment of
the theoretical uncertainties. The consistency of the results from
these two different approaches that are rooted in the same framework 
is highly encouraging, promising many opportunities for cross-validating 
and interpreting nuclear structure results in future applications
(cf.~sections \ref{sec:groundstate} and \ref{sec:sm}). The capability 
to confront nuclear interactions from Chiral EFT with a wealth of 
new many-body data will be of great importance in the ongoing effort 
of understanding and improving these interactions, which are currently 
the dominant source of uncertainty in all \emph{ab initio} many-body 
calculations (see section \ref{sec:groundstate}).

Of course, there is much work to be done. A variety of efforts are 
underway to further extend the capabilities of the 
MR-IMSRG and IMSRG+SM approaches. The MR-IMSRG allows
us to explore both dynamic and static correlations. The
former are due to the dynamics of correlated nucleon pairs, triples,
etc.~in the nucleus, which is captured well by the (generalized)
particle-hole expansion that underlies the MR-IMSRG flow equations.
In contrast, static correlations are collective and would require
us to treat up to $ApAh$ correlations, in exact and numerically
unfeasible IMSRG($A$) or MR-IMSRG($A$) truncation schemes. However, collective
correlations can be treated efficiently by breaking and restoring 
symmetries of the nuclear wave function, and using the Generator
Coordinate Method to mix various projected configurations (see, e.g.,
\cite{Ring:1980bb}). We can calculate the density matrices of such
many-body wave functions, and use them as reference states for the 
MR-IMSRG(2), combining dynamic and static correlation. The use of 
PNP reference states in our applications of the MR-IMSRG(2) 
to open-shell nuclei (see section \ref{sec:groundstate}) is the 
simplest possible example of such a combined approach, and the use
of GCM reference states with richer collective structures is being 
explored now.

Another important new development is the successful use of Magnus
expansion techniques to explicitly construct the unitary transformation 
that is generated by the IMSRG \cite{Morris:2015ve}. This greatly 
simplifies the evaluation of general observables,
which can be obtained with a simple application of $U(s)$ to the operator
of interest rather than
a concurrent evolution alongside the Hamiltonian by means of additional
sets of flow equations. Moreover, the Magnus methods make it possible
to construct systematic approximations to the complete IMSRG(3) flow, 
analogous in many-body content and computational efficiency to 
non-iterative triples methods in CC \cite{Taube:2008kx,Taube:2008vn,Piecuch:2005dp,Binder:2013fk}.
The extension of this approach to the MR-IMSRG is in progress.

The triples corrections to IMSRG Hamiltonians will also be extremely
valuable for the IMSRG+SM approach, allowing us to test the many-body
convergence of the valence-space interaction and operators. The
construction of valence-space transition operators is now in full 
swing, with the prospect of shedding new light on the emergence of
the phenomenological effective charges. A new refinement of the 
targeted normal ordering procedure was presented in Ref.~\cite{Stroberg:2016th}, 
improving once again the agreement between IMSRG+SM and
large-scale MR-IMSRG calculations.

As an alternative to the exact valence-space diagonalization in the
IMSRG+SM approach, we are working on combining the IMSRG with 
Equation-of-Motion (EoM) methods (cf.~\cite{Shavitt:2009}). The basic
framework has been developed and tested for the single-reference
case \cite{Parzuchowski:2016pi}, and we aim to generalize it to
multireference applications as a next step, incorporating triples
corrections to the (MR-)IMSRG evolved Hamiltonian at the same time.

Last but not least, the IMSRG+SM and IMSRG based EoM approaches
are, in essence, established techniques for many-body calculations
whose results are (possibly) enhanced through the use of an IMSRG
improved effective Hamiltonian. We have begun to explore the use of
such IMSRG and MR-IMSRG improved Hamiltonians as input for other
methods. A promising combination of MR-IMSRG and NCSM is discussed
in \cite{Gebrerufael:2016rp}. 

The RG perspective is a key element that is woven into all of 
the applications discussed in this review, and the future directions 
mentioned in our look ahead. In my (admittedly biased) view, this
is a unique feature of the IMSRG framework that sets it apart
from the other many-body methods that we touched upon during this
work. When the comparison with those other methods and experimental
data is our first and foremost concern, we are primarily interested 
in the $s\to\infty$ limit of the IMSRG or MR-IMSRG evolution, but
the flow trajectory offers a wealth of additional insight. By considering
points along the trajectory, we can gain a new understanding of
how many-body correlations are reshuffled between the wave function
and the Hamiltonian, or different pieces of the Hamiltonian, making
transparent what is only implicitly assumed in other methods. Like 
in the free-space SRG (or other RG methods), we have the freedom
to work at intermediate values of $s$ if this is more practical than
working at $s=0$ (in appropriate units) or in the limit $s\to\infty$,
especially if we would incur unacceptable numerical errors at either
of these extremes (see, e.g., \cite{Gebrerufael:2016rp,Li:2015nq,Li:2016rm}). 
This is the inherent power of a framework that 
integrates many-body and renormalization group techniques, and the
reason why the IMSRG is an extremely versatile and valuable
tool for quantum many-body theory.

\section*{Acknowledgments}
Special acknowledgments are due to all my collaborators on the IMSRG 
framework presented in this work: In chronological order, they are 
S.~K.~Bogner, K.~Tsukiyama, A.~Schwenk, T.~D.~Morris, N.~M.~Parzuchowski, 
J.~D.~Holt, and S.~R.~Stroberg. Furthermore, I am grateful to C.~Barbieri, 
S.~Binder, A.~Calci, T.~Duguet, F.~Evangelista, R.~J.~Furnstahl, E.~Gebrerufael, 
G.~Hagen, K.~Hebeler, M.~Hjorth-Jensen, G.~R.~Jansen, R.~Roth, J.~Simonis, 
V.~Som\`{a}, and K.~A.~Wendt for many useful discussions on the subjects 
discussed in this work. I am also grateful to A.~Calci, S.~Binder, and 
R.~Roth for providing matrix elements of the chiral $3N$ interactions, 
and to S.~R.~Stroberg for sharing his scripts for the 
visualization of Shell model results.

I would like to thank the National Superconducting Cyclotron Laboratory 
(NSCL)/Facility for Rare Isotope Beams (FRIB) and Michigan State University for 
startup support during the preparation of this work.
Computing resources were provided by the Ohio Supercomputing Center (OSC), 
the Michigan State University High Performance Computing Center (HPCC)/Institute 
for Cyber-Enabled Research (iCER), and the National Energy Research Scientific Computing 
Center (NERSC), a DOE Office of Science User Facility supported by the Office of 
Science of the U.S.~Department of Energy under Contract No. DE-AC02-05CH11231.

\appendix
\section{\label{app:NO}Products and Commutators of Normal-Ordered Operators}
We introduce the permutation symbol $P_{ij}$ to interchange the attached indices 
in any expression, i.e.,
\begin{equation}\label{eq:def_Pij}
  P_{ij} g(\ldots,i,\ldots,j) \equiv g(\ldots,j,\ldots,i)\,.
\end{equation}

\subsection{Operator Products}
\begin{align}
  :\AO^{a}_{b}::\AO^{k}_{l}:
  &=:\AO^{ak}_{bl}:-\lambda^{a}_{l}:\AO^{k}_{b}:-\xi^{k}_{b}:\AO^{a}_{l}:-\lambda^{a}_{l}\xi^{k}_{b}+\lambda^{ak}_{bl}
\end{align}

\begin{align}
  :\AO^{a}_{b}::\AO^{kl}_{mn}:
  &=:\AO^{akl}_{bmn}: 
    +\left(1-P_{mn}\right)\lambda^{a}_{n}:\AO^{kl}_{bm}:
    +\left(1-P_{kl}\right)\xi^{l}_{b}:\AO^{ak}_{mn}:\notag\\
  &\hphantom{=}  
    +\left(1-P_{kl}\right)\left(1-P_{mn}\right)\lambda^{al}_{bn}:\AO^{k}_{m}:
    \notag\\
  &\hphantom{=}  
    +\left(1-P_{kl}\right)\lambda^{ak}_{mn}:\AO^{l}_{b}:
    +\left(1-P_{mn}\right)\lambda^{kl}_{bm}:\AO^{a}_{n}:
    \notag\\
  &\hphantom{=}  
    +\left(1-P_{kl}\right)\left(1-P_{mn}\right)\lambda^{a}_{m}\xi^{l}_{b}:\AO^{k}_{n}:
    \notag\\
  &\hphantom{=}  
    +(1-P_{mn})\lambda^{a}_{n}\lambda^{kl}_{bm}
    +(1-P_{kl})\lambda^{ak}_{mn}\xi^{l}_{b}
    +\lambda^{akl}_{bmn}
\end{align}

\begin{align}
  &:\AO^{ab}_{cd}::\AO^{kl}_{mn}:
  \notag\\
  &=:\AO^{abkl}_{cdmn}: 
    \notag\\
  &\hphantom{=}
    +(1-P_{ab})(1-P_{mn})\lambda^{a}_{m}:\AO^{bkl}_{cdn}: 
    +(1-P_{cd})(1-P_{kl})\xi^{k}_{c}:\AO^{abl}_{dmn}:
  \notag\\
  &\hphantom{=}
    +\left(\lambda^{ab}_{mn}+(1-P_{mn})\lambda^{a}_{m}\lambda^{b}_{n}\right):\AO^{kl}_{cd}:
    +\left(\lambda^{kl}_{cd}+(1-P_{cd})\xi^{k}_{c}\xi^{l}_{d}\right):\AO^{ab}_{mn}:
  \notag\\
  &\hphantom{=}
    +(1-P_{ab})(1-P_{cd})(1-P_{kl})(1-P_{mn})\left(\lambda^{ak}_{cm}-\lambda^{a}_{m}\xi^{k}_{c}\right):\AO^{bl}_{dn}:
  \notag\\
  &\hphantom{=}
   +(1-P_{ab})(1-P_{kl})\lambda^{al}_{mn}:\AO^{bk}_{cd}:
   +(1-P_{ab})(1-P_{kl})\lambda^{bk}_{cd}:\AO^{al}_{mn}:
  \notag\\
  &\hphantom{=}
   +(1-P_{cd})(1-P_{mn})\lambda^{kl}_{cn}:\AO^{ab}_{dm}:
   +(1-P_{cd})(1-P_{mn})\lambda^{ab}_{dm}:\AO^{kl}_{cn}:
  \notag\\
  &\hphantom{=}
  +(1-P_{ab})(1-P_{mn})
    \left(\lambda^{bkl}_{cdn}
        -(1-P_{cd})(1-P_{kl})\xi^{k}_{c}\lambda^{bl}_{dn}
        +\lambda^{b}_{n}\left(\lambda^{kl}_{cd}+\xi^{k}_{c}\xi^{l}_{d}-\xi^{k}_{d}\xi^{l}_{c}\right) 
    \right):\AO^{a}_{m}:
  \notag\\
  &\hphantom{=}
   +(1-P_{cd})(1-P_{kl})
    \left(\lambda^{abl}_{dmn}
        -(1-P_{ab})(1-P_{mn})\lambda^{a}_{m}\lambda^{bl}_{dn}
        +\xi^{l}_{d}\left(\lambda^{ab}_{mn}+\lambda^{a}_{m}\lambda^{b}_{n}-\lambda^{a}_{n}\lambda^{b}_{m}\right)
  \right):\AO^{k}_{c}:
  \notag\\
  &\hphantom{=}
   +(1-P_{ab})(1-P_{cd})
    \left(
      \lambda^{bkl}_{dmn}
        +(1-P_{mn})\lambda^{b}_{n}\lambda^{kl}_{dm}
        +(1-P_{kl})\xi^{l}_{d}\lambda^{bl}_{mn}
    \right):\AO^{a}_{c}:
  \notag\\
  &\hphantom{=}
   +(1-P_{kl})(1-P_{mn})
    \left(
      \lambda^{abl}_{cdn}
        +(1-P_{ab})\lambda^{a}_{n}\lambda^{bl}_{cd}
        +(1-P_{cd})\xi^{l}_{c}\lambda^{ab}_{dn}
    \right):\AO^{k}_{m}:
  \notag\\
  &\hphantom{=}
  +\lambda^{abkl}_{cdmn}
  +(1-P_{ab})(1-P_{mn})\lambda^{a}_{m}\lambda^{bkl}_{cdn} 
  +(1-P_{cd})(1-P_{kl})\xi^{k}_{c}\lambda^{abl}_{dmn}
  \notag\\[3pt]
  &\hphantom{=}
    -(1-P_{cd})(1-P_{mn})\lambda^{ab}_{cm}\lambda^{kl}_{dn}
    +(1-P_{cd})(1-P_{kl})(1-P_{mn})\lambda^{ak}_{cm}\lambda^{bl}_{dn}
  \notag\\[3pt]
  &\hphantom{=}
    +(1-P_{kl})\left(\lambda^{al}_{mn}\lambda^{bk}_{cd}-\lambda^{ak}_{cd}\lambda^{bl}_{mn}\right)
  \notag\\[3pt]
  &\hphantom{=}
  +(1-P_{ab})(1-P_{cd})(1-P_{kl})(1-P_{mn})\lambda^{b}_{m}\xi^{k}_{c}\lambda^{al}_{dn}
  \notag\\[3pt]
  &\hphantom{=}
    +\left(\lambda^{ab}_{mn}+\lambda^{a}_{m}\lambda^{b}_{n}-\lambda^{a}_{n}\lambda^{b}_{m}\right)
     \left(\lambda^{kl}_{cd}+\xi^{k}_{c}\xi^{l}_{d}-\xi^{k}_{d}\xi^{l}_{c}\right)
\end{align}

\subsection{Commutators}

\begin{align}
  \comm{:\AO^{a}_{b}:}{:\AO^{k}_{l}:}
  &=\delta^{k}_{b}:\AO^{a}_{l}:-\delta^{a}_{l}:\AO^{k}_{b}:+\lambda^{a}_{l}\delta^{k}_{b}-\lambda^{k}_{b}\delta^{a}_{l}
\end{align}

\begin{align}
  \comm{:\AO^{a}_{b}:}{:\AO^{kl}_{mn}:}
  &
    =\left(1-P_{kl}\right)\delta^{k}_{b}:\AO^{al}_{mn}:
    -\left(1-P_{mn}\right)\delta^{a}_{m}:\AO^{kl}_{bn}:
    \notag\\
  &\hphantom{=}  
    +\left(1-P_{kl}\right)\left(1-P_{mn}\right)
    \left(\delta^{l}_{b}\lambda^{a}_{n}-\delta^{a}_{n}\lambda^{l}_{b}\right):\AO^{k}_{m}:
    \notag\\
  &\hphantom{=}  
    +\left(1-P_{kl}\right)\delta^{k}_{b}\lambda^{al}_{mn}
    -\left(1-P_{mn}\right)\delta^{a}_{m}\lambda^{kl}_{bn}
\end{align}

\begin{align}
  &\comm{:\AO^{ab}_{cd}:}{:\AO^{kl}_{mn}:}
  \notag\\
  &=
    \left(1-P_{ab}\right)\left(1-P_{mn}\right)\delta^{a}_{m}:\AO^{bkl}_{cdn}: 
    -\left(1-P_{cd}\right)\left(1-P_{kl}\right)\delta^{k}_{c}:\AO^{abl}_{dmn}: 
  \notag\\
  &\hphantom{=}
    +(1-P_{cd})\left(\xi^{k}_{c}\xi^{l}_{d}-\lambda^{k}_{c}\lambda^{l}_{d}\right):\AO^{ab}_{mn}:
    +(1-P_{ab})\left(\lambda^{a}_{m}\lambda^{b}_{n}-\xi^{a}_{m}\xi^{b}_{n}\right):\AO^{kl}_{cd}:
  \notag\\
  &\hphantom{=}
    +(1-P_{ab})(1-P_{cd})(1-P_{kl})(1-P_{mn})
      \left(\delta^{l}_{d}\lambda^{b}_{n}-\delta^{b}_{n}\lambda^{l}_{d}\right):\AO^{ak}_{cm}:
  \notag\\
  &\hphantom{=}
   +(1-P_{ab})(1-P_{mn})\left(
      \delta^{b}_{n}\lambda^{kl}_{cd} 
      +(1-P_{cd})\left((1-P_{kl})\delta^{k}_{c}\lambda^{bl}_{dn}
        +\lambda^{b}_{n}\xi^{k}_{c}\xi^{l}_{d}-\xi^{b}_{n}\lambda^{k}_{c}\lambda^{l}_{d}\right)
  \right):\AO^{a}_{m}:
  \notag\\
  &\hphantom{=}
   -(1-P_{cd})(1-P_{kl})\left(
      \delta^{l}_{d}\lambda^{ab}_{mn} 
      +(1-P_{ab})\left((1-P_{mn})\delta^{a}_{m}\lambda^{bl}_{dn}
      +\lambda^{l}_{d}\xi^{a}_{m}\xi^{b}_{n}-\xi^{l}_{d}\lambda^{a}_{m}\lambda^{b}_{n}\right)
  \right):\AO^{k}_{c}:
  \notag\\
  &\hphantom{=}
   -(1-P_{ab})(1-P_{cd})\left(
        (1-P_{mn})\delta^{b}_{m}\lambda^{kl}_{dn}
       -(1-P_{kl})\delta^{k}_{d}\lambda^{bl}_{mn}
    \right):\AO^{a}_{c}:
  \notag\\
  &\hphantom{=}
   +(1-P_{kl})(1-P_{mn})\left(
        (1-P_{ab})\delta^{a}_{n}\lambda^{bl}_{cd}
       -(1-P_{cd})\delta^{l}_{c}\lambda^{ab}_{dn}
    \right):\AO^{k}_{m}:
  \notag\\
  &\hphantom{=}
  +(1-P_{ab})(1-P_{mn})\delta^{a}_{m}\lambda^{bkl}_{cdn} 
  -(1-P_{cd})(1-P_{kl})\delta^{k}_{c}\lambda^{abl}_{dmn} 
  \notag\\
  &\hphantom{=}
  +(1-P_{cd})\lambda^{ab}_{mn}\left(\xi^{k}_{c}\xi^{l}_{d}-\lambda^{k}_{c}\lambda^{l}_{d}\right)
  +(1-P_{ab})\lambda^{kl}_{cd}\left(\lambda^{a}_{m}\lambda^{b}_{n}-\xi^{a}_{m}\xi^{b}_{n}\right)
  \notag\\
  &\hphantom{=}
  +(1-P_{ab})(1-P_{cd})(1-P_{kl})(1-P_{mn})
       \left(\delta^{l}_{d}\lambda^{b}_{n}-\delta^{b}_{n}\lambda^{l}_{d}\right)\lambda^{ak}_{cm}
  \notag\\
  &\hphantom{=}
  +(1-P_{ab})(1-P_{cd})
    \left(
       \lambda^{a}_{m}\lambda^{b}_{n}\xi^{k}_{c}\xi^{l}_{d}
      -\lambda^{k}_{c}\lambda^{l}_{d}\xi^{a}_{m}\xi^{b}_{n}
    \right)
\end{align}

\section{\label{app:PNP}Particle-Number Projected HFB Reference States}
In this appendix, we summarize the essential properties of particle-number
projected Hartree-Fock-Bogoliubov (HFB) states. More details can be found,
e.g., in \cite{Ring:1980bb,Sheikh:2000xx}. We introduce fermionic 
quasiparticle operators $\alphaO_{\tau k}, \aalphaO_{\tau k}$ that are 
superpositions of creation and annihilation operators by means of a 
Bogoliubov-Valatin transformation:
\begin{align}\label{eq:bogoliubov}
    \aalphaO_{ k}        &=u_{ k}\aaO_{ k} - v_{ k}\aO_{\bar{k}}\,,\\
    \aalphaO_{\bar{k}}   &=u_{ k}\aaO_{\bar{k}} + v_{ k}\aO_{ k}\,,\\
    \alphaO_{ k}         &=u_{ k}\aO_{ k}  - v_{ k}\aaO_{\bar{k}}\,,\\
    \alphaO_{\bar{k}}    &=u_{ k}\aO_{\bar{k}} + v_{ k}\aaO_{ k}\,.
\end{align}
Here $k$ is a collective index for the 
single-particle states in the so-called \emph{canonical basis}, i.e., the eigenbasis of 
the HFB density matrix. The bars indicate time-reversed states. The occupation coefficients 
can be chosen to be real if only like-particle (i.e., proton-proton and 
neutron-neutron) pairing is considered, and they satisfy
\begin{equation}
  u_{k}^2 + v_{k}^2 = 1\,.
\end{equation}
In terms of these coefficients, a solution of the HFB equations can be written as
\begin{equation}\label{eq:hfb_state}
  \ket{\Phi}=\prod_{k}\alphaO_{k}\ket{\text{vac}}
    =\prod_{k>0}\left(u_{ k}+v_{ k}
      \aaO_{ k}\aaO_{\bar{k}}\right)\ket{\text{vac}}\,,
\end{equation}
where $\ket{\text{vac}}$ refers to the particle vacuum state. It is clear
from equation \eqref{eq:hfb_state} that $\ket{\Phi}$ is a superposition of states 
with even proton and neutron number, and therefore not an eigenstate of the 
corresponding proton, neutron, or nucleon number operators. The HFB equations 
are solved under the constraint that the \emph{expectation values} of these
number operators match a given nucleus.

The broken particle-number symmetry can be restored by projecting $\ket{\Phi}$
on good $Z$ and $N$ with the operator
\begin{equation}\label{eq:projector}
  \PO_{ZN}
    =\PO_{Z}\PO_{N}
    =\frac{1}{(2\pi)^2}\int_0^{2\pi} d\phi_p\int_0^{2\pi} d\phi_n\, e^{i\phi_p(\hat\ZO-Z)}e^{i\phi_n(\hat\NO-N)}\,.
\end{equation}
Expectation values in the projected HFB states are formally given by
\begin{equation}\label{eq:pnp_expect_abstract}
  \matrixe{\Phi_{ZN}}{\OO}{\Phi_{ZN}}
    =\frac{\dmatrixe{\Phi}{\OO\PO_{ZN}}}{\dmatrixe{\Phi}{\PO_{ZN}}}\,.
\end{equation}
Under the unitary transformations generated by the number operators, the 
particle creation and annihilation operators transform as
\begin{align}
  e^{i\phi_k\hat\AO_k}\aaO_{k}e^{-i\phi_k\hat\AO_k}&=e^{i\phi_k}\aaO_{k}\,,\\
  e^{i\phi_k\hat\AO_k}\aO_{k}e^{-i\phi_k\hat\AO_k}&=e^{-i\phi_k}\aO_{k}\,,
\end{align}
where $\phi_k\in\{\phi_p,\phi_n\}$ and $\hat{A}_k\in\{\hat{Z},\hat{N}\}$ are 
the appropriate gauge angle and number operator for the single-particle state
$k$. Using these relations, we can write the gauge-rotated quasi-particle state 
as
\begin{align}\label{eq:rotated_state}
  \ket{\Phi(\phi_p,\phi_n)}\equiv e^{i\left(\phi_p\hat\ZO+\phi_n\hat\NO\right)}\ket{\Phi}&=
      \prod_{k>0}\left(u_{k}+v_{k}e^{2i\phi_k}\aaO_{k}\aaO_{\bar k}\right)\ket{\text{vac}}
\end{align}
and introduce the proton and neutron norm kernels ($\tau=p,n$)
\begin{equation}
  x_\tau(\phi_\tau)\equiv\dmatrixe{\Phi}{e^{i\phi_\tau\left(\hat\AO_\tau-A_\tau\right)}}
  =e^{-i\phi_\tau A_\tau}\prod_{k>0}\left(u^2_k + v^2_k e^{2i\phi_k}\right)\,.
\end{equation}
The overlap between the particle-number projected state and the initial
HFB state is now given by
\begin{align}
  \dmatrixe{\Phi}{\PO_{ZN}}&=
    \frac{1}{(2\pi)^2}\int_0^{2\pi} d\phi_p\int_0^{2\pi} d\phi_n\,
      \dmatrixe{\Phi}{e^{i\phi_p(\hat\ZO-Z)}e^{i\phi_n(\hat\NO-N)}}\notag\\
  &=\frac{1}{(2\pi)^2}\left(\int_0^{2\pi} d\phi_p\,x_p(\phi_p)\right)
                     \left(\int_0^{2\pi} d\phi_n\,x_n(\phi_n)\right)\,.\label{eq:pnp_expect_denominator}
\end{align}
The numerator of equation \eqref{eq:pnp_expect_abstract} can be expressed in
terms of the norm kernel and the gauge-rotated state as
\begin{align}
  \dmatrixe{\Phi}{\OO\PO_{ZN}}
  &=\frac{1}{(2\pi)^2}\int^{2\pi}_{0}d\phi_p\int^{2\pi}_{0}d\phi_n\,
    e^{-i(\phi_pZ+\phi_nN)} \dmatrixe{\Phi}{\OO e^{i(\phi_p\hat\ZO+i\phi_n\hat\NO)}}\notag\\
  &=\frac{1}{(2\pi)^2}\int^{2\pi}_{0}d\phi_p\int^{2\pi}_{0}d\phi_n\,
    \dmatrixe{\Phi}{e^{i(\phi_p(\hat\ZO-Z)+i\phi_n(\hat\NO-N))}}
    \frac{\dmatrixe{\Phi}{\OO e^{i(\phi_p\hat\ZO+i\phi_n\NO)}}}{\dmatrixe{\Phi}{e^{i(\phi_p\hat\ZO+i\phi_n\NO)}}}\notag\\
  &=\frac{1}{(2\pi)^2}\int^{2\pi}_{0}d\phi_p\int^{2\pi}_{0}d\phi_n\,
    x_p(\phi_p)x_n(\phi_n)
    \frac{\matrixe{\Phi}{\OO}{\Phi(\phi_p,\phi_n)}}{\braket{\Phi}{\Phi(\phi_p,\phi_n)}}\,.\label{eq:pnp_expect_numerator}
\end{align}
Defining the operator kernel 
\begin{equation}\label{eq:pnp_operator_kernel}
  O(\phi_p,\phi_n)\equiv \frac{\matrixe{\Phi}{\OO}{\Phi(\phi_p,\phi_n)}}{\braket{\Phi}{\Phi(\phi_p,\phi_n)}}
\end{equation}
and the auxiliaries
\begin{equation}
  y_\tau(\phi_\tau) = \frac{x_\tau(\phi_\tau)}{\int_0^{2\pi}d\phi_\tau\,x_\tau(\phi_\tau)}\,,
  \quad 
  \int_0^{2\pi} d\phi_\tau y_\tau(\phi_\tau) = 1\,,
\end{equation}
we can combine equations \eqref{eq:pnp_expect_denominator} and \eqref{eq:pnp_expect_numerator} 
into the following compact expression for the expectation value of $\OO$
in the particle-number projected state:
\begin{align}
  \dmatrixe{\Phi_{ZN}}{\OO}
  &=\int^{2\pi}_{0}d\phi_p\int^{2\pi}_{0}d\phi_n\,
    y_p(\phi_p)y_n(\phi_n)O(\phi_p,\phi_n)\,.
\end{align}
As we have seen above, the inital and gauge-rotated HFB states have essentially the same
structure, because the transformation of the basis operators only introduces simple phase
factors. This implies that a simple extension of Wick's theorem for non-orthogonal product
states can be applied to express the operator kernel \eqref{eq:pnp_operator_kernel}
in terms of the basic contractions \cite{Ring:1980bb}, the so-called transition density 
matrix and transition pairing tensors:
\begin{align}
  \rho_{kl}(\phi_k)&=\frac{\matrixe{\Phi}{\aaO_l\aO_k}{\Phi(\phi_p,\phi_n)}}{\braket{\Phi}{\Phi(\phi_p,\phi_n)}}\,,\\
  \overline{\kappa}_{kl}(\phi_k)&=\frac{\matrixe{\Phi}{\aaO_k\aaO_l}{\Phi(\phi_p,\phi_n)}}{\braket{\Phi}{\Phi(\phi_p,\phi_n)}}\,,\\
  \kappa_{kl}(\phi_k)&=\frac{\matrixe{\Phi}{\aO_l\aO_k}{\Phi(\phi_p,\phi_n)}}{\braket{\Phi}{\Phi(\phi_p,\phi_n)}}\,.
\end{align}
For $\phi_p=\phi_n=0$, the standard density matrix and pairing tensor of non-projected HFB 
theory are obtained. Since we do not allow proton-neutron pairing here, the contractions only 
depend on the gauge angle matching the isospin projection contained in the collective indices, 
because the HFB product state factorizes into proton and neutron parts 
(cf.~equations \eqref{eq:hfb_state},
\eqref{eq:rotated_state}):
\begin{equation}
  \ket{\Phi(\phi_p,\phi_n)}=\ket{\Phi_p(\phi_p)}\otimes\ket{\Phi_n(\phi_n)}\,.
\end{equation}
For instance, if $k$ and $l$ are proton indices, we have
\begin{equation}
  \rho_{kl}(\phi_p)=
  \frac{\matrixe{\Phi}{\aaO_{l}\aO_k}{\Phi(\phi_p,\phi_n)}}{\braket{\Phi}{\Phi(\phi_p,\phi_n)}}
  =\frac{\matrixe{\Phi_p}{\aaO_{l}\aO_k}{\Phi_p(\phi_p)}\braket{\Phi_n}{\Phi_n(\phi_n)}}{\braket{\Phi_p}{\Phi_p(\phi_p)}\braket{\Phi_n}{\Phi_n(\phi_n)}}
  =\frac{\matrixe{\Phi_p}{\aaO_{l}\aO_k}{\Phi_p(\phi_p)}}{\braket{\Phi_p}{\Phi_p(\phi_p)}}\,.
\end{equation}

Switching to the tensorial notation we use for the MR-IMSRG, the transition density matrices 
and pairing tensors in the canonical basis representation are given by
\begin{align}
  \rho^{k}_{l}(\phi_\tau) & = \frac{v_k^2 e^{2i\phi_{\tau}}}{u_k^2 + v_k^2 e^{2i\phi_\tau}} \delta^{k}_{l}\,,\label{eq:pnp_rtrans_tensor}\\
  \bar{\kappa}^{kl}(\phi_\tau) & = \frac{u_k v_k}{u_k^2 + v_k^2 e^{2i\phi_\tau}} \delta^{k\bar{l}}\,,\\
  \kappa_{kl}(\phi_\tau) & = \frac{u_k v_k e^{2i\phi_{\tau}}}{u_k^2 + v_k^2 e^{2i\phi_\tau}} \delta_{k\bar{l}}\,.
\end{align}
The one-body density matrix of the projected state is obtained by integration:
\begin{equation}\label{eq:pnp_occupations}
  \rho^{k}_{l} = \int^{2\pi}_{0}d\phi_k\,y_k(\phi_k)\rho^{k}_{l}(\phi_k)
               = \int^{2\pi}_{0}d\phi_k\,y_k(\phi_k)\frac{v_k^2 e^{2i\phi_{\tau}}}{u_k^2 + v_k^2 e^{2i\phi_\tau}} \delta^{k}_{l}\,.
\end{equation}
We see that all of the density matrices are diagonal in the canonical basis,
which is identical to the natural orbital basis that is most convenient for the
formulation of the MR-IMSRG flow. We can also directly read off the projected
occupation numbers $n_k$ from equation \eqref{eq:pnp_occupations}.

The full two-body and three-body density matrices are given by
\begin{align}
  \rho^{kl}_{mn}
  &= \int^{2\pi}_{0}\!\!\!d\phi_p\int^{2\pi}_{0}\!\!\!d\phi_n\, y_p(\phi_p)y_n(\phi_n) 
    \left(
      \frac{v_k^2 e^{2i\phi_{k}}}{u_k^2 + v_k^2 e^{2i\phi_k}} \frac{v_l^2 e^{2i\phi_l}}{u_l^2 + v_l^2 e^{2i\phi_l}}
      \left(\delta^{k}_{m}\delta^{l}_{n} - \delta^{k}_{n}\delta^{l}_{m}\right)
    \right.\notag\\
  &\hphantom{=}\hspace{13em}\left.  
       + \frac{u_kv_k}{u_k^2 + v_k^2 e^{2i\phi_{k}}} \frac{u_m v_m e^{2i\phi_{m}}}{u_m^2 + v_m^2 e^{2i\phi_{m}}}
        \delta^{k\bar{l}}\delta_{m\bar{n}}
    \right)
\end{align}
and
\begin{align}
  \rho^{pqr}_{stu}
  &=\int^{2\pi}_{0}\!\!\!d\phi_p\int^{2\pi}_{0}\!\!\!d\phi_n\, y_p(\phi_p)y_n(\phi_n)
    \notag\\
  &\hphantom{=}\qquad\times  
    \left(
      \frac{v_p^2e^{2i\phi_p}}{u_p^2+v_p^2e^{2i\phi_p}}
      \frac{v_q^2e^{2i\phi_q}}{u_q^2+v_q^2e^{2i\phi_q}}
      \frac{v_r^2e^{2i\phi_r}}{u_r^2+v_r^2e^{2i\phi_r}}
    \right.
    \notag\\
  &\hphantom{=}\qquad\qquad\times  
      \left(
        \delta^{p}_{s}\delta^{q}_{t}\delta^{r}_{u} 
       +\delta^{p}_{t}\delta^{q}_{u}\delta^{r}_{s} 
       +\delta^{p}_{u}\delta^{q}_{s}\delta^{r}_{t} 
       -\delta^{p}_{t}\delta^{q}_{s}\delta^{r}_{u} 
       -\delta^{p}_{s}\delta^{q}_{u}\delta^{r}_{t} 
       -\delta^{p}_{u}\delta^{q}_{t}\delta^{r}_{s} 
      \right)
    \notag\\
  &\hphantom{=}\qquad\qquad
      +\frac{v_p^2e^{2i\phi_p}}{u_p^2+v_p^2e^{2i\phi_p}}
       \frac{u_qv_q}{u_q^2+v_q^2e^{2i\phi_q}}
       \frac{u_tv_te^{2i\phi_t}}{u_t^2+v_t^2e^{2i\phi_t}}
       \delta^{p}_{s}\delta^{q\bar{r}}\delta_{t\bar{u}}
    \notag\\
  &\hphantom{=}\qquad\qquad
      -\frac{v_p^2e^{2i\phi_p}}{u_p^2+v_p^2e^{2i\phi_p}}
       \frac{u_qv_q}{u_q^2+v_q^2e^{2i\phi_q}}
       \frac{u_sv_se^{2i\phi_s}}{u_s^2+v_s^2e^{2i\phi_s}}
       \delta^{p}_{t}\delta^{q\bar{r}}\delta_{s\bar{u}}
    \notag\\
  &\hphantom{=}\qquad\qquad
      +\frac{v_p^2e^{2i\phi_p}}{u_p^2+v_p^2e^{2i\phi_p}}
       \frac{u_qv_q}{u_q^2+v_q^2e^{2i\phi_q}}
       \frac{u_sv_se^{2i\phi_s}}{u_s^2+v_s^2e^{2i\phi_s}}
       \delta^{p}_{u}\delta^{q\bar{r}}\delta_{s\bar{t}}
    \notag\\
  &\hphantom{=}\qquad\qquad
      -\frac{v_q^2e^{2i\phi_q}}{u_q^2+v_q^2e^{2i\phi_q}}
       \frac{u_pv_p}{u_p^2+v_p^2e^{2i\phi_p}}
       \frac{u_tv_te^{2i\phi_t}}{u_t^2+v_t^2e^{2i\phi_t}}
       \delta^{q}_{s}\delta^{p\bar{r}}\delta_{t\bar{u}}
    \notag\\
  &\hphantom{=}\qquad\qquad
      +\frac{v_q^2e^{2i\phi_q}}{u_q^2+v_q^2e^{2i\phi _q}}
       \frac{u_pv_p}{u_p^2+v_p^2e^{2i\phi_p}}
       \frac{u_sv_se^{2i\phi_s}}{u_s^2+v_s^2e^{2i\phi_s}}
       \delta^{q}_{t}\delta^{p\bar{r}}\delta_{s\bar{u}}
    \notag\\
  &\hphantom{=}\qquad\qquad
      -\frac{v_q^2e^{2i\phi_q}}{u_q^2+v_q^2e^{2i\phi_q}}
       \frac{u_pv_p}{u_p^2+v_p^2e^{2i\phi_p}}
       \frac{u_sv_se^{2i\phi_s}}{u_s^2+v_s^2e^{2i\phi_s}}
       \delta^{q}_{u}\delta^{p\bar{r}}\delta_{s\bar{t}}
    \notag\\
  &\hphantom{=}\qquad\qquad
      +\frac{v_r^2e^{2i\phi_r}}{u_r^2+v_r^2e^{2i\phi_r}}
       \frac{u_pv_p}{u_p^2+v_p^2e^{2i\phi_p}}
       \frac{u_tv_te^{2i\phi_t}}{u_t^2+v_t^2e^{2i\phi_t}}
       \delta^{r}_{s}\delta^{p\bar{q}}\delta_{t\bar{u}}
    \notag\\
  &\hphantom{=}\qquad\qquad
      -\frac{v_r^2e^{2i\phi_r}}{u_r^2+v_r^2e^{2i\phi_r}}
       \frac{u_pv_p}{u_p^2+v_p^2e^{2i\phi_p}}
       \frac{u_sv_se^{2i\phi_s}}{u_s^2+v_s^2e^{2i\phi_s}}
       \delta^{r}_{t}\delta^{p\bar{q}}\delta_{s\bar{u}}
    \notag\\
  &\hphantom{=}\qquad\qquad
    \left.
      +\frac{v_r^2e^{2i\phi_r}}{u_r^2+v_r^2e^{2i\phi_r}}
       \frac{u_pv_p}{u_p^2+v_p^2e^{2i\phi_p}}
       \frac{u_sv_se^{2i\phi_s}}{u_s^2+v_s^2e^{2i\phi_s}}
       \delta^{r}_{u}\delta^{p\bar{q}}\delta_{s\bar{t}}
    \right)\,.
\end{align}
From these, the irreducible two-body and three-body density matrices are 
obtained by subtacting all antisymmetrized products of lower-rank density 
matrices:
\begin{gather}
   \lambda^{kl}_{mn} = \rho^{kl}_{mn} - \lambda^{k}_{m}\lambda^{l}_{n} + \lambda^{k}_{n}\lambda^{l}_{m}\,,\\
    \lambda^{pqr}_{stu}=\rho^{pqr}_{stu}
        -\AC\left(\lambda^{pq}_{st}\lambda^{r}_{u}\right)
        -\AC\left(\lambda^{p}_{s}\lambda^{q}_{t}\lambda^{r}_{u}\right)\,.
\end{gather}

\bibliography{2016_physica_scripta}

\providecommand{\newblock}{}
\begin{thebibliography}{100}
\expandafter\ifx\csname url\endcsname\relax
  \def\url#1{{\tt #1}}\fi
\expandafter\ifx\csname urlprefix\endcsname\relax\def\urlprefix{URL }\fi
\providecommand{\eprint}[2][]{\url{#2}}

\bibitem{Rainwater:1950ly}
Rainwater J 1950 {\em Phys. Rev.\/} {\bf 79}(3) 432--434
  \urlprefix\url{http://link.aps.org/doi/10.1103/PhysRev.79.432}

\bibitem{Bohr:1951eu}
Bohr A 1951 {\em Phys. Rev.\/} {\bf 81}(1) 134--138
  \urlprefix\url{http://link.aps.org/doi/10.1103/PhysRev.81.134}

\bibitem{Bohr:1953fy}
Bohr A and Mottelson B~R 1953 {\em Dan. Mat. Fys. Medd.\/} {\bf 27} 1

\bibitem{Bohr:1953pd}
Bohr A and Mottelson B~R 1953 {\em Phys. Rev.\/} {\bf 89}(1) 316--317
  \urlprefix\url{http://link.aps.org/doi/10.1103/PhysRev.89.316}

\bibitem{Bohr:1953qd}
Bohr A and Mottelson B~R 1953 {\em Phys. Rev.\/} {\bf 90}(4) 717--719
  \urlprefix\url{http://link.aps.org/doi/10.1103/PhysRev.90.717.2}

\bibitem{Brown:2001rg}
Brown B~A 2001 {\em Prog. Part. Nucl. Phys.\/} {\bf 47} 517--599
  \urlprefix\url{http://www.sciencedirect.com/science/article/pii/S0146641001001594}

\bibitem{Caurier:2005qf}
Caurier E, Mart\'{\i}nez-Pinedo G, Nowacki F, Poves A and Zuker A~P 2005 {\em
  Rev. Mod. Phys.\/} {\bf 77}(2) 427--488
  \urlprefix\url{http://link.aps.org/doi/10.1103/RevModPhys.77.427}

\bibitem{Brown:2006fk}
Brown B~A and Richter W~A 2006 {\em Phys. Rev. C\/} {\bf 74}(3) 034315
  \urlprefix\url{http://link.aps.org/doi/10.1103/PhysRevC.74.034315}

\bibitem{Erler:2012dz}
Erler J, Birge N, Kortelainen M, Nazarewicz W, Olsen E, Perhac A~M and Stoitsov
  M 2012 {\em Nature\/} {\bf 486} 509--512
  \urlprefix\url{http://dx.doi.org/10.1038/nature11188}

\bibitem{Kortelainen:2010ad}
Kortelainen M, Lesinski T, Mor\'e J, Nazarewicz W, Sarich J, Schunck N,
  Stoitsov M~V and Wild S 2010 {\em Phys. Rev. C\/} {\bf 82}(2) 024313
  \urlprefix\url{http://link.aps.org/doi/10.1103/PhysRevC.82.024313}

\bibitem{Kortelainen:2012fu}
Kortelainen M, McDonnell J, Nazarewicz W, Reinhard P~G, Sarich J, Schunck N,
  Stoitsov M~V and Wild S~M 2012 {\em Phys. Rev. C\/} {\bf 85}(2) 024304
  \urlprefix\url{http://link.aps.org/doi/10.1103/PhysRevC.85.024304}

\bibitem{McDonnell:2015di}
McDonnell J~D, Schunck N, Higdon D, Sarich J, Wild S~M and Nazarewicz W 2015
  {\em Phys. Rev. Lett.\/} {\bf 114}(12) 122501
  \urlprefix\url{http://link.aps.org/doi/10.1103/PhysRevLett.114.122501}

\bibitem{Detmold:2015xw}
Detmold W 2015 {\em Nuclear Physics from Lattice QCD\/} (Cham: Springer
  International Publishing) pp 153--194 ISBN 978-3-319-08022-2
  \urlprefix\url{http://dx.doi.org/10.1007/978-3-319-08022-2_5}

\bibitem{Epelbaum:2009ve}
Epelbaum E, Hammer H~W and Mei\ss{}ner U~G 2009 {\em Rev. Mod. Phys.\/} {\bf
  81}(4) 1773--1825
  \urlprefix\url{http://link.aps.org/doi/10.1103/RevModPhys.81.1773}

\bibitem{Machleidt:2011bh}
Machleidt R and Entem D 2011 {\em Phys. Rept.\/} {\bf 503} 1 -- 75 ISSN
  0370-1573
  \urlprefix\url{http://www.sciencedirect.com/science/article/pii/S0370157311000457}

\bibitem{Epelbaum:2015gf}
Epelbaum E, Krebs H and Mei\ss{}ner U~G 2015 {\em Phys. Rev. Lett.\/} {\bf
  115}(12) 122301
  \urlprefix\url{http://link.aps.org/doi/10.1103/PhysRevLett.115.122301}

\bibitem{Entem:2015qf}
Entem D~R, Kaiser N, Machleidt R and Nosyk Y 2015 {\em Phys. Rev. C\/} {\bf
  91}(1) 014002
  \urlprefix\url{http://link.aps.org/doi/10.1103/PhysRevC.91.014002}

\bibitem{Entem:2015hl}
Entem D~R, Kaiser N, Machleidt R and Nosyk Y 2015 {\em Phys. Rev. C\/} {\bf
  92}(6) 064001
  \urlprefix\url{http://link.aps.org/doi/10.1103/PhysRevC.92.064001}

\bibitem{Gezerlis:2014zr}
Gezerlis A, Tews I, Epelbaum E, Freunek M, Gandolfi S, Hebeler K, Nogga A and
  Schwenk A 2014 {\em Phys. Rev. C\/} {\bf 90}(5) 054323
  \urlprefix\url{http://link.aps.org/doi/10.1103/PhysRevC.90.054323}

\bibitem{Lynn:2016ec}
Lynn J~E, Tews I, Carlson J, Gandolfi S, Gezerlis A, Schmidt K~E and Schwenk A
  2016 {\em Phys. Rev. Lett.\/} {\bf 116}(6) 062501
  \urlprefix\url{http://link.aps.org/doi/10.1103/PhysRevLett.116.062501}

\bibitem{Pastore:2009zr}
Pastore S, Girlanda L, Schiavilla R, Viviani M and Wiringa R~B 2009 {\em Phys.
  Rev. C\/} {\bf 80}(3) 034004
  \urlprefix\url{http://link.aps.org/doi/10.1103/PhysRevC.80.034004}

\bibitem{Pastore:2011dq}
Pastore S, Girlanda L, Schiavilla R and Viviani M 2011 {\em Phys. Rev. C\/}
  {\bf 84}(2) 024001
  \urlprefix\url{http://link.aps.org/doi/10.1103/PhysRevC.84.024001}

\bibitem{Piarulli:2013vn}
Piarulli M, Girlanda L, Marcucci L~E, Pastore S, Schiavilla R and Viviani M
  2013 {\em Phys. Rev. C\/} {\bf 87}(1) 014006
  \urlprefix\url{http://link.aps.org/doi/10.1103/PhysRevC.87.014006}

\bibitem{Kolling:2009yq}
K\"olling S, Epelbaum E, Krebs H and Mei\ss{}ner U~G 2009 {\em Phys. Rev. C\/}
  {\bf 80}(4) 045502
  \urlprefix\url{http://link.aps.org/doi/10.1103/PhysRevC.80.045502}

\bibitem{Kolling:2011bh}
K\"olling S, Epelbaum E, Krebs H and Mei\ss{}ner U~G 2011 {\em Phys. Rev. C\/}
  {\bf 84}(5) 054008
  \urlprefix\url{http://link.aps.org/doi/10.1103/PhysRevC.84.054008}

\bibitem{Ekstrom:2015fk}
Ekstr\"om A, Jansen G~R, Wendt K~A, Hagen G, Papenbrock T, Carlsson B~D,
  Forss\'en C, Hjorth-Jensen M, Navr\'atil P and Nazarewicz W 2015 {\em Phys.
  Rev. C\/} {\bf 91}(5) 051301
  \urlprefix\url{http://link.aps.org/doi/10.1103/PhysRevC.91.051301}

\bibitem{Shirokov:2004fe}
Shirokov A~M, Mazur A~I, Zaytsev S~A, Vary J~P and Weber T~A 2004 {\em Phys.
  Rev. C\/} {\bf 70}(4) 044005
  \urlprefix\url{http://link.aps.org/doi/10.1103/PhysRevC.70.044005}

\bibitem{Shirokov:2005bv}
Shirokov A~M, Vary J~P, Mazur A~I, Zaytsev S~A and Weber T~A 2005 {\em Physics
  Letters B\/} {\bf 621} 96--101
  \urlprefix\url{http://www.sciencedirect.com/science/article/pii/S0370269305008518}

\bibitem{Shirokov:2007by}
Shirokov A~M, Vary J~P, Mazur A~I and Weber T~A 2007 {\em Physics Letters B\/}
  {\bf 644} 33--37
  \urlprefix\url{http://www.sciencedirect.com/science/article/pii/S0370269306014158}

\bibitem{Shirokov:2016wo}
Shirokov A~M, Shin I~J, Kim Y, Sosonkina M, Maris P and Vary J~P 2016 {\em
  Physics Letters B\/} {\bf 761} 87--91
  \urlprefix\url{http://www.sciencedirect.com/science/article/pii/S0370269316304269}

\bibitem{Bogner:2003os}
Bogner S~K, Kuo T~T~S and Schwenk A 2003 {\em Phys. Rept.\/} {\bf 386} 1--27
  (\textit{Preprint} \eprint{nucl-th/0305035})

\bibitem{Bogner:2007od}
Bogner S~K, Furnstahl R~J and Perry R~J 2007 {\em Phys. Rev. C\/} {\bf 75}
  061001(R) (\textit{Preprint} \eprint{nucl-th/0611045})

\bibitem{Bogner:2010pq}
Bogner S~K, Furnstahl R~J and Schwenk A 2010 {\em Prog. Part. Nucl. Phys.\/}
  {\bf 65} 94--147 (\textit{Preprint} \eprint{0912.3688})

\bibitem{Furnstahl:2013zt}
Furnstahl R~J and Hebeler K 2013 {\em Rept. Prog. Phys.\/} {\bf 76} 126301
  \urlprefix\url{http://stacks.iop.org/0034-4885/76/i=12/a=126301}

\bibitem{Bethe:1971qf}
Bethe H~A 1971 {\em Ann. Rev. Nucl. Sci.\/} {\bf 21} 93--244
  \urlprefix\url{http://dx.doi.org/10.1146/annurev.ns.21.120171.000521}

\bibitem{Polyzou:1990fk}
Polyzou W and Gl{\"o}ckle W 1990 {\em Few-Body Systems\/} {\bf 9} 97--121 ISSN
  0177-7963 \urlprefix\url{http://dx.doi.org/10.1007/BF01091701}

\bibitem{Kuo:1967qf}
Kuo T 1967 {\em Nucl. Phys. A\/} {\bf 90} 199 -- 208 ISSN 0375-9474
  \urlprefix\url{http://www.sciencedirect.com/science/article/pii/037594746790749X}

\bibitem{Kuo:1968ty}
Kuo T~T~S and Brown G~E 1968 {\em Nuclear Physics A\/} {\bf 114} 241--279
  \urlprefix\url{http://www.sciencedirect.com/science/article/pii/0375947468903539}

\bibitem{Brueckner:1954qf}
Brueckner K~A, Levinson C~A and Mahmoud H~M 1954 {\em Phys. Rev.\/} {\bf 95}(1)
  217--228 \urlprefix\url{http://link.aps.org/doi/10.1103/PhysRev.95.217}

\bibitem{Brueckner:1955rw}
Brueckner K~A and Levinson C~A 1955 {\em Phys. Rev.\/} {\bf 97} 1344--1352

\bibitem{Day:1967zl}
Day B~D 1967 {\em Rev. Mod. Phys.\/} {\bf 39} 719--744

\bibitem{Barrett:1970jl}
Barrett B~R and Kirson M~W 1970 {\em Nuclear Physics A\/} {\bf 148} 145 -- 180
  ISSN 0375-9474
  \urlprefix\url{http://www.sciencedirect.com/science/article/pii/0375947470906172}

\bibitem{Kirson:1971la}
Kirson M~W 1971 {\em Annals of Physics\/} {\bf 66} 624--650
  \urlprefix\url{http://www.sciencedirect.com/science/article/pii/0003491671900728}

\bibitem{Barrett:1972bs}
Barrett B~R 1972 {\em Physics Letters B\/} {\bf 38} 371--375
  \urlprefix\url{http://www.sciencedirect.com/science/article/pii/0370269372901591}

\bibitem{Kirson:1974oq}
Kirson M~W 1974 {\em Annals of Physics\/} {\bf 82} 345 -- 368 ISSN 0003-4916
  \urlprefix\url{http://www.sciencedirect.com/science/article/pii/0003491674901171}

\bibitem{Goode:1974pi}
Goode P and Kirson M~W 1974 {\em Physics Letters B\/} {\bf 51} 221--224
  \urlprefix\url{http://www.sciencedirect.com/science/article/pii/0370269374902780}

\bibitem{Vary:1973dn}
Vary J~P, Sauer P~U and Wong C~W 1973 {\em Phys. Rev. C\/} {\bf 7}(5)
  1776--1785 \urlprefix\url{http://link.aps.org/doi/10.1103/PhysRevC.7.1776}

\bibitem{Glazek:1993il}
Glazek S~D and Wilson K~G 1993 {\em Phys. Rev. D\/} {\bf 48} 5863--5872

\bibitem{Wegner:1994dk}
Wegner F 1994 {\em Ann. Phys. (Leipzig)\/} {\bf 3} 77

\bibitem{Bogner:2006qf}
Bogner S~K, Furnstahl R~J, Ramanan S and Schwenk A 2006 {\em Nucl. Phys. A\/}
  {\bf 773} 203--220
  \urlprefix\url{http://www.sciencedirect.com/science/article/B6TVB-4K4WG34-2/2/218ed5e2e993065d0432f1ffeafdc50e}

\bibitem{Tichai:2016vl}
Tichai A, Langhammer J, Binder S and Roth R 2016 {\em Physics Letters B\/} {\bf
  756} 283--288
  \urlprefix\url{http://www.sciencedirect.com/science/article/pii/S0370269316002008}

\bibitem{Roth:2010ys}
Roth R and Langhammer J 2010 {\em Phys. Lett. B\/} {\bf 683} 272 -- 277 ISSN
  0370-2693
  \urlprefix\url{http://www.sciencedirect.com/science/article/pii/S037026930901507X}

\bibitem{Barrett:2013oq}
Barrett B~R, Navr\'{a}til P and Vary J~P 2013 {\em Prog. Part. Nucl. Phys.\/}
  {\bf 69} 131 -- 181 ISSN 0146-6410
  \urlprefix\url{http://www.sciencedirect.com/science/article/pii/S0146641012001184}

\bibitem{Jurgenson:2013fk}
Jurgenson E~D, Maris P, Furnstahl R~J, Navr\'atil P, Ormand W~E and Vary J~P
  2013 {\em Phys. Rev. C\/} {\bf 87}(5) 054312
  \urlprefix\url{http://link.aps.org/doi/10.1103/PhysRevC.87.054312}

\bibitem{Hergert:2013ij}
Hergert H, Binder S, Calci A, Langhammer J and Roth R 2013 {\em Phys. Rev.
  Lett.\/} {\bf 110}(24) 242501
  \urlprefix\url{http://link.aps.org/doi/10.1103/PhysRevLett.110.242501}

\bibitem{Roth:2014fk}
Roth R, Calci A, Langhammer J and Binder S 2014 {\em Phys. Rev. C\/} {\bf
  90}(2) 024325
  \urlprefix\url{http://link.aps.org/doi/10.1103/PhysRevC.90.024325}

\bibitem{Binder:2014fk}
Binder S, Langhammer J, Calci A and Roth R 2014 {\em Phys. Lett. B\/} {\bf 736}
  119 -- 123 ISSN 0370-2693
  \urlprefix\url{http://www.sciencedirect.com/science/article/pii/S0370269314004961}

\bibitem{Hagen:2014ve}
Hagen G, Papenbrock T, Hjorth-Jensen M and Dean D~J 2014 {\em Rept. Prog.
  Phys.\/} {\bf 77} 096302
  \urlprefix\url{http://stacks.iop.org/0034-4885/77/i=9/a=096302}

\bibitem{Hagen:2016rb}
Hagen G, Hjorth-Jensen M, Jansen G~R and Papenbrock T 2016 {\em Phys.
  Scripta\/} {\bf 91} 063006
  \urlprefix\url{http://stacks.iop.org/1402-4896/91/i=6/a=063006}

\bibitem{Tsukiyama:2011uq}
Tsukiyama K, Bogner S~K and Schwenk A 2011 {\em Phys. Rev. Lett.\/} {\bf 106}
  222502

\bibitem{Hergert:2013mi}
Hergert H, Bogner S~K, Binder S, Calci A, Langhammer J, Roth R and Schwenk A
  2013 {\em Phys. Rev. C\/} {\bf 87}(3) 034307
  \urlprefix\url{http://link.aps.org/doi/10.1103/PhysRevC.87.034307}

\bibitem{Hergert:2016jk}
Hergert H, Bogner S~K, Morris T~D, Schwenk A and Tsukiyama K 2016 {\em Physics
  Reports\/} {\bf 621} 165--222
  \urlprefix\url{http://www.sciencedirect.com/science/article/pii/S0370157315005414}

\bibitem{Brown:2014fk}
Brown B and Rae W 2014 {\em Nuclear Data Sheets\/} {\bf 120} 115 -- 118 ISSN
  0090-3752
  \urlprefix\url{http://www.sciencedirect.com/science/article/pii/S0090375214004748}

\bibitem{Kehrein:2006kx}
Kehrein S 2006 {\em The Flow Equation Approach to Many-Particle Systems\/}
  ({\em Springer Tracts in Modern Physics\/} vol 237) (Springer Berlin /
  Heidelberg)

\bibitem{Heidbrink:2002kx}
Heidbrink C and Uhrig G 2002 {\em Eur. Phys. J. B\/} {\bf 30}(4) 443--459 ISSN
  1434-6028 10.1140/epjb/e2002-00401-9
  \urlprefix\url{http://dx.doi.org/10.1140/epjb/e2002-00401-9}

\bibitem{Drescher:2011kx}
Drescher N~A, Fischer T and Uhrig G~S 2011 {\em Eur. Phys. J. B\/} {\bf 79}(2)
  225--240 ISSN 1434-6028 10.1140/epjb/e2010-10723-6
  \urlprefix\url{http://dx.doi.org/10.1140/epjb/e2010-10723-6}

\bibitem{Krull:2012bs}
Krull H, Drescher N~A and Uhrig G~S 2012 {\em Phys. Rev. B\/} {\bf 86}(12)
  125113 \urlprefix\url{http://link.aps.org/doi/10.1103/PhysRevB.86.125113}

\bibitem{Fauseweh:2013zv}
Fauseweh B and Uhrig G~S 2013 {\em Phys. Rev. B\/} {\bf 87}(18) 184406
  \urlprefix\url{http://link.aps.org/doi/10.1103/PhysRevB.87.184406}

\bibitem{Krones:2015ft}
Krones J and Uhrig G~S 2015 {\em Phys. Rev. B\/} {\bf 91}(12) 125102
  \urlprefix\url{http://link.aps.org/doi/10.1103/PhysRevB.91.125102}

\bibitem{Shavitt:2009}
Shavitt I and Bartlett R~J 2009 {\em Many-Body Methods in Chemistry and
  Physics: MBPT and Coupled-Cluster Theory\/} (Cambridge University Press)

\bibitem{White:2002fk}
White S~R 2002 {\em J. Chem. Phys.\/} {\bf 117} 7472--7482 ISSN 00219606
  \urlprefix\url{http://dx.doi.org/doi/10.1063/1.1508370}

\bibitem{Yanai:2006uq}
Yanai T and Chan G~K~L 2006 {\em J. Chem. Phys.\/} {\bf 124} 194106 (pages~16)
  \urlprefix\url{http://link.aip.org/link/?JCP/124/194106/1}

\bibitem{Yanai:2007kx}
Yanai T and Chan G~K~L 2007 {\em J. Chem. Phys.\/} {\bf 127} 104107 (pages~14)
  \urlprefix\url{http://link.aip.org/link/?JCP/127/104107/1}

\bibitem{Nakatsuji:1976yq}
Nakatsuji H 1976 {\em Phys. Rev. A\/} {\bf 14}(1) 41--50
  \urlprefix\url{http://link.aps.org/doi/10.1103/PhysRevA.14.41}

\bibitem{Valdemoro:1987zl}
Valdemoro C 1987 {\em Theory and Practice of the Spin-Adapted Reduced
  Hamiltonians (SRH)\/} (Dordrecht: Springer Netherlands) pp 275--288 ISBN
  978-94-009-3855-7
  \urlprefix\url{http://dx.doi.org/10.1007/978-94-009-3855-7_14}

\bibitem{Mukherjee:2001uq}
Mukherjee D and Kutzelnigg W 2001 {\em J. Chem. Phys.\/} {\bf 114} 2047--2061
  \urlprefix\url{http://link.aip.org/link/?JCP/114/2047/1}

\bibitem{Kutzelnigg:2002kx}
Kutzelnigg W and Mukherjee D 2002 {\em J. Chem. Phys.\/} {\bf 116} 4787--4801
  \urlprefix\url{http://link.aip.org/link/?JCP/116/4787/1}

\bibitem{Kutzelnigg:2004vn}
Kutzelnigg W and Mukherjee D 2004 {\em J. Chem. Phys.\/} {\bf 120} 7340--7349
  \urlprefix\url{http://link.aip.org/link/?JCP/120/7340/1}

\bibitem{Kutzelnigg:2004ys}
Kutzelnigg W and Mukherjee D 2004 {\em J. Chem. Phys.\/} {\bf 120} 7350--7368
  \urlprefix\url{http://link.aip.org/link/?JCP/120/7350/1}

\bibitem{Mazziotti:2006fk}
Mazziotti D~A 2006 {\em Phys. Rev. Lett.\/} {\bf 97}(14) 143002
  \urlprefix\url{http://link.aps.org/doi/10.1103/PhysRevLett.97.143002}

\bibitem{Mazziotti:2007qe}
Mazziotti D~A 2007 {\em Reduced-Density-Matrix Mechanics: With Applications to
  Many-Electron Atoms and Molecules\/} ({\em Advances in Chemical Physics\/}
  vol 134) (Wiley)

\bibitem{Evangelista:2014rq}
Evangelista F~A 2014 {\em J. Chem. Phys.\/} {\bf 141} 054109
  \urlprefix\url{http://scitation.aip.org/content/aip/journal/jcp/141/5/10.1063/1.4890660}

\bibitem{Li:2015nq}
Li C and Evangelista F~A 2015 {\em J. Chem. Theory Comput.\/} {\bf 11}
  2097--2108 \urlprefix\url{http://dx.doi.org/10.1021/acs.jctc.5b00134}

\bibitem{Li:2016rm}
Li C and Evangelista F~A 2016 {\em The Journal of Chemical Physics\/} {\bf 144}
  164114
  \urlprefix\url{http://scitation.aip.org/content/aip/journal/jcp/144/16/10.1063/1.4947218}

\bibitem{Hannon:2016lp}
Hannon K~P, Li C and Evangelista F~A 2016 {\em The Journal of Chemical
  Physics\/} {\bf 144} 204111
  \urlprefix\url{http://scitation.aip.org/content/aip/journal/jcp/144/20/10.1063/1.4951684}

\bibitem{Dyson:1949fj}
Dyson F~J 1949 {\em Phys. Rev.\/} {\bf 75}(11) 1736--1755
  \urlprefix\url{http://link.aps.org/doi/10.1103/PhysRev.75.1736}

\bibitem{Blanes:2009fk}
Blanes S, Casas F, Oteo J and Ros J 2009 {\em Phys. Rept.\/} {\bf 470} 151 --
  238 ISSN 0370-1573
  \urlprefix\url{http://www.sciencedirect.com/science/article/pii/S0370157308004092}

\bibitem{Brockett:1991kx}
Brockett R 1991 {\em Linear Algebra and its Applications\/} {\bf 146} 79 -- 91
  ISSN 0024-3795
  \urlprefix\url{http://www.sciencedirect.com/science/article/pii/002437959190021N}

\bibitem{Chu:1994vn}
Chu M~T 1994 {\em Fields Institute Communications\/} {\bf 3} 87
  \urlprefix\url{http://dx.doi.org/10.1090/fic/003}

\bibitem{Chu:1995ys}
Chu M~T 1995 {\em Linear Algebra and its Applications\/} {\bf 215} 261 -- 273
  ISSN 0024-3795
  \urlprefix\url{http://www.sciencedirect.com/science/article/pii/002437959300091D}

\bibitem{Roth:2011kx}
Roth R, Langhammer J, Calci A, Binder S and Navr\'atil P 2011 {\em Phys. Rev.
  Lett.\/} {\bf 107}(7) 072501
  \urlprefix\url{http://link.aps.org/doi/10.1103/PhysRevLett.107.072501}

\bibitem{Hergert:2014vn}
Hergert H, Bogner S~K, Morris T~D, Binder S, Calci A, Langhammer J and Roth R
  2014 {\em Phys. Rev. C\/} {\bf 90}(4) 041302
  \urlprefix\url{http://link.aps.org/doi/10.1103/PhysRevC.90.041302}

\bibitem{Hagen:2010uq}
Hagen G, Papenbrock T, Dean D~J and Hjorth-Jensen M 2010 {\em Phys. Rev. C\/}
  {\bf 82} 034330
  \urlprefix\url{http://link.aps.org/doi/10.1103/PhysRevC.82.034330}

\bibitem{Roth:2012qf}
Roth R, Binder S, Vobig K, Calci A, Langhammer J and Navr\'atil P 2012 {\em
  Phys. Rev. Lett.\/} {\bf 109}(5) 052501
  \urlprefix\url{http://link.aps.org/doi/10.1103/PhysRevLett.109.052501}

\bibitem{Binder:2013zr}
Binder S, Langhammer J, Calci A, Navr\'atil P and Roth R 2013 {\em Phys. Rev.
  C\/} {\bf 87}(2) 021303
  \urlprefix\url{http://link.aps.org/doi/10.1103/PhysRevC.87.021303}

\bibitem{Soma:2011vn}
Som\`a V, Duguet T and Barbieri C 2011 {\em Phys. Rev. C\/} {\bf 84}(6) 064317
  \urlprefix\url{http://link.aps.org/doi/10.1103/PhysRevC.84.064317}

\bibitem{Soma:2013ys}
Som\`a V, Barbieri C and Duguet T 2013 {\em Phys. Rev. C\/} {\bf 87}(1) 011303
  \urlprefix\url{http://link.aps.org/doi/10.1103/PhysRevC.87.011303}

\bibitem{Soma:2014fu}
Som\`a V, Barbieri C and Duguet T 2014 {\em Phys. Rev. C\/} {\bf 89}(2) 024323
  \urlprefix\url{http://link.aps.org/doi/10.1103/PhysRevC.89.024323}

\bibitem{Soma:2014eu}
Som\`a V, Cipollone A, Barbieri C, Navr\'atil P and Duguet T 2014 {\em Phys.
  Rev. C\/} {\bf 89}(6) 061301
  \urlprefix\url{http://link.aps.org/doi/10.1103/PhysRevC.89.061301}

\bibitem{Jurgenson:2009bs}
Jurgenson E~D, Navr\'{a}til P and Furnstahl R~J 2009 {\em Phys. Rev. Lett.\/}
  {\bf 103} 082501

\bibitem{Hebeler:2012ly}
Hebeler K 2012 {\em Phys. Rev. C\/} {\bf 85}(2) 021002
  \urlprefix\url{http://link.aps.org/doi/10.1103/PhysRevC.85.021002}

\bibitem{Griesshammer:2015dp}
Griesshammer H~W 2015 {Assessing Theory Uncertainties in EFT Power Countings
  from Residual Cutoff Dependence} {\em in proceedings of the "8th
  International Workshop on Chiral Dynamics"\/} vol PoS(CD15) p 104
  (\textit{Preprint} \eprint{1511.00490})

\bibitem{Jurgenson:2011zr}
Jurgenson E~D, Navr\'atil P and Furnstahl R~J 2011 {\em Phys. Rev. C\/} {\bf
  83}(3) 034301
  \urlprefix\url{http://link.aps.org/doi/10.1103/PhysRevC.83.034301}

\bibitem{Wendt:2013uq}
Wendt K~A 2013 {\em Advances in the Application of the Similarity
  Renormalization Group to Strongly Interacting Systems\/} Ph.D. thesis The
  Ohio State University

\bibitem{Calci:2014xy}
Calci A 2014 {\em Evolved Chiral Hamiltonians at the Three-Body Level and
  Beyond\/} Ph.D. thesis TU Darmstadt

\bibitem{Epelbaum:2002nr}
Epelbaum E, Nogga A, Gl\"ockle W, Kamada H, Mei\ss{}ner U~G and Wita\l{}a H
  2002 {\em Phys. Rev. C\/} {\bf 66}(6) 064001
  \urlprefix\url{http://link.aps.org/doi/10.1103/PhysRevC.66.064001}

\bibitem{Epelbaum:2006mo}
Epelbaum E 2006 {\em Prog. Part. Nucl. Phys.\/} {\bf 57} 654--741
  \urlprefix\url{http://www.sciencedirect.com/science/article/B6TJC-4HJ47W9-1/2/0c11dbe2861bbc64aa53c3f16416717a}

\bibitem{Gazit:2009qf}
Gazit D, Quaglioni S and Navr\'atil P 2009 {\em Phys. Rev. Lett.\/} {\bf
  103}(10) 102502
  \urlprefix\url{http://link.aps.org/doi/10.1103/PhysRevLett.103.102502}

\bibitem{Wendt:2013ys}
Wendt K~A 2013 {\em Phys. Rev. C\/} {\bf 87}(6) 061001
  \urlprefix\url{http://link.aps.org/doi/10.1103/PhysRevC.87.061001}

\bibitem{Wang:2012uq}
Wang M, Audi G, Wapstra A, Kondev F, MacCormick M, Xu X and Pfeiffer B 2012
  {\em Chin. Phys. C\/} {\bf 36} 1603
  \urlprefix\url{http://stacks.iop.org/1674-1137/36/i=12/a=003}

\bibitem{Madrid:2005rt}
de~la Madrid R 2005 {\em European Journal of Physics\/} {\bf 26} 287
  \urlprefix\url{http://stacks.iop.org/0143-0807/26/i=2/a=008}

\bibitem{Michel:2009oz}
Michel N, Nazarewicz W, P{\l}oszajczak M and Vertse T 2009 {\em J. Phys. G\/}
  {\bf 36} 013101
  \urlprefix\url{http://stacks.iop.org/0954-3899/36/i=1/a=013101}

\bibitem{Jaganathen:2014zp}
Jaganathen Y, Michel N and P\l{}oszajczak M 2014 {\em Phys. Rev. C\/} {\bf
  89}(3) 034624
  \urlprefix\url{http://link.aps.org/doi/10.1103/PhysRevC.89.034624}

\bibitem{Fossez:2015if}
Fossez K, Michel N, P\l{}oszajczak M, Jaganathen Y and Id~Betan R~M 2015 {\em
  Phys. Rev. C\/} {\bf 91}(3) 034609
  \urlprefix\url{http://link.aps.org/doi/10.1103/PhysRevC.91.034609}

\bibitem{Roth:2010vp}
Roth R, Neff T and Feldmeier H 2010 {\em Prog. Part. Nucl. Phys.\/} {\bf 65}
  50--93 (\textit{Preprint} \eprint{1003.3624})

\bibitem{Kutzelnigg:1997fk}
Kutzelnigg W and Mukherjee D 1997 {\em J. Chem. Phys.\/} {\bf 107} 432--449
  \urlprefix\url{http://link.aip.org/link/?JCP/107/432/1}

\bibitem{Kong:2010kx}
Kong L, Nooijen M and Mukherjee D 2010 {\em J. Chem. Phys.\/} {\bf 132} 234107
  (pages~8) \urlprefix\url{http://link.aip.org/link/?JCP/132/234107/1}

\bibitem{Hergert:2009wh}
Hergert H and Roth R 2009 {\em Phys. Lett. B\/} {\bf 682} 27--32
  (\textit{Preprint} \eprint{0908.1334})

\bibitem{Hagen:2007zc}
Hagen G, Papenbrock T, Dean D~J, Schwenk A, Nogga A, W\l{}och M and Piecuch P
  2007 {\em Phys. Rev. C\/} {\bf 76}(3) 034302
  \urlprefix\url{http://link.aps.org/doi/10.1103/PhysRevC.76.034302}

\bibitem{Binder:2013fk}
Binder S, Piecuch P, Calci A, Langhammer J, Navr\'atil P and Roth R 2013 {\em
  Phys. Rev. C\/} {\bf 88}(5) 054319
  \urlprefix\url{http://link.aps.org/doi/10.1103/PhysRevC.88.054319}

\bibitem{Gebrerufael:2016fe}
Gebrerufael E, Calci A and Roth R 2016 {\em Phys. Rev. C\/} {\bf 93}(3) 031301
  \urlprefix\url{http://link.aps.org/doi/10.1103/PhysRevC.93.031301}

\bibitem{Tsukiyama:2012fk}
Tsukiyama K, Bogner S~K and Schwenk A 2012 {\em Phys. Rev. C\/} {\bf 85}(6)
  061304 \urlprefix\url{http://link.aps.org/doi/10.1103/PhysRevC.85.061304}

\bibitem{Morris:2015ve}
Morris T~D, Parzuchowski N~M and Bogner S~K 2015 {\em Phys. Rev. C\/} {\bf
  92}(3) 034331
  \urlprefix\url{http://link.aps.org/doi/10.1103/PhysRevC.92.034331}

\bibitem{Brandow:1967tg}
Brandow B~H 1967 {\em Rev. Mod. Phys.\/} {\bf 39}(4) 771--828
  \urlprefix\url{http://link.aps.org/doi/10.1103/RevModPhys.39.771}

\bibitem{Fetter:2003ve}
Fetter A and Walecka J 2003 {\em Quantum Theory of Many-particle Systems\/}
  Dover Books on Physics (Dover Publications) ISBN 9780486428277
  \urlprefix\url{http://books.google.com/books?id=0wekf1s83b0C}

\bibitem{Dickhoff:2004fk}
Dickhoff W and Barbieri C 2004 {\em Prog. Part. Nucl. Phys.\/} {\bf 52} 377 --
  496 ISSN 0146-6410
  \urlprefix\url{http://www.sciencedirect.com/science/article/pii/S0146641004000535}

\bibitem{Barbieri:2007fk}
Barbieri C, Van~Neck D and Dickhoff W~H 2007 {\em Phys. Rev. A\/} {\bf 76}(5)
  052503 \urlprefix\url{http://link.aps.org/doi/10.1103/PhysRevA.76.052503}

\bibitem{Cipollone:2013uq}
Cipollone A, Barbieri C and Navr\'atil P 2013 {\em Phys. Rev. Lett.\/} {\bf
  111}(6) 062501
  \urlprefix\url{http://link.aps.org/doi/10.1103/PhysRevLett.111.062501}

\bibitem{Kutzelnigg:1982ly}
Kutzelnigg W 1982 {\em J. Chem. Phys.\/} {\bf 77} 3081--3097
  \urlprefix\url{http://link.aip.org/link/?JCP/77/3081/1}

\bibitem{Kutzelnigg:1983ve}
Kutzelnigg W and Koch S 1983 {\em J. Chem. Phys.\/} {\bf 79} 4315--4335
  \urlprefix\url{http://link.aip.org/link/?JCP/79/4315/1}

\bibitem{Kutzelnigg:1984qf}
Kutzelnigg W 1984 {\em J. Chem. Phys.\/} {\bf 80} 822--830
  \urlprefix\url{http://link.aip.org/link/?JCP/80/822/1}

\bibitem{Epstein:1926fp}
Epstein P~S 1926 {\em Phys. Rev.\/} {\bf 28}(4) 695--710
  \urlprefix\url{http://link.aps.org/doi/10.1103/PhysRev.28.695}

\bibitem{Nesbet:1955lq}
Nesbet R~K 1955 {\em Proceedings of the Royal Society of London A:
  Mathematical, Physical and Engineering Sciences\/} {\bf 230} 312--321 ISSN
  0080-4630 (\textit{Preprint}
  \eprint{http://rspa.royalsocietypublishing.org/content/230/1182/312.full.pdf})
  \urlprefix\url{http://rspa.royalsocietypublishing.org/content/230/1182/312}

\bibitem{Carlson:2015lq}
Carlson J, Gandolfi S, Pederiva F, Pieper S~C, Schiavilla R, Schmidt K~E and
  Wiringa R~B 2015 {\em Rev. Mod. Phys.\/} {\bf 87}(3) 1067--1118
  \urlprefix\url{http://link.aps.org/doi/10.1103/RevModPhys.87.1067}

\bibitem{Entem:2003th}
Entem D~R and Machleidt R 2003 {\em Phys. Rev. C\/} {\bf 68} 041001
  \urlprefix\url{http://link.aps.org/doi/10.1103/PhysRevC.68.041001}

\bibitem{Ekstrom:2013uq}
Ekstr\"om A, Baardsen G, Forss\'en C, Hagen G, Hjorth-Jensen M, Jansen G~R,
  Machleidt R, Nazarewicz W, Papenbrock T, Sarich J and Wild S~M 2013 {\em
  Phys. Rev. Lett.\/} {\bf 110}(19) 192502
  \urlprefix\url{http://link.aps.org/doi/10.1103/PhysRevLett.110.192502}

\bibitem{Hebeler:2015qr}
Hebeler K, Krebs H, Epelbaum E, Golak J and Skibi\ifmmode~\acute{n}\else
  \'{n}\fi{}ski R 2015 {\em Phys. Rev. C\/} {\bf 91}(4) 044001
  \urlprefix\url{http://link.aps.org/doi/10.1103/PhysRevC.91.044001}

\bibitem{Epelbaum:2015fb}
Epelbaum E, Krebs H and Mei{\ss}ner U~G 2015 {\em Eur. Phys. J. A\/} {\bf 51}
  1--29 \urlprefix\url{http://dx.doi.org/10.1140/epja/i2015-15053-8}

\bibitem{Binder:2016la}
Binder S, Calci A, Epelbaum E, Furnstahl R~J, Golak J, Hebeler K, Kamada H,
  Krebs H, Langhammer J, Liebig S, Maris P, Mei\ss{}ner U~G, Minossi D, Nogga
  A, Potter H, Roth R, Skibi\ifmmode~\acute{n}\else \'{n}\fi{}ski R, Topolnicki
  K, Vary J~P and Wita\l{}a H (LENPIC Collaboration) 2016 {\em Phys. Rev. C\/}
  {\bf 93}(4) 044002
  \urlprefix\url{http://link.aps.org/doi/10.1103/PhysRevC.93.044002}

\bibitem{Hagen:2015ve}
Hagen G, Ekstr\"{o}m A, Forss\'{e}n C, Jansen G~R, Nazarewicz W, Papenbrock T,
  Wendt K~A, Bacca S, Barnea N, Carlsson B, Drischler C, Hebeler K,
  Hjorth-Jensen M, Miorelli M, Orlandini G, Schwenk A and Simonis J 2015 {\em
  Nat. Phys.\/} {\bf 12} 186
  \urlprefix\url{http://dx.doi.org/10.1038/nphys3529}

\bibitem{Garcia-Ruiz:2016fk}
Garcia~Ruiz R~F, Bissell M~L, Blaum K, Ekstrom A, Frommgen N, Hagen G, Hammen
  M, Hebeler K, Holt J~D, Jansen G~R, Kowalska M, Kreim K, Nazarewicz W,
  Neugart R, Neyens G, Nortershauser W, Papenbrock T, Papuga J, Schwenk A,
  Simonis J, Wendt K~A and Yordanov D~T 2016 {\em Nat. Phys.\/} {\bf 12}
  594--598 \urlprefix\url{http://dx.doi.org/10.1038/nphys3645}

\bibitem{Coon:2012uq}
Coon S~A, Avetian M~I, Kruse M~K~G, van Kolck U, Maris P and Vary J~P 2012 {\em
  Phys. Rev. C\/} {\bf 86}(5) 054002
  \urlprefix\url{http://link.aps.org/doi/10.1103/PhysRevC.86.054002}

\bibitem{Furnstahl:2012ys}
Furnstahl R~J, Hagen G and Papenbrock T 2012 {\em Phys. Rev. C\/} {\bf 86}(3)
  031301 \urlprefix\url{http://link.aps.org/doi/10.1103/PhysRevC.86.031301}

\bibitem{More:2013bh}
More S~N, Ekstr\"om A, Furnstahl R~J, Hagen G and Papenbrock T 2013 {\em Phys.
  Rev. C\/} {\bf 87}(4) 044326
  \urlprefix\url{http://link.aps.org/doi/10.1103/PhysRevC.87.044326}

\bibitem{Furnstahl:2014vn}
Furnstahl R~J, More S~N and Papenbrock T 2014 {\em Phys. Rev. C\/} {\bf 89}(4)
  044301 \urlprefix\url{http://link.aps.org/doi/10.1103/PhysRevC.89.044301}

\bibitem{Furnstahl:2015sf}
Furnstahl R~J, Hagen G, Papenbrock T and Wendt K~A 2015 {\em Journal of Physics
  G: Nuclear and Particle Physics\/} {\bf 42} 034032
  \urlprefix\url{http://stacks.iop.org/0954-3899/42/i=3/a=034032}

\bibitem{Wendt:2015zl}
Wendt K~A, Forss\'en C, Papenbrock T and S\"a\"af D 2015 {\em Phys. Rev. C\/}
  {\bf 91}(6) 061301
  \urlprefix\url{http://link.aps.org/doi/10.1103/PhysRevC.91.061301}

\bibitem{Odell:2016qq}
Odell D, Papenbrock T and Platter L 2016 {\em Phys. Rev. C\/} {\bf 93}(4)
  044331 \urlprefix\url{http://link.aps.org/doi/10.1103/PhysRevC.93.044331}

\bibitem{Hergert:2009zn}
Hergert H and Roth R 2009 {\em Phys. Rev. C\/} {\bf 80}(2) 024312
  \urlprefix\url{http://link.aps.org/doi/10.1103/PhysRevC.80.024312}

\bibitem{Calci:2016hb}
Calci A and Roth R 2016 {\em Phys. Rev. C\/} {\bf 94}(1) 014322
  \urlprefix\url{http://link.aps.org/doi/10.1103/PhysRevC.94.014322}

\bibitem{Angeli:2013rz}
Angeli I and Marinova K~P 2013 {\em Atomic Data and Nuclear Data Tables\/} {\bf
  99} 69--95
  \urlprefix\url{http://www.sciencedirect.com/science/article/pii/S0092640X12000265}

\bibitem{Gebrerufael:2016rp}
Gebrerufael E, Vobig K, Hergert H and Roth R 2016
  \urlprefix\url{https://arxiv.org/abs/1610.05254}

\bibitem{Taube:2008kx}
Taube A~G and Bartlett R~J 2008 {\em J. Chem. Phys.\/} {\bf 128} 044110
  (pages~13) \urlprefix\url{http://link.aip.org/link/?JCP/128/044110/1}

\bibitem{Taube:2008vn}
Taube A~G and Bartlett R~J 2008 {\em J. Chem. Phys.\/} {\bf 128} 044111
  (pages~9) \urlprefix\url{http://link.aip.org/link/?JCP/128/044111/1}

\bibitem{Roth:2006lr}
Roth R, Papakonstantinou P, Paar N, Hergert H, Neff T and Feldmeier H 2006 {\em
  Phys. Rev. C\/} {\bf 73} 044312 (\textit{Preprint} \eprint{nucl-th/0510036})
  \urlprefix\url{http://link.aps.org/doi/10.1103/PhysRevC.73.044312}

\bibitem{Guenther:2010ge}
G{\"u}nther A, Roth R, Hergert H and Reinhardt S 2010 {\em Phys. Rev. C\/} {\bf
  82} 024319 (\textit{Preprint} \eprint{1005.1599})
  \urlprefix\url{http://link.aps.org/doi/10.1103/PhysRevC.82.024319}

\bibitem{Hebeler:2011dq}
Hebeler K, Bogner S~K, Furnstahl R~J, Nogga A and Schwenk A 2011 {\em Phys.
  Rev. C\/} {\bf 83}(3) 031301
  \urlprefix\url{http://link.aps.org/doi/10.1103/PhysRevC.83.031301}

\bibitem{Langhammer:2012uq}
Langhammer J, Roth R and Stumpf C 2012 {\em Phys. Rev. C\/} {\bf 86}(5) 054315
  \urlprefix\url{http://link.aps.org/doi/10.1103/PhysRevC.86.054315}

\bibitem{Cipollone:2015fk}
Cipollone A, Barbieri C and Navr\'atil P 2015 {\em Phys. Rev. C\/} {\bf 92}(1)
  014306 \urlprefix\url{http://link.aps.org/doi/10.1103/PhysRevC.92.014306}

\bibitem{Otsuka:2010cr}
Otsuka T, Suzuki T, Holt J~D, Schwenk A and Akaishi Y 2010 {\em Phys. Rev.
  Lett.\/} {\bf 105}(3) 032501
  \urlprefix\url{http://link.aps.org/doi/10.1103/PhysRevLett.105.032501}

\bibitem{Hagen:2012oq}
Hagen G, Hjorth-Jensen M, Jansen G~R, Machleidt R and Papenbrock T 2012 {\em
  Phys. Rev. Lett.\/} {\bf 108}(24) 242501
  \urlprefix\url{http://link.aps.org/doi/10.1103/PhysRevLett.108.242501}

\bibitem{Epelbaum:2014kx}
Epelbaum E, Krebs H, L\"ahde T~A, Lee D, Mei\ss{}ner U~G and Rupak G 2014 {\em
  Phys. Rev. Lett.\/} {\bf 112}(10) 102501
  \urlprefix\url{http://link.aps.org/doi/10.1103/PhysRevLett.112.102501}

\bibitem{Holt:2013fk}
Holt J, Men\'{e}ndez J and Schwenk A 2013 {\em Eur. Phys. J. A\/} {\bf 49} 1--6
  ISSN 1434-6001 \urlprefix\url{http://dx.doi.org/10.1140/epja/i2013-13039-2}

\bibitem{Bogner:2014tg}
Bogner S~K, Hergert H, Holt J~D, Schwenk A, Binder S, Calci A, Langhammer J and
  Roth R 2014 {\em Phys. Rev. Lett.\/} {\bf 113}(14) 142501
  \urlprefix\url{http://link.aps.org/doi/10.1103/PhysRevLett.113.142501}

\bibitem{Stroberg:2016fk}
Stroberg S~R, Hergert H, Holt J~D, Bogner S~K and Schwenk A 2016 {\em Phys.
  Rev. C\/} {\bf 93}(5) 051301
  \urlprefix\url{http://link.aps.org/doi/10.1103/PhysRevC.93.051301}

\bibitem{Roth:2007fk}
Roth R and Navr\'atil P 2007 {\em Phys. Rev. Lett.\/} {\bf 99}(9) 092501
  \urlprefix\url{http://link.aps.org/doi/10.1103/PhysRevLett.99.092501}

\bibitem{Roth:2009eu}
Roth R 2009 {\em Phys. Rev. C\/} {\bf 79} 064324
  \urlprefix\url{http://link.aps.org/doi/10.1103/PhysRevC.79.064324}

\bibitem{Piecuch:2005dp}
Piecuch P and W\l{}och M 2005 {\em J. Chem. Phys.\/} {\bf 123} 224105
  \urlprefix\url{http://scitation.aip.org/content/aip/journal/jcp/123/22/10.1063/1.2137318}

\bibitem{Lee:2009bh}
Lee D 2009 {\em Prog. Part. Nucl. Phys.\/} {\bf 63} 117 -- 154 ISSN 0146-6410
  \urlprefix\url{http://www.sciencedirect.com/science/article/pii/S014664100800094X}

\bibitem{Hoffman:2008ly}
Hoffman C~R, Baumann T, Bazin D, Brown J, Christian G, DeYoung P~A, Finck J~E,
  Frank N, Hinnefeld J, Howes R, Mears P, Mosby E, Mosby S, Reith J, Rizzo B,
  Rogers W~F, Peaslee G, Peters W~A, Schiller A, Scott M~J, Tabor S~L,
  Thoennessen M, Voss P~J and Williams T 2008 {\em Phys. Rev. Lett.\/} {\bf
  100}(15) 152502
  \urlprefix\url{http://link.aps.org/doi/10.1103/PhysRevLett.100.152502}

\bibitem{Kohley:2015ay}
Kohley Z, Baumann T, Christian G, DeYoung P~A, Finck J~E, Frank N, Luther B,
  Lunderberg E, Jones M, Mosby S, Smith J~K, Spyrou A and Thoennessen M 2015
  {\em Phys. Rev. C\/} {\bf 91}(3) 034323
  \urlprefix\url{http://link.aps.org/doi/10.1103/PhysRevC.91.034323}

\bibitem{Caesar:2013qf}
Caesar C, Simonis J, Adachi T, Aksyutina Y, Alcantara J, Altstadt S,
  Alvarez-Pol H, Ashwood N, Aumann T, Avdeichikov V, Barr M, Beceiro S,
  Bemmerer D, Benlliure J, Bertulani C~A, Boretzky K, Borge M~J~G, Burgunder G,
  Caamano M, Casarejos E, Catford W, Cederk\"all J, Chakraborty S, Chartier M,
  Chulkov L, Cortina-Gil D, Datta~Pramanik U, Diaz~Fernandez P, Dillmann I,
  Elekes Z, Enders J, Ershova O, Estrade A, Farinon F, Fraile L~M, Freer M,
  Freudenberger M, Fynbo H~O~U, Galaviz D, Geissel H, Gernh\"auser R, Golubev
  P, Gonzalez~Diaz D, Hagdahl J, Heftrich T, Heil M, Heine M, Heinz A,
  Henriques A, Holl M, Holt J~D, Ickert G, Ignatov A, Jakobsson B, Johansson
  H~T, Jonson B, Kalantar-Nayestanaki N, Kanungo R, Kelic-Heil A, Kn\"obel R,
  Kr\"oll T, Kr\"ucken R, Kurcewicz J, Labiche M, Langer C, Le~Bleis T, Lemmon
  R, Lepyoshkina O, Lindberg S, Machado J, Marganiec J, Maroussov V, Men\'endez
  J, Mostazo M, Movsesyan A, Najafi A, Nilsson T, Nociforo C, Panin V, Perea A,
  Pietri S, Plag R, Prochazka A, Rahaman A, Rastrepina G, Reifarth R, Ribeiro
  G, Ricciardi M~V, Rigollet C, Riisager K, R\"oder M, Rossi D, Sanchez~del Rio
  J, Savran D, Scheit H, Schwenk A, Simon H, Sorlin O, Stoica V, Streicher B,
  Taylor J, Tengblad O, Terashima S, Thies R, Togano Y, Uberseder E, Van~de
  Walle J, Velho P, Volkov V, Wagner A, Wamers F, Weick H, Weigand M, Wheldon
  C, Wilson G, Wimmer C, Winfield J~S, Woods P, Yakorev D, Zhukov M~V, Zilges
  A, Zoric M and Zuber K (R3B collaboration) 2013 {\em Phys. Rev. C\/} {\bf
  88}(3) 034313
  \urlprefix\url{http://link.aps.org/doi/10.1103/PhysRevC.88.034313}

\bibitem{Lunderberg:2012cr}
Lunderberg E, DeYoung P~A, Kohley Z, Attanayake H, Baumann T, Bazin D,
  Christian G, Divaratne D, Grimes S~M, Haagsma A, Finck J~E, Frank N, Luther
  B, Mosby S, Nagi T, Peaslee G~F, Schiller A, Snyder J, Spyrou A, Strongman
  M~J and Thoennessen M 2012 {\em Phys. Rev. Lett.\/} {\bf 108}(14) 142503
  \urlprefix\url{http://link.aps.org/doi/10.1103/PhysRevLett.108.142503}

\bibitem{Schuster:2014oq}
Schuster M~D, Quaglioni S, Johnson C~W, Jurgenson E~D and Navr\'atil P 2014
  {\em Phys. Rev. C\/} {\bf 90}(1) 011301
  \urlprefix\url{http://link.aps.org/doi/10.1103/PhysRevC.90.011301}

\bibitem{Lapoux:2016xu}
Lapoux V, Som\`a V, Barbieri C, Hergert H, Holt J~D and Stroberg S~R 2016 {\em
  Phys. Rev. Lett.\/} {\bf 117}(5) 052501
  \urlprefix\url{http://link.aps.org/doi/10.1103/PhysRevLett.117.052501}

\bibitem{Wienholtz:2013bh}
Wienholtz F, Beck D, Blaum K, Borgmann C, Breitenfeldt M, Cakirli R~B, George
  S, Herfurth F, Holt J~D, Kowalska M, Kreim S, Lunney D, Manea V, Menendez J,
  Neidherr D, Rosenbusch M, Schweikhard L, Schwenk A, Simonis J, Stanja J, Wolf
  R~N and Zuber K 2013 {\em Nature\/} {\bf 498} 346--349
  \urlprefix\url{http://dx.doi.org/10.1038/nature12226}

\bibitem{Hagen:2012nx}
Hagen G, Hjorth-Jensen M, Jansen G~R, Machleidt R and Papenbrock T 2012 {\em
  Phys. Rev. Lett.\/} {\bf 109}(3) 032502
  \urlprefix\url{http://link.aps.org/doi/10.1103/PhysRevLett.109.032502}

\bibitem{Holt:2012fk}
Holt J~D, Otsuka T, Schwenk A and Suzuki T 2012 {\em J. Phys. G\/} {\bf 39}
  085111 \urlprefix\url{http://stacks.iop.org/0954-3899/39/i=8/a=085111}

\bibitem{Holt:2014vn}
Holt J~D, Men\'endez J, Simonis J and Schwenk A 2014 {\em Phys. Rev. C\/} {\bf
  90}(2) 024312
  \urlprefix\url{http://link.aps.org/doi/10.1103/PhysRevC.90.024312}

\bibitem{Gallant:2012kx}
Gallant A~T, Bale J~C, Brunner T, Chowdhury U, Ettenauer S, Lennarz A,
  Robertson D, Simon V~V, Chaudhuri A, Holt J~D, Kwiatkowski A~A, Man\'e E,
  Men\'endez J, Schultz B~E, Simon M~C, Andreoiu C, Delheij P, Pearson M~R,
  Savajols H, Schwenk A and Dilling J 2012 {\em Phys. Rev. Lett.\/} {\bf
  109}(3) 032506
  \urlprefix\url{http://link.aps.org/doi/10.1103/PhysRevLett.109.032506}

\bibitem{Steppenbeck:2013dq}
Steppenbeck D, Takeuchi S, Aoi N, Doornenbal P, Matsushita M, Wang H, Baba H,
  Fukuda N, Go S, Honma M, Lee J, Matsui K, Michimasa S, Motobayashi T,
  Nishimura D, Otsuka T, Sakurai H, Shiga Y, Soderstrom P~A, Sumikama T, Suzuki
  H, Taniuchi R, Utsuno Y, Valiente-Dobon J~J and Yoneda K 2013 {\em Nature\/}
  {\bf 502} 207--210 \urlprefix\url{http://dx.doi.org/10.1038/nature12522}

\bibitem{Kreim:2014cr}
Kreim K, Bissell M, Papuga J, Blaum K, Rydt M~D, Ruiz R~G, Goriely S, Heylen H,
  Kowalska M, Neugart R, Neyens G, N{\"o}rtersh{\"a}user W, Rajabali M,
  Alarc{\'o}n R~S, Stroke H and Yordanov D 2014 {\em Phys. Lett. B\/} {\bf 731}
  97 -- 102 ISSN 0370-2693
  \urlprefix\url{http://www.sciencedirect.com/science/article/pii/S0370269314001038}

\bibitem{Chiara:2012ys}
Chiara C~J, Broda R, Walters W~B, Janssens R~V~F, Albers M, Alcorta M, Bertone
  P~F, Carpenter M~P, Hoffman C~R, Lauritsen T, Rogers A~M, Seweryniak D, Zhu
  S, Kondev F~G, Fornal B, Kr\'olas W, Wrzesi\ifmmode~\acute{n}\else
  \'{n}\fi{}ski J, Larson N, Liddick S~N, Prokop C, Suchyta S, David H~M and
  Doherty D~T 2012 {\em Phys. Rev. C\/} {\bf 86}(4) 041304
  \urlprefix\url{http://link.aps.org/doi/10.1103/PhysRevC.86.041304}

\bibitem{Recchia:2013kx}
Recchia F, Chiara C~J, Janssens R~V~F, Weisshaar D, Gade A, Walters W~B, Albers
  M, Alcorta M, Bader V~M, Baugher T, Bazin D, Berryman J~S, Bertone P~F, Brown
  B~A, Campbell C~M, Carpenter M~P, Chen J, Crawford H~L, David H~M, Doherty
  D~T, Hoffman C~R, Kondev F~G, Korichi A, Langer C, Larson N, Lauritsen T,
  Liddick S~N, Lunderberg E, Macchiavelli A~O, Noji S, Prokop C, Rogers A~M,
  Seweryniak D, Stroberg S~R, Suchyta S, Williams S, Wimmer K and Zhu S 2013
  {\em Phys. Rev. C\/} {\bf 88}(4) 041302
  \urlprefix\url{http://link.aps.org/doi/10.1103/PhysRevC.88.041302}

\bibitem{Suchyta:2014vn}
Suchyta S, Liddick S~N, Tsunoda Y, Otsuka T, Bennett M~B, Chemey A, Honma M,
  Larson N, Prokop C~J, Quinn S~J, Shimizu N, Simon A, Spyrou A, Tripathi V,
  Utsuno Y and VonMoss J~M 2014 {\em Phys. Rev. C\/} {\bf 89}(2) 021301
  \urlprefix\url{http://link.aps.org/doi/10.1103/PhysRevC.89.021301}

\bibitem{Marinova:2011lo}
Marinova K, Geithner W, Kowalska M, Blaum K, Kappertz S, Keim M, Kloos S,
  Kotrotsios G, Lievens P, Neugart R, Simon H and Wilbert S 2011 {\em Phys.
  Rev. C\/} {\bf 84}(3) 034313
  \urlprefix\url{http://link.aps.org/doi/10.1103/PhysRevC.84.034313}

\bibitem{Gibelin:2007jw}
Gibelin J, Beaumel D, Motobayashi T, Aoi N, Baba H, Blumenfeld Y, Dombr\'adi Z,
  Elekes Z, Fortier S, Frascaria N, Fukuda N, Gomi T, Ishikawa K, Kondo Y, Kubo
  T, Lima V, Nakamura T, Saito A, Satou Y, Takeshita E, Takeuchi S, Teranishi
  T, Togano Y, Vinodkumar A~M, Yanagisawa Y and Yoshida K 2007 {\em Phys. Rev.
  C\/} {\bf 75}(5) 057306
  \urlprefix\url{http://link.aps.org/doi/10.1103/PhysRevC.75.057306}

\bibitem{Lepailleur:2013kh}
Lepailleur A, Sorlin O, Caceres L, Bastin B, Borcea C, Borcea R, Brown B~A,
  Gaudefroy L, Gr\'evy S, Grinyer G~F, Hagen G, Hjorth-Jensen M, Jansen G~R,
  Llidoo O, Negoita F, de~Oliveira F, Porquet M~G, Rotaru F, Saint-Laurent M~G,
  Sohler D, Stanoiu M and Thomas J~C 2013 {\em Phys. Rev. Lett.\/} {\bf 110}(8)
  082502 \urlprefix\url{http://link.aps.org/doi/10.1103/PhysRevLett.110.082502}

\bibitem{Caceres:2015fk}
C\'{a}ceres L, Lepailleur A, Sorlin O, Stanoiu M, Sohler D, Dombr\'{a}di Z,
  Bogner S~K, Brown B~A, Hergert H, Holt J~D, Schwenk A, Azaiez F, Bastin B,
  Borcea C, Borcea R, Bourgeois C, Elekes Z, F\"{u}l\"{o}p Z, Gr\'{e}vy S,
  Gaudefroy L, Grinyer G~F, Guillemaud-Mueller D, Ibrahim F, Kerek A,
  Krasznahorkay A, Lewitowicz M, Lukyanov S~M, Mr\'{a}zek J, Negoita F,
  de~Oliveira F, Penionzhkevich Y~E, Podoly\'{a}k Z, Porquet M~G, Rotaru F,
  Roussel-Chomaz P, Saint-Laurent M~G, Savajols H, Sletten G, Thomas J~C,
  Tim\`{a}r J, Timis C and Vajta Z 2015 {\em Phys. Rev. C\/} {\bf 92}(1) 014327
  \urlprefix\url{http://link.aps.org/doi/10.1103/PhysRevC.92.014327}

\bibitem{Jansen:2014qf}
Jansen G~R, Engel J, Hagen G, Navratil P and Signoracci A 2014 {\em Phys. Rev.
  Lett.\/} {\bf 113}(14) 142502
  \urlprefix\url{http://link.aps.org/doi/10.1103/PhysRevLett.113.142502}

\bibitem{Lisetskiy:2008fk}
Lisetskiy A~F, Barrett B~R, Kruse M~K~G, Navr\'{a}til P, Stetcu I and Vary J~P
  2008 {\em Phys. Rev. C\/} {\bf 78}(4) 044302
  \urlprefix\url{http://link.aps.org/doi/10.1103/PhysRevC.78.044302}

\bibitem{Dikmen:2015fk}
Dikmen E, Lisetskiy A~F, Barrett B~R, Maris P, Shirokov A~M and Vary J~P 2015
  {\em Phys. Rev. C\/} {\bf 91}(6) 064301
  \urlprefix\url{http://link.aps.org/doi/10.1103/PhysRevC.91.064301}

\bibitem{NuDat:2014}
National Nuclear Data Center, information extracted from the NuDat 2 database,
  http://www.nndc.bnl.gov/nudat2/
  \urlprefix\url{http://www.nndc.bnl.gov/nudat2/}

\bibitem{Jansen:2016kq}
Jansen G~R, Schuster M~D, Signoracci A, Hagen G and Navr\'atil P 2016 {\em
  Phys. Rev. C\/} {\bf 94}(1) 011301
  \urlprefix\url{http://link.aps.org/doi/10.1103/PhysRevC.94.011301}

\bibitem{Jansen:2013zr}
Jansen G~R 2013 {\em Phys. Rev. C\/} {\bf 88}(2) 024305
  \urlprefix\url{http://link.aps.org/doi/10.1103/PhysRevC.88.024305}

\bibitem{Oertzen:2006ec}
von Oertzen W, Freer M and Kanada-En'yo Y 2006 {\em Physics Reports\/} {\bf
  432} 43--113
  \urlprefix\url{http://www.sciencedirect.com/science/article/pii/S0370157306002626}

\bibitem{Ring:1980bb}
Ring P and Schuck P 1980 {\em The Nuclear Many-Body Problem\/} 1st ed
  (Springer)

\bibitem{Stroberg:2016th}
Stroberg S~R, Calci A, Hergert H, Holt J~D, Bogner S~K, Roth R and Schwenk A
  2016  (\textit{Preprint} \eprint{1607.03229})

\bibitem{Parzuchowski:2016pi}
Parzuchowski N~M, Morris T~D and Bogner S~K 2016  (\textit{Preprint}
  \eprint{1611.00661})

\bibitem{Sheikh:2000xx}
Sheikh J~A and Ring P 2000 {\em Nucl. Phys. A\/} {\bf 665} 71--91
  (\textit{Preprint} \eprint{nucl-th/9907065})

\end{thebibliography}

\end{document}